\documentclass[12pt,preprint]{aastex}

\newcommand\Rsun{\hbox{$R_\odot$}}

\newcommand\Rstar{\hbox{$R_*$}}

\newcommand\Teff{\hbox{$T_{eff}$}}
\newcommand\TL{\hbox{$T_L$}}
\newcommand\TX{\hbox{$T_X$}}
\newcommand\THHe{\hbox{$T_{HHe}$}}

\newcommand\VP{\hbox{$V_{\rm P}$}}
\newcommand\HWHM{\hbox{$HWHM$}}
\newcommand\EMX{\hbox{$EM_{\rm X}$}}
\newcommand\fir{\hbox{$fir$}}

 \newcommand\etal{\hbox{et~al.}}
\newcommand\ROSAT{\hbox{\it ROSAT}}

\newcommand\Chandra{\hbox{\it Chandra}}
\newcommand\Copernicus{\hbox{\it Copernicus}}
\newcommand\Einstein{\hbox{\it Einstein}}

\newcommand\NVII{\hbox{N {\sc vii}}}

\newcommand\OVI{\hbox{O {\sc vi}}}
\newcommand\OVII{\hbox{O {\sc vii}}}
\newcommand\OVIII{\hbox{O {\sc viii}}}
\newcommand\MgXI{\hbox{Mg {\sc xi}}}
\newcommand\MgXII{\hbox{Mg {\sc xii}}}
\newcommand\FeXVII{\hbox{Fe {\sc xvii}}}
\newcommand\NeIX{\hbox{Ne {\sc ix}}}
\newcommand\NeX{\hbox{Ne {\sc x}}}
\newcommand\SiXIII{\hbox{Si {\sc xiii}}}
\newcommand\SiXIV{\hbox{Si {\sc xiv}}}
\newcommand\SXV{\hbox{S {\sc xv}}}
\newcommand\SXVI{\hbox{S {\sc xvi}}}
\newcommand\ArXVII{\hbox{Ar {\sc xvii}}}

\newcommand\ftoi{\hbox{$f/i$}}
\newcommand\PV{\hbox{P {\sc v}}}
\newcommand\HtoHe{\hbox{$H/He$}}
\newcommand\RRR{\hbox{\sf R}}
\newcommand\zpup{\hbox{$\zeta\ ${\rm Pup}}}
\newcommand\zori{\hbox{$\zeta\ ${\rm Ori}}}
\newcommand\zoriA{\hbox{$\zeta\ ${\rm Ori A}}}

\newcommand\cygate{\hbox{Cyg OB2 No. 8a}}
\newcommand\cygnin{\hbox{Cyg OB2 No. 9}}
\newcommand\dori{\hbox{$\delta\ ${\rm Ori}}}
\newcommand\doriA{\hbox{$\delta\ ${\rm Ori A}}}
\newcommand\tsco{\hbox{$\tau\ ${\rm Sco}}}
\newcommand\toriC{\hbox{$\theta^1$ Ori C}}
\newcommand\eori{\hbox{$\epsilon\ ${\rm Ori}}}
\newcommand\sori{\hbox{$\sigma\ ${\rm Ori}}}
\newcommand\iori{\hbox{$\iota\ ${\rm Ori}}}
\newcommand\zoph{\hbox{$\zeta\ ${\rm Oph}}}

\newcommand\bcru{\hbox{$\beta\ ${\rm Cru}}}
\newcommand\xper{\hbox{$\xi\ ${\rm Per}}}
\newcommand\Berghofer{\hbox{Bergh\"{o}fer}}
\newcommand\vinf{\hbox{$v_\infty$}}
\newcommand\Mdot{{\hbox{$\dot M$}}}
\newcommand\Msunyr{\hbox{$M_\odot\,$yr$^{-1}$}}
\newcommand\kms{\hbox{km$\,$s$^{-1}$}}
\newcommand\LXLBol{{\hbox{$L_X/L_{Bol}$}}}

\newcommand\lamz{\hbox{$\lambda_{\circ}$}}

\newcommand\Rfir{\hbox{$R_{fir}$}}
\newcommand\Rtau{\hbox{$R_{\tau = 1}$}}
\newcommand\UO{\hbox{$U_{rel}$}}

\slugcomment{Published in 2007, ApJ, 668, 456; ERRATUM (see Appendix) scheduled for
2008, ApJ, 680}
\shorttitle{X-ray Line Emission from OB Stars}
\shortauthors{Waldron and Cassinelli}

\begin{document}

\title{An Extensive Collection of Stellar Wind X-ray Source Region Emission Line
Parameters, Temperatures, Velocities, and Their Radial Distributions as Obtained from Chandra
Observations of 17 OB Stars\\ }

\author {W.~L. Waldron$^{1}$ and J.~P. Cassinelli$^{2}$\\
$^{1}$Eureka Scientific Inc., 2452 Delmer St., Oakland CA, 94602;
wwaldron@satx.rr.com\\
$^{2}$Dept. of Astronomy, University of Wisconsin-Madison, Madison, WI 53711;
cassinelli@astro.wisc.edu}

\begin{abstract}
{\it Chandra} high energy resolution observations have now been obtained from numerous
non-peculiar O and early B stars. The observed X-ray emission line properties differ from
pre-launch predictions, and the interpretations are still problematic. We present a straightforward
analysis of a broad collection of OB stellar line profile data to search for morphological trends.
X-ray line emission parameters and the spatial distributions of derived quantities are
examined with respect to luminosity class. The X-ray source locations and their corresponding
temperatures are extracted by using the He-like $f/i$ line ratios and the H-like to He-like line
ratios respectively. Our luminosity class study reveals line widths increasing with luminosity. 
Although the majority of the OB emission lines are found to be symmetric, with little central line
displacement, there is evidence for small, but finite, blue-ward line-shifts that also increase with
luminosity. The spatial X-ray temperature distributions indicate that the highest temperatures
occur near the star and steadily decrease outward. This trend is most pronounced in the OB
supergiants. For the lower density wind stars, both high and low X-ray source temperatures exist
near the star. However, we find no evidence of any high temperature X-ray emission in the outer
wind regions for any OB star. Since the temperature distributions are counter to basic shock
model predictions, we call this the {\it ``near-star high-ion problem''} for OB stars. By invoking
the traditional OB stellar mass loss rates, we find a good correlation between the $fir$-inferred
radii and their associated X-ray continuum optical depth unity radii. We conclude by presenting
some possible explanations to the X-ray source problems that have been revealed by this study.
\end{abstract}


\section{Introduction}

The \Chandra\ Satellite has provided astronomers the ability to study the high energy resolution
X-ray emission line spectra of numerous stars other than the Sun.  With regards to 
early-type stars (hereafter OB stars), these line profiles allow us to study the complex stellar wind
distribution of the X-ray source regions. The X-ray line profile shapes provide information on the
Doppler velocities associated with the line formation regions, and allow us to comment on the
shocks embedded in these stellar winds.  In addition, emission line flux ratios allow us to
explore the radial distributions of the source regions and their associated X-ray temperatures.


The main goal of this paper is to present the X-ray data from a large collection of ``normal``
OB stars and use the data to search for trends in the X-ray emission line parameters as a
function of spectral type and luminosity class. Morphological trends in the X-ray emission line
characteristics can help define important properties of X-ray emission that will be used to
constrain the subsequent X-ray source modeling efforts.  Although this type of study involves
fewer assumptions than used in the modeling of each specific star, we suggest that it
provides a broader, more comprehensive view of the X-ray source region characteristics. 

In our analysis we will also be interested in comparing how X-ray parameters that can be
expressed in terms of velocities (i.e., line widths, line shifts, and X-ray temperatures) relate to the
stellar wind terminal velocities and radial velocity structures. There are a variety of ways to
discuss the data within the framework of several different shock pictures: 1) an ambient stellar
wind impinging on dense blobs (Lucy \& White 1980); 2) a distribution of ``saw-tooth`` forward
shocks (Lucy 1982); 3) a wind running into a driven wave which produces forward and reverse
shocks (MacFarlane \& Cassinelli 1989), and; 4) a detailed hydrodynamic numerical simulation of
the line-driven instability which incorporates the complex time-dependence of the line driving
force which also produces forward and reverse shocks (Owocki, Castor, \& Rybicki 1988;
Feldmeier 1995).  The primary difference between these models is that the line-driven instability,
or more appropriately the ``de-shadowing instability``, accelerates the pre-shock wind plasma to
velocities that are larger than the ambient wind velocities resulting in a highly rarefied wind
structure prior to the shock jump, and the post-shock velocity (i.e., the velocity of the X-ray
emitting plasma) is predicted to be comparable to the ambient wind velocity. In the other shock
models, the post-shock velocity is some fraction ($< 1$, with a typical value of $\sim 0.5$) of the
ambient wind velocity. Correspondingly, the post-shock density (i.e., the density of the X-ray
emitting plasma) of the de-shadowing model is then also comparable to the ambient wind density,
whereas, for the other shock models, the post-shock density is a factor of four times larger than
the ambient wind density as determined by the Rankine-Hugoniot equations when the pre-shock
Mach number is very large. Since our goal is to provide a global view of the OB stellar X-ray
emission line properties and their dependencies on stellar parameters, we choose to use what we
will refer to as a {\it ``basic shock model``} description in our discussion of the relationships
between the X-ray and wind velocity parameters. The premise of this model is in-line with the
blob, saw-tooth, and driven wave shock models where the pre-shock velocity at any given radius
is equal to the local ambient wind velocity as defined  by the standard $\beta$-law velocity
structure.  

In \S 2 we provide a summary of the general OB stellar X-ray emission properties as determined
from previous studies.  In \S 3 we present our selected stars, data reduction procedures, the X-ray
lines that are used, and a description of the basic X-ray emission line parameters for our line
fitting procedure. Using our \Chandra\ High Energy Transmission Grating Spectrometer
(HETGS) data along with the available archived \Chandra\ data, a total of 17 OB stars are studied
in this analysis. The observed velocity information (half-width-half-maximum and line-shifts of the
line emission peak) displayed as histogram distributions, illustrating their dependence on
luminosity class, are presented in \S 4. In \S 5 we present our discussion and analysis of X-ray
emission line ratios and their application towards estimating the values of the source temperatures
and the predominant stellar wind X-ray source locations.  In \S 6 we focus on deriving the radial
temperature distributions of the OB X-ray sources. We find that this confirms the {\it "near-star
high-ion problem"} found in earlier isolated stellar studies. The results and conclusions are
summarized in \S 7.

\section{Summary of OB Stellar X-ray Emission}

Spectral lines and their profiles have been providing crucial information about OB stellar X-ray
astronomy since the beginning of the field. Broad UV spectral lines of superionization stages such
as \OVI\ were discovered in \Copernicus\ observations (Lamers \& Morton 1976).  These broad
profiles and the persistence of the superionization stages over a well delimited range in spectral
types led to the realization that these ions were produced by the Auger Effect in which two
electrons are removed from the dominant stages of ionization following K-shell absorption of
X-rays (Cassinelli \& Olson 1979).  These authors suggested that the X-rays responsible for the
superionization could come from a hot corona at the base of the cool wind. Soon thereafter the
\Einstein\ Observatory discovered X-rays from OB stars (Harnden \etal\ 1979; Seward \etal\
1979).  However, the observations did not show the expected large attenuation of soft X-rays by
the overlying cool wind. This led to the realization that the X-ray emission must arise from shock
structures embedded within the stellar wind (Lucy \& White 1980). The origin of these shocks is 
associated with the instability of line driven winds to velocity distribution disturbances (Lucy \&
Solomon 1970). In the Lucy \& White case, the shocks are bow-shocks around radiative-driven
clumps, whereas in the model of Lucy (1982), the X-rays are formed in a periodic shock
structure. Spectral lines at X-ray wavelengths were detected by the \Einstein\ Sold State
Spectrometer (SSS) (Cassinelli \& Swank 1983). They found evidence of high energy ion {\it line
emission} (\SiXIII\ \& \SXV) from the late O-supergiant, \zori. Since these lines are formed at
such high temperatures, they suggested that these ions are located in magnetically confined
regions at the base of the wind. 
In addition, since the overall X-ray fluxes from OB stars are less variable than would be
expected from spherical shells of shocked wind material, Cassinelli \& Swank also suggested that
the shocks were fragmentary in form, such that there would be a statistically steady supply of
X-ray emission from these wind distributed ``shock fragments``. Some of these early ideas
regarding X-ray source regions have persisted to the present time.

The \ROSAT\ Satellite had a greater sensitivity than the \Einstein\ Satellite, although not the
energy resolution of the SSS instrument. In a survey of the Bright Star Catalogue OB stars,
\Berghofer\ \etal\ (1996 \& 1997) confirmed that all O-stars are X-rays sources, with X-ray
luminosities following the ``hot-star law`` $\LXLBol = 10^{-7}$ until about B1. For the later B
stars, Cohen et al. (1996) found a rapid decrease in X-ray luminosity with spectral type which
could be explained primarily by the slower speeds of the B star winds and the reduced mass loss
rates.

With \Chandra, high spectral resolution of X-ray line emission from OB stars has now become
observable with the HETGS Medium Energy Grating (MEG) and High Energy Grating (HEG)
detectors. The impact of this diagnostic capability became very clear in the first detailed study of
HETGS data from an O-star (\zori) by Waldron \& Cassinelli (2001).  Their analysis found three
fundamental unexpected results associated with: 1) X-ray line profile shape characteristics; 2) the
correlation between the wind X-ray source locations and their respective X-ray continuum
optical depth unity radii, and; 3) the presence of deeply embedded high energy ionization stages.

One of the biggest surprises from \Chandra\ and {\it XMM-Newton} high energy resolution
observations of OB stars is the absence of blue-shifted, asymmetric X-ray emission lines.  These
resolved spectral lines had been predicted by MacFarlane \etal\ (1991) to be broad and skewed
towards the blue, owing to the fact that the long-ward (``red-ward``) radiation from the back side
of the star is more strongly attenuated than the short-ward (``blue-ward``) radiation from the near
side. In the Waldron \& Cassinelli (2001) analysis of \zori\, broad line profiles were seen, but
these lines lacked the predicted skewness and blue-shifts. In the case of the O4f star \zpup\ the
lines are somewhat similar to the predicted blue-shifts (Kahn et al. 2001; Cassinelli et al. 2001). 
However, nearly all other OB stars show minimal blue-shifted lines (within the resolving 
power of the instruments), and the lines are essentially symmetric (Waldron \& Cassinelli
2002; Miller et al. 2002; Cohen et al. 2003; Mewe et al. 2003; Schulz et al. 2003; Waldron et al.
2004; Gagne et al. 2005). Hence, explaining the lack of substantial blue-shifted lines that are
presumably formed in an outflowing wind has become one of the most perplexing problems to
emerge from high spectral resolution X-ray observations.

Waldron \& Cassinelli (2001) were the first to demonstrate that a stellar wind distribution of
X-ray sources distributed above 1.5 stellar radii could explain the observed lines profile shapes
seen in \zori\, {\it if the stellar wind mass loss rate is at least an order of magnitude smaller
than previously thought}. Since then, several studies (e.g., Kramer, Cohen \& Owocki 2003;
Leutenegger et al. 2006; Cohen et al. 2006) have found that reduced mass loss rates are
needed to explain the observed line profile shapes if the X-ray sources are originating from a
distribution of stellar wind shocks (e.g., the shock model developed by Feldmeier 1995), and the
majority of the observed X-ray emission is found to arise from a stellar wind location between 1.5
to 2 stellar radii (Leutenegger et al. 2006; Cohen et al. 2006). At first this apparent mass loss rate
reduction requirement was addressed as evidence that these winds are highly clumped, since a
clumped wind would provide channels to allow the deeply embedded X-rays to escape more
freely. Furthermore, a clumped wind would also be consistent with the mass loss determination
results from IR and radio observations, since a lower mass loss rate could lead to the same
emergent flux as an un-clumped wind of a higher mass loss rate. A key effect of wind clumping is
to enhance the escape probability of X-ray photons, and this has also become known as the
``porosity`` of a stellar wind. However, as we will see later, for some of our stars, the reduced
mass loss idea runs counter to a key X-ray observational result, the observed X-ray sources are
located at their respective X-ray continuum optical depth unity wind radii as determined from
using traditional mass loss rates (see discussions in Sec. 5.1 \& 7).

The perception that these stars do have reduced mass loss rates is based primarily on
FUSE observations and analyses of the \PV\ P-Cygni line profiles from several O-stars (Fullerton,
Massa, \& Prinja 2006).  From this study, it now appears that either the mass loss rates are a
factor of 10 or more lower than previously thought, or these winds are severely clumped over
small spatial scales. Fullerton et al. argue that UV lines from the ion \PV\ (1118, 1128 \AA)
should give line depths that are independent of clumping, consequently, their results imply lower
mass loss rates for these stars. 
However, if there are clumps in these winds and these are
surrounded by X-ray producing shocks, it is questionable to assume that the ionization balance of 
phosphorus is not shifted to higher states by the Auger effect, or whether recombination in the
high clump density decreases the ion abundance of \PV. So the assumption that \PV\ is dominant
everywhere is arguable.
With regards to
X-rays, some authors find that clumping over small spatial scales appears to be able to explain the
observed X-ray line profile shapes (Feldmeier, Oskinova, \& Hamann 2003; Oskinova, Feldmeier,
\& Hamann 2004). However, Owocki \& Cohen (2006) argue that the required porosity lengths
are unlikely, hence, they conclude in favor of reduced mass loss rates. We find it ironic that 25
years ago the possibility of reduced mass loss rates would have permitted base coronae models to
be an acceptable explanation of the X-ray emission from OB stars (Cassinelli \etal\ 1981; Waldron
1984). However, base coronae models were rejected because the needed mass loss rates disagreed
with values derived from the available radio fluxes. Nevertheless, a better picture involving
fragmented shocks in the winds emerged from the observations, and it is now abundantly clear
that shocks are responsible for a major fraction of the X-ray emission from OB stars. However,
there remain a large number of problems that we address in this paper that have led to questions
about the overall nature of these wind distributed X-ray sources.  

The \Chandra\ high energy resolution data has allowed us to exploit a major line emission
diagnostic, the ratio of the forbidden to intercombination emission lines (\ftoi) arising from
He-like ions (ranging from \OVII\ to \SXV).  This ratio had long been known as a very useful
diagnostic tool for determining solar X-ray electron densities (Gabriel \& Jordan 1969). However,
it was clear that the densities derived with this interpretation turned out to be far too high
($>10^{13}$ cm$^{-3}$) for \zpup\ and \zori. Kahn \etal\ (2001) and Waldron \& Cassinelli
(2001) quickly realized that the cause of this weakening of the $f-$line relative to the $i-$line is
not from collisional excitation, but rather radiative excitation from the presence of the strong 
UV/EUV flux from the photospheres of OB stars. The effect of radiation fields on the \ftoi\
ratio had been accounted for in the calculations of Blumenthal, Drake, \& Tucker (1972). The
intense photospheric radiation near OB stars causes a depopulation of the $2^3S_1$ level ($f-
$line) and a higher population of the $2^3P_J$ levels (J = 0, 1, 2) ($i-$lines) by photo-excitation.
This better understanding of the excitation process meant that the \ftoi\ ratio could be used to
derive the radial distances of the predominant X-ray sources using the geometric dilution factor of
the UV/EUV radiation. Basically, the smaller the \ftoi\ ratio, the closer the X-ray source is to the
star.

Waldron \& Cassinelli (2001) found that the {\it fir-inferred} radii (\Rfir) derived from analyses
of the observed \ftoi\ ratios indicated that the He-like ions are present over a wide range in radial
distances from the star. This means that there are X-ray sources distributed throughout the stellar
wind as would be expected from a distributed of stellar wind shocks. In
addition,
these \Rfir\ were found to correlate with their respective X-ray continuum optical depth unity
radii, \Rtau, i.e., the wind location where the X-ray continuum optical depth has a value
approximately equal to unity.  This observed correlation means that the highest energy He-like
ions are located deep in the wind, with a steady progression outward of the lower energy He-like
ions. This spatial distribution has now been observed in other stars with dense winds, \zpup\
(Cassinelli et al. 2001), \dori\ (Miller et al. 2002), and \eori\ (Waldron 2005). Somewhat
surprisingly, this correlated behavior has also been observed in the highly luminous O-star,
\cygate, a star which is believed to have a mass loss rate at least 5 times larger than \zpup\
(Waldron et al. 2004). As discussed by Waldron et al. (2004), although the relationship between
the \Rfir\ and corresponding \Rtau\ is not exact, the general conformance is rather good. Our
understanding of this behavior is related to the strength of the stellar wind opacity at a given
energy.  Since the opacity scales as $\approx \lambda^3$, at long wavelengths the wind can be
very optically thick to X-rays. Hence, we can only see the long wavelength line radiation, i.e., the
lower ion stages, that is emerging from the outer layers of the wind. In contrast, at short
wavelengths, the wind becomes more transparent to X-rays and we can detect line emission that
originates from deep within the wind. At these short wavelengths it is the higher ion stages
that are producing the line emission. Since the X-ray emissivity depends on the square of the
electron density, even if the line emission for a given ion is arising from all depths in a wind, we
would predominantly only see the emission from the highest density regions, which tend to be
those as close to the star as possible. Because of attenuation by X-ray continuum opacity we
expect to be able to detect radiation that is produced predominantly near optical depth unity. At
this corresponding radial depth and farther out, the radiation can escape freely, but radiation from
sources deeper in the wind 
is attenuated by the overlying material. The fact that there is
agreement between \Rfir\ derived from emission line ratios, and \Rtau\ derived from a
consideration of the ambient wind optical depth, is certainly a relation that is to be expected. We
ourselves had not predicted it, but found that it provided an explanation of the wide range of radii
that are inferred from the observed \ftoi\ ratios. However, even this favorable relation has come
into question since there is a call for the mass loss rates to be reduced by a large
factor. We intend to see whether the \Rfir\ and \Rtau\ are correlated in this sample of many more
stars than had been in our earlier studies of one or a few stars. 

This spatial distribution of He-like ions leads to what is perhaps the most significant problem
in OB stellar X-ray astronomy. Very high He-like ionization stages such as \ArXVII\ (3.95 \AA)
seen in \cygate\ (Waldron et al. 2004) and \SXV\ (5.04 \AA) seen in \zpup\ (Cassinelli et al. 2001)
occur at wavelengths where the continuum opacity is very low, hence, any line emission from
these ions can be detected from very deep within the wind. The observed \ftoi\ ratios from
these ions have confirmed that these high ion stages are in fact forming very
close to the stellar surface ($< 1.2$ stellar radii). From the distribution of ionization stages, and
the H-like to He-like line ratios to diagnose temperatures, Waldron (2005) found that the
spatial distribution of X-ray temperatures (\TX) for OB supergiants steadily decreases outward
from 20 MK near the surface to, 10 MK at 1.5 \Rstar, 5 MK at 3 \Rstar, and 2.5 MK at around 8
\Rstar. {\it Such a decreasing temperature distribution poses a problem in current X-ray studies
because the temperatures required for the high ionization stages are higher than should be
producible by shocks at such small radial heights in the wind!} For shocks there is a maximal
temperature which is determined by the jump in velocity across the shock front. One would
expect that this velocity jump should be no more than the local wind speed, which is small near
the base of the wind. Some authors (Leutenegger et al. 2006; Cohen et al. 2006) have presented
arguments that this is not a problem because shocks can in fact form at the radii in question.
However, that is not the full extent of the problem, not only are shocks needed, but the shocks
must have the velocity jump sufficient to produce the hot temperatures and high ions that are
observed to be originating near the star. 
Shock model predictions (e.g., Feldmeier, Puls, \& Pauldrach 1997a; Runacres \& Owocki
2002) show that shocks can form at and above 1.5 \Rstar, and the derived X-ray shock
temperatures (\TX) at these low radial locations are found to be highly dependent on the
line-driven instability triggering mechanism. For wind structures that are either self-excited or
includes explicit photospheric perturbations, {\it the predicted shock \TX\ at these low radial
locations are $< 2$ MK; well below the temperature required to produce the observed high
ionization stages}. However, as shown by Feldmeier et al. (1997a), photospheric turbulence can
generate a shock \TX\ of $\sim 10$ MK at $\geq 1.5$ \Rstar, but the contribution from this shock
with regards to the overall observed X-ray emission appears to be quite small.  Furthermore,
although these models indicate that weak shocks can form below 1.5 \Rstar, these shocks cannot
generate X-ray temperatures.

Since the expected maximum shock temperatures are found to be significantly smaller than the
temperatures required for these high ion stages, we call this the {\it ``near-star high-ion
problem``} (hereafter abbreviated as $NSHIP$). Possible explanations of the $NSHIP$ have been
proposed, but a consensus has not yet been reached. For the specific case of \tsco\ (B0V), Howk
\etal\ (2000) suggested that clumps form from the density enhancements in the embedded wind
shocks and these clumps become decoupled from the ambient wind velocity law. The clumps
follow trajectories that have them fall back toward the star which could lead to high relative
velocities between the clumps and the wind even at relatively small radial distances from the star. 
Each clump would also have a range of temperatures distributed over the frontal bow-shock
extending from low \TX\ in the wings of the bow-shock to a maximum \TX\ at the frontal apex of
the bow-shock. Howk \etal\ suggested that the conditions allowing for an in-fall of clump material
might only be present in main sequence stars such as \tsco, because wind clump drag forces would
prevent the in-fall in more luminous stars with denser and faster winds. However, in light of the
observational demand for clumpy winds, this model warrants a closer look. Very high \TX\ are
seen in some OB stars such as \toriC\ (which has a field strength of several kilogauss; see Gagne
et al. 2005) which are believed to have magnetically controlled outflows analogous to Bp stars
(ud-Doula \& Owocki 2002). Even with more moderate fields there could also be solar-like
phenomena occurring in these stars which may help explain the $NSHIP$.  Studies have found
that moderate magnetic field structures could rise buoyantly through the radiative envelope of OB
stars (MacGregor \& Cassinelli 2003; Mullan \& MacDonald 2005), and, in fact, the first complex
magnetic topological map of the early B main sequence star, \tsco, has been revealed (Donati et
al. 2006). Hence, as had been proposed by Cassinelli \& Swank (1983), one could envision that
these high ion stages may reside in magnetically confined loops close to the stellar surface.
Alternatively there could be``coronal bullets`` or ``plasmoids`` of fast moving material that might
be ejected from the surface of the star owing perhaps to magnetic reconnection in the
sub-photospheric region as proposed for the sun by Cargill \& Pneuman (1984). Although our
study will primarily focus on the OB stars that do not show the extreme kilogauss magnetic fields
with the hope of understanding the X-ray emission problems of ordinary OB stars, we include
\toriC\ in our study so that comparisons can be made between "normal" stars and a
highly magnetic one.

\section{Determining X-ray Emission Line Characteristics}

Existing detailed studies of individual OB stars have provided many interesting results about the
stellar X-ray sources. However, a study of many stars, using identical analysis
techniques is called for and is needed to search for general trends in the OB stellar X-ray source
properties. In particular, we explore the line characteristics as a function of stellar luminosity class
since stellar luminosity is considered to be the dominant contributor in the driving of these OB
stellar winds. In this section, we present the OB stars used in our study, their relevant stellar
parameters, and the X-ray emission lines used in our analysis. We also discuss our line fitting
approach and the emission line parameters that can be extracted.

\subsection{The Stellar Sources and Observed Lines }

We have compiled the available archived HETGS data for 17 OB stars.  Our analysis includes
both the MEG and HEG spectral data.  Although for most of our stars the HEG spectral lines
have much lower signal-to-noise ($S/N$) than their respective MEG lines, we use both data sets
to check the consistency in derived parameters and possibly add support to the conclusions. 
Our program stars are listed in Table \ref{tab:PARM}, along with the relevant adopted stellar
parameters, and the \Chandra\ observation identification numbers (Obs ID).  For a few stars,
their observations were carried out over two or more separate time segments. The MEG and
HEG spectra and relevant spectral response files (ARFs \& RMFs) were extracted using the
standard CIAO software (version 3.2.2). For stars with multiple observations, these observations
were co-added. Our study will focus only on the H-like and He-like lines and two Fe XVII lines. 
For the H-like and Fe XVII lines we consider only those lines with signal-to-noise ratio of $S/N >
5 $. For the He-like \fir\ (forbidden, intercombination, resonance) lines, we require that the total
flux from the three lines must have $S/N > 5$. In a few cases, primarily for the high energy lines,
if a reasonable flux has been established (i.e., $\ge$ 3), we use these results only for estimating
line ratios that provide interesting limits. The quantity $N_{WO}$ listed in Table
\ref{tab:PARM} is the scale factor associated with the stellar wind column density (see Table
\ref{tab:PARM} notes for a definition). Table \ref{tab:LINES} lists the emission lines used, their
rest wavelengths, and the temperature (\TL) associated with the peak emission for each line.  Also
included in the table are the wavelength dependent X-ray continuum absorption cross sections for
the stellar wind ($\sigma_W$), the ISM ($\sigma_{ISM}$), and several atomic parameters that
are needed to calculate the He-like \ftoi\ ratio.  The $\sigma_{ISM}$ cross sections represent the
``cold`` gas limit, i.e., ISM absorption cross sections. However, it has long been known that for
all O and early B stars, the value of $\sigma_W$ is always less than $\sigma_{ISM}$ for energies
$<$ 1.5 keV. This is because the intense UV/EUV radiation near the star increases the ionization
state in the wind relative to the cold ISM. The difference in cross sections is especially large at
low energies (e.g., see Fig. 2 in Waldron et al. 1998).  While for energies above 1.5 keV ($<$ 8.3
\AA), $\sigma_W$ is $\approx$ $\sigma_{ISM}$. For example, from Table \ref{tab:LINES} we
see that the difference between $\sigma_{ISM}$ and $\sigma_{W}$ first becomes less than 20\%
at wavelengths short-ward of 12 \AA ($>$ 1 keV).  The values of $\sigma_W$ listed in Table
\ref{tab:LINES} are representative of a typical O-star at a $\Teff = 35000 K$.  In general,
$\sigma_W$ at energies $< 1.5$ keV will be smaller for stars with larger \Teff, and conversely for
stars with lower \Teff. For each star, the product $N_{WO} \times \sigma_W$ gives the
commonly used scaling parameter, $\tau_*$. This optical depth parameter has often been used in
various emission line profile modeling efforts (e.g., Owocki \& Cohen 2001).  To obtain the radial
and wavelength dependent X-ray continuum optical depth, we consider a standard ``$\beta$-law``
velocity structure with $\beta=0.8$ (Groenewegen, Lamers, \& Pauldrach 1989). This gives
$\tau(r,\lambda) = 5 \sigma_W(\lambda) N_{WO} (1 - w(r)^{0.25})$, where w(r) is the radial
wind velocity normalized by the terminal velocity (the procedure for determining \Rtau\ is
discussed in the Appendix of Waldron et al. 2004).

\subsection{Description of X-ray Emission Line Parameters}

We examine four basic X-ray emission line parameters: the total {\it ``observed``} line flux, the
{\it ``observed``} line emission measure (\EMX), the line width or more specifically the
half-width-at-half-maximum (\HWHM), and for the line shift, we introduce a new terminology
which we will refer to as the ``peak line shift velocity`` (\VP) as discussed below. Line widths and
line shifts are expressed either in physical units (\kms) or as normalized to the terminal velocity of
the wind, \vinf. In search studies for global trends, the important velocity parameters are those
relative to the ambient wind, and not the actual physically velocities.  We have to emphasize that
the line fluxes and their associated \EMX\ are the {\it observed} values since, in principle, the
actual line fluxes and \EMX\ of the X-ray emitting sources may be influenced by the presence of
wind absorption.

a) The total observed line flux is the total energy integrated line flux determined from our
statistical best-fit modeling, and provides the line fluxes needed in our study of line ratio
diagnostics. The extraction of a line flux for a given wavelength region must be handled with care
if one wishes to compare various observed line flux ratios with theoretical predictions.  In
particular, we find that it is extremely important to account for all of the lines that could be
contributing to a given wavelength sector of the line, such as other strong lines and satellite lines. 
The importance of including other lines was clearly demonstrated by Ness et al. (2003) in their
analysis of the He-like \NeIX\ \fir\ lines from Capella. Also, appropriate adjustments for
contaminating lines must be made in the model line flux ratios before comparisons with observed
ratios can be made as discussed in Section 5. 

b) In general astrophysical studies, if distance and the temperature (which provides a line
emissivity) are known, the total line flux ($F_X$) can be used to extract an {\it "observed"}  line
emission measure using 

\begin{equation}
\EMX\ (cm^{-3}) = \frac {4 \pi d^2 F_X} {\epsilon (\TL)}
\label{eq:EMX} 
\end{equation}
where $d$ is the stellar distance, and $\epsilon (\TL)$ is the maximum line emissivity (e.g., see
Kahn et al. 2001) evaluated at \TL.  However, in the case of OB stars, the interpretation of the
total observed line flux can be somewhat ambiguous, especially for OB stars with massive winds,
because there can be unobservable sources located deep within the wind. To account for the total
amount of hot material in a wind, one would need to find the attenuation of the line flux by the
overlying wind. In addition, this definition of \EMX\ assumes that all the emission for a given line
is at the same temperature throughout the wind, and as discussed in Section 6, we now know that
there is a radial dependent X-ray temperature structure. However, these two effects would require
a significant amount of modeling of the wind structure, and the introduction of significant
uncertainties. In this paper we want to focus on a uniform discussion of observational properties. 
Therefore, we choose to deal only with the ``observable`` \EMX. Of course one who is interested
in the total X-ray productivity of the wind needs to understand that our derived \EMX\ are to be
strictly treated as lower limits.

c) The breadth of a line is expressed as a \HWHM. This provides information regarding the
dispersion in the velocity of the line emitting plasma. For the general case in which there is a
spatial distribution of X-ray sources, the derived \HWHM\ represent integrals over both depth and
impact parameters within each small range in the line-of-sight (LOS) velocity. However, even in
the absence of details regarding the  line formation process, we find useful information can be
derived from the line widths. For example, if a line is only produced very close to the star, one
would expect it to be narrow, with a width perhaps comparable to the expected thermal or
turbulent speed at the base of the wind. In fact, this is often seen in the X-ray observations of cool
stars such as Capella. In an OB star, a narrow line could also result from certain asymmetric wind
structures such as X-ray material confined to volume sectors which are seen inclined relative to
the observer's LOS. Even if the total outflow velocity in these sectors may be large, the line width
could be highly dependent on the inclination angle. For example, an observer's LOS perpendicular
to the flow would essentially see a near zero line width. We find that almost all OB stellar
X-ray emission lines have moderate to large \HWHM, although a few stars do have very narrow
lines (e.g., $\tau$ Sco; Cohen et al. 2003).

d) We have introduced the phrase, ``peak-line shift velocity`` (\VP) to represent the required
Gaussian line profile model velocity-shift needed to obtain a best-fit to the given observed line
profile. We use this terminology to emphasize that, in general, \VP\ does not correspond to a
Doppler shift of any specific part of the wind or atmosphere. 
The observed shift is affected both by the spatial distribution of the X-ray sources in the wind, and
by the degree to which the wind absorption attenuates the radiation from each X-ray source
region. The observed peak-line shift is weighted by the dominant source locations, from which the
emitted X-ray line radiation can escape through the overlying wind. Consider the simplest case of
a {\it single} spherically symmetric shell of X-ray emitting material moving at some velocity $V$.
For an optically thin wind, the observed line profile would be symmetric and flat topped extending
from about $-V$ to $+V$ (neglecting stellar occultation effects), with negligible line shift, $\VP
\approx 0$, and a \HWHM\ $ \le V$.  However, for the case of an optically thick wind, the
red-ward side of the line is more heavily attenuated, and thus the peak in the emission, \VP, would
occur at a blue-ward shift. This velocity could be slightly smaller than $V$, and the line shape
would be sloped down long-ward from this \VP, producing a triangular shaped line as had been
predicted by MacFarlane \etal\ (1991).  For this thick wind case, the \HWHM\ can be significantly
less than $V$.  For even more complicated scenarios, such as a stellar wind with many discrete
X-ray source regions, the observed line shift \VP\ represents an ``average`` over all the source
regions that are both capable of contributing to the observed line emission which are located at an
X-ray continuum optical depth of about unity or less. 

\subsection{Line Fitting Procedure}

To obtain a totally unbiased collection of emission line parameters we want to use a
model-independent extraction method. Two methods have been considered: 1) Gaussian fits to
the line profiles, and; 2) the ``moment`` method discussed by Cohen et al. (2006).  In
the latter, one calculates the first three moments of a line profile using only the actual observed
count spectrum (no ARF and RMF corrections) which in turn provide information on the line
shift, line width, and asymmetry of the line profile. 
Although this would appear to be a good unbiased way to describe a line, there are limitations
as
discussed by Cohen et al. (2006): 1) with regards to obtaining the \HWHM\ from the $2^{nd}$
moment, the moment analysis does not allow one to separate the effects of instrumental
broadening from physical broadening, and; 2) the moment method cannot extract reliable
information from blended lines (e.g., the He-like \fir\ lines). Furthermore, the moment method
cannot be used to extract the total line flux. Since our goal is to provide a  collection of physical
line emission parameters and line fluxes for single and blended lines in a self-consistent manner,
we chose to use the Gaussian line fitting procedure described by Waldron \etal\ (2004).
However, we do find that the $1^{st}$ moment method, which is used to extract \VP, provides
almost identical results as compared to those obtained using the Gaussian method. 

For the Gaussian line fitting procedure, we assume that all lines within a given wavelength region
have Gaussian line profiles superimposed on a bremsstrahlung continuum.  We use $\chi^2$
statistics to determine the best fit parameters (line flux, \HWHM, \VP, \& \EMX). We also
considered the C-Stat (Cash 1979) approach to line fitting but did not find any advantage
over the $\chi^2$ approach used in this analysis, basically because the C-Stat method was
developed for handling weak lines and in this paper we are considering only the strongest lines.
The error bars for each parameter are determined from the 90\% confidence regions.  For all line
fits we assume a continuum temperature of 10 MK. The actual value assumed is not critical to our
results since we are only interested in fitting the shape of the line and the strength of the line
emission relative to the continuum. All line flux emissivities and line rest wavelengths are taken
from the Astrophysical Plasma Emission Database (APED; Smith \& Brickhouse 2000; Smith et
al. 2001). It is worth stressing again that even though we have only chosen to analyze the
strongest observed lines, many of the line regions contain overlaps with other fairly strong lines.
We find it important to include all extraneous lines that may be contributing in a given wavelength
region in order to get an accurate fit to the line profile of interest.  In the case of the three He-like
\fir\ lines, all three lines are likely to be formed under the same physical conditions, and our model
fitting procedure assumes that the \HWHM\ and \VP\ are the same for all three lines, and only the
line strengths are different. The model count spectrum is determined for the default wavelength
binning of the MEG (0.005 \AA) and HEG (0.0025 \AA), using the appropriate $\pm 1^{st}$
order ARFs and RMFs. To obtain the best fit model parameters and associated errors, both the
model and MEG and HEG $\pm 1^{st}$ order spectra are re-binned to bin sizes of 0.02
\AA\ and 0.01 \AA, which are respectively the approximate resolution limit of the MEG (0.023
\AA) and HEG (0.012 \AA). The main advantage of this approach is that it provides tighter
constraints on the model parameters. The determination of line ratios (discussed in \S 6) are based
on the resultant best-fit model line fluxes.

\section{Distributions Versus Luminosity for the \HWHM, \VP, and \EMX}

We provide histograms of the \HWHM\ and \VP\ in physical units (\kms) for the case of
comparing all OB stars, and \HWHM\ and \VP\ normalized to their respective star's terminal
velocity (\vinf) in our study of luminosity class dependence.  It is these normalized distributions
that are of interest for developing an understanding of the behavior of the X-ray emission line
parameters versus OB spectral types and luminosity classes, but is not necessarily relevant when
looking at the overall OB distributions for all classes. The \EMX\ histograms represent the
physical observed emission measure ($cm^{-3}$).

First we show the line parameter distributions for all 17 OB stars, then we examine how both
the MEG and HEG derived parameter distributions depend on the stellar luminosity class. We
consider three luminosity class groupings: 1) main sequence, MS (luminosity class V); 2) giants
(luminosity classes IV \& III), and; 3) supergiants (luminosity classes II \& I). For all histograms,
the bins represent the percentage of lines within a given range of \HWHM, \VP, or \EMX.  The
\HWHM\ and \VP\ bin sizes are 200 \kms\ for the actual velocity values, and 0.1 for the
normalized case.  For the \VP\ histograms, the bin spacing is set up to center on \VP\ = 0.  

The observed \EMX\ are derived from the total line flux by using the approach discussed by Kahn
et al. (2001), using the APED emissivities and eq. \ref{eq:EMX}.  The major assumption used in
extracting \EMX\ is that we have assumed each line is at its maximum emissivity level (i.e.,
maximum temperature \TL).  In our study of X-ray temperatures (see Sec. 5.2) we find that this is
not a bad assumption since all extracted temperatures are very close to their maximum values. 
We present the \EMX\ histogram distributions in units of the log \EMX\ in increments of 0.5.

\subsection{\HWHM, \VP, and \EMX\ Distributions for All OB Stars}

\begin{figure}
\includegraphics[height=20cm,angle=180]{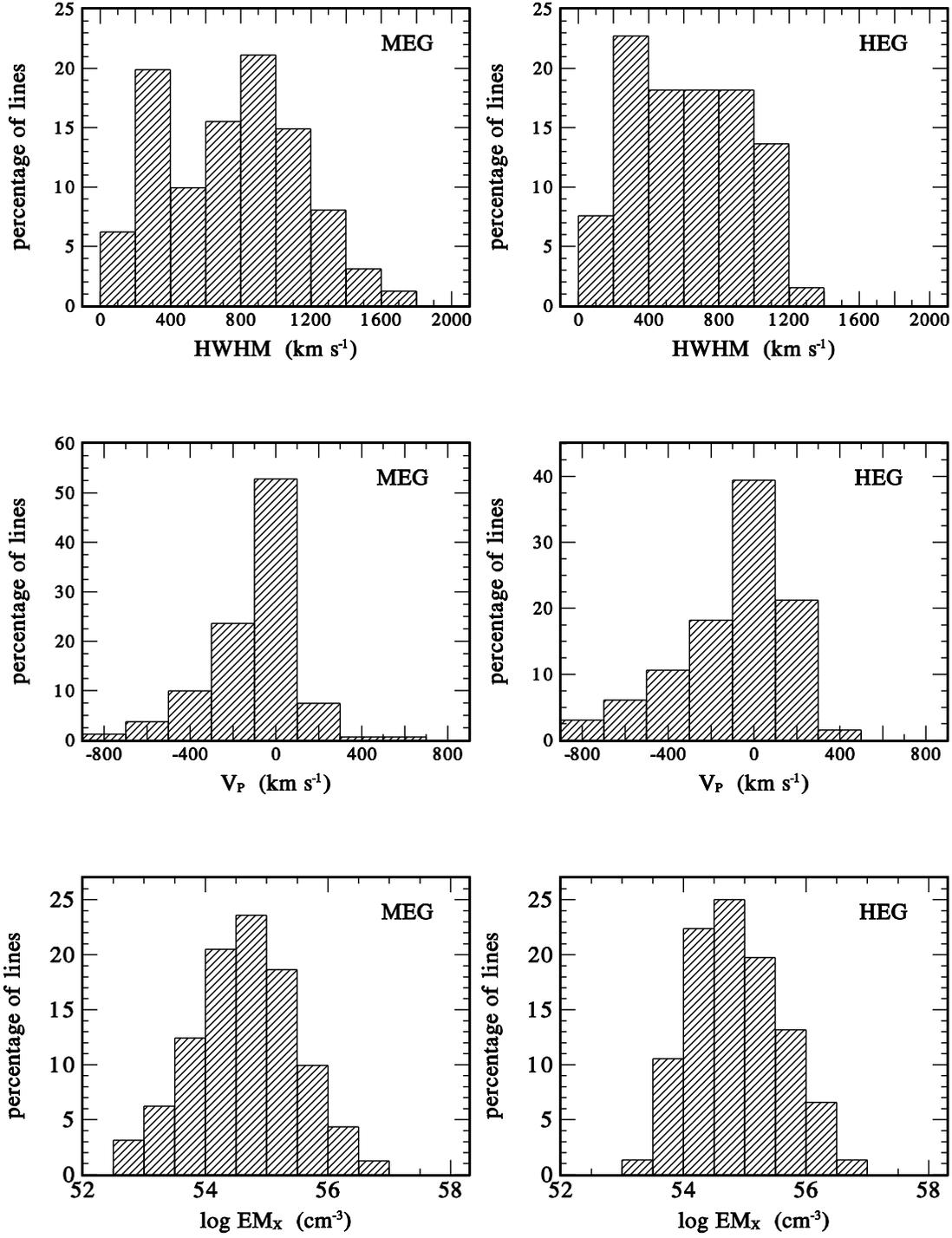}
\caption{The MEG and HEG \HWHM, \VP, and \EMX\ histograms for all OB stars illustrating
the percentage of lines within a given parameter bin range.
\label{fig:VHWEM_OB}}
\end{figure}

The \HWHM, \VP, and \EMX\ MEG and HEG histograms for all OB stars are shown in Figure
\ref{fig:VHWEM_OB}.  Several key features are noted: 1) the MEG and
HEG \VP\ distributions are nearly symmetric (both show slight asymmetry blue-ward) around
\VP\ = 0 with $\sim$ 80\% of all lines lying between $\pm 250$ \kms; 2) the MEG and HEG
\EMX\ distributions indicate that more than half of all lines lie within a relatively small range of
$\sim 1.5$ dex in log \EMX; 3) the MEG and HEG \HWHM\ distributions show a large
range from 0 to 1800 \kms, and; 4) the MEG \HWHM\ shows a double peaked distribution
with peaks at 350 \kms\ and 850 \kms\ which is not seen in the HEG distribution which only
shows a peak at 350 \kms.  The explanation is related to the sensitivity of the instruments in that
the MEG 850 \kms\ peak is due primarily to lines from low ion stages which are inaccessible or
very weak in HEG spectra.

The \VP\ distributions illustrate one of the most surprising results arising from HETGS
data analyses, {\it the majority of all OB lines show essentially no line shifts}. We emphasize that
almost all \VP\ are typically within the wavelength resolution limits of the MEG (0.024 \AA)
and HEG (0.012 \AA). For example, the MEG velocity resolutions at wavelengths of 25, 20, 15,
10, and 5 \AA\ are respectively 276, 345, 460, 690, and 1380 \kms.  
Correspondingly, the HEG limits can be obtained by taking half of the MEG determined
values (long-ward of 20 \AA\ is out of the HEG wavelength range).

These OB stellar distributions illustrate a basic dilemma in early attempts to understand the X-ray
emission from OB stars.  As in pre-launch expectations (e.g., MacFarlane \etal\ 1991),
the expected line broadness is observed, but there is clearly a lack of substantial blue-shifted
line profiles. From this sample, it is now clear that this non-shifted line behavior holds for
essentially all OB stars. Hence there must be some common property related to X-ray production
that needs to be determined.  In the following subsections we explore the \HWHM, \VP, and
\EMX\ distributions versus luminosity class to search for a better understanding of the non-shifted
behavior seen in the distributions of Figure \ref{fig:VHWEM_OB}.

\subsection{\HWHM\ Dependence on Luminosity Class}

\begin{figure}
\includegraphics[height=20cm,angle=180]{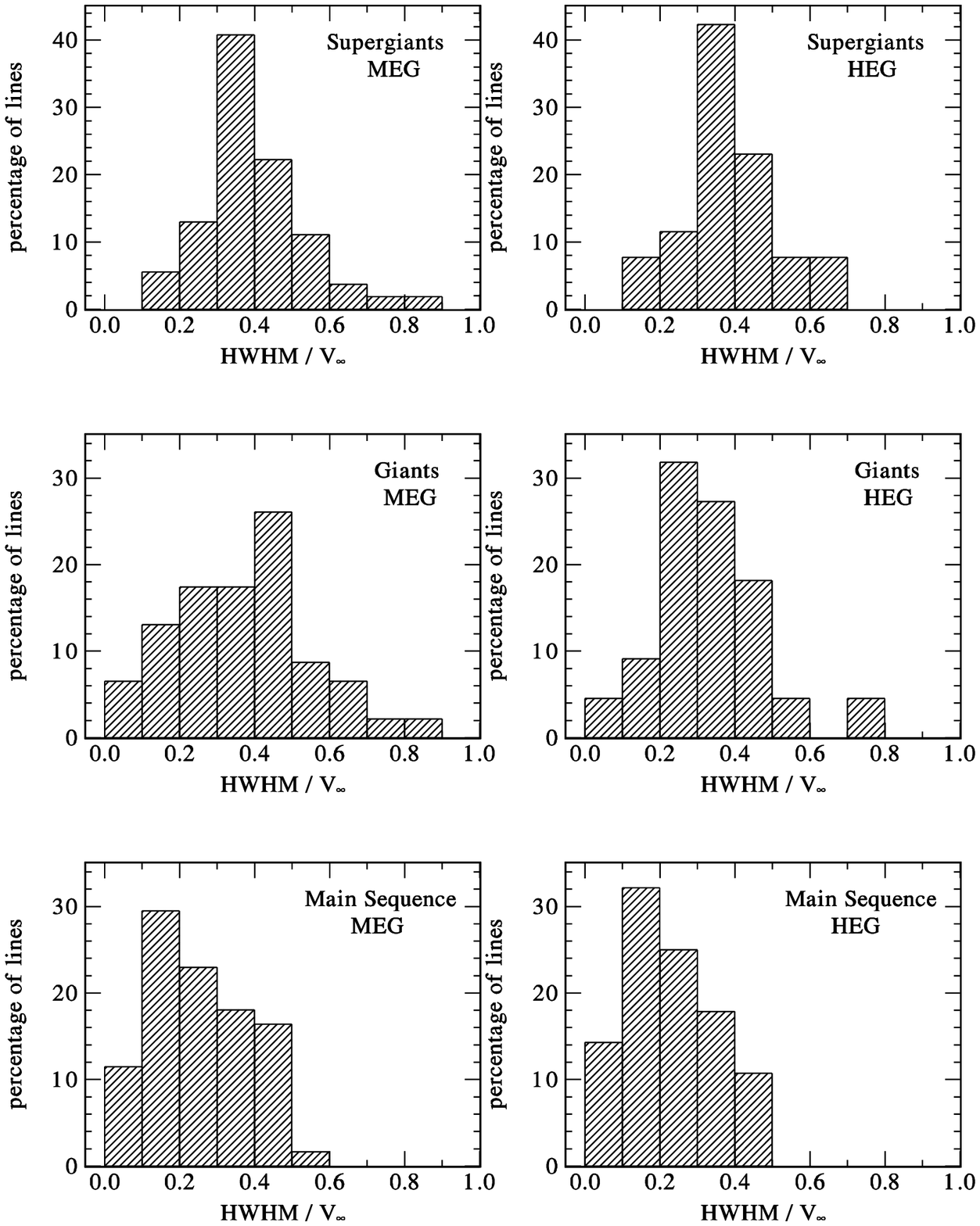}
\caption{The MEG and HEG $\HWHM / \vinf$ histograms illustrating the luminosity class
dependence of the percentage of lines within a given bin range. 
\label{fig:HWV8_LUM}}
\end{figure}

The MEG and HEG histograms showing the \HWHM\ dependence on the three luminosity class
groups are shown in Figure \ref{fig:HWV8_LUM} (normalized to \vinf). 
With regards to all luminosity classes, a key finding is that all \HWHM\ are $<$ \vinf. 
We see that the
histogram distributions are similar for supergiant and giants with peaks between 0.35 to 0.45, and
both show asymmetries towards lower values (except in their HEG distributions). The most
notable difference is the \HWHM/\vinf\ histogram for the MS stars where there is no sharp peak.
Instead, a rather a large range of 0.1 to 0.5 $\times$ \vinf\ is evident.  A partial explanation of this
can be seen from an inspection of the observed \vinf\ and $N_{WO}$ listed in Table
\ref{tab:PARM}. We see that the supergiant stars all have similar \vinf\ and $N_{WO}$, but the
MS stars have a rather large range in \vinf.  For the MS stars the broad peak may be due
to the fact that their $N_{WO}$ are weak, and thus the wind absorption effects do not play a
major role in determining the line profile shapes.  The giants also show a larger range in both
\vinf\ and $N_{WO}$.

The actual velocities from the \HWHM\ values are consistent with X-rays arising from shocks
located in the accelerating parts of the wind and/or in the case of rotationally distorted flows,
from sectors of the wind with low velocities along the line of sight. Surprisingly, both the
supergiants and giants show significant line emission at $\HWHM/\vinf < 0.3$. At least for the
supergiants, we would have expected to see very little emission at these values since, due
to their dense winds, one simply cannot see to the base of the wind owing to the strong wind
attenuation. A possible explanation is that if one cannot see to the back side of the star owing to
wind attenuation, this would also tend to make the \HWHM\ smaller, but this would also require
a large blue-shift in \VP, which is not observed. For the MS stars, although they also have a large
range in \vinf, their \HWHM/\vinf\ histogram is not affected. 
It may be that the low mass loss rates of the MS stars relative to the giant and supergaint stars
leads to lower wind column densities, hence, the observer can see deeper into the MS winds. 
Thus we are primarily detecting X-rays from
regions of low velocities instead of regions with speeds near \vinf. Whereas for the supergaint and
giant stars, depending on the line wavelength and the X-ray opacity, we can observe to a broader
range of depths and have a \HWHM\ that depends on \vinf.  Furthermore, since none of the
groups show significant number of lines with $\HWHM/\vinf > 0.5$, the majority of the observed
X-ray emission must arise relatively close to the star ($< 2$ stellar radii) which means that strong
wind shocks in the outer wind regions do not exist as predicted by early shock models.

\subsection{Peak Line-Shift Dependence on Luminosity Class}

\begin{figure}
\includegraphics[height=20cm,angle=180]{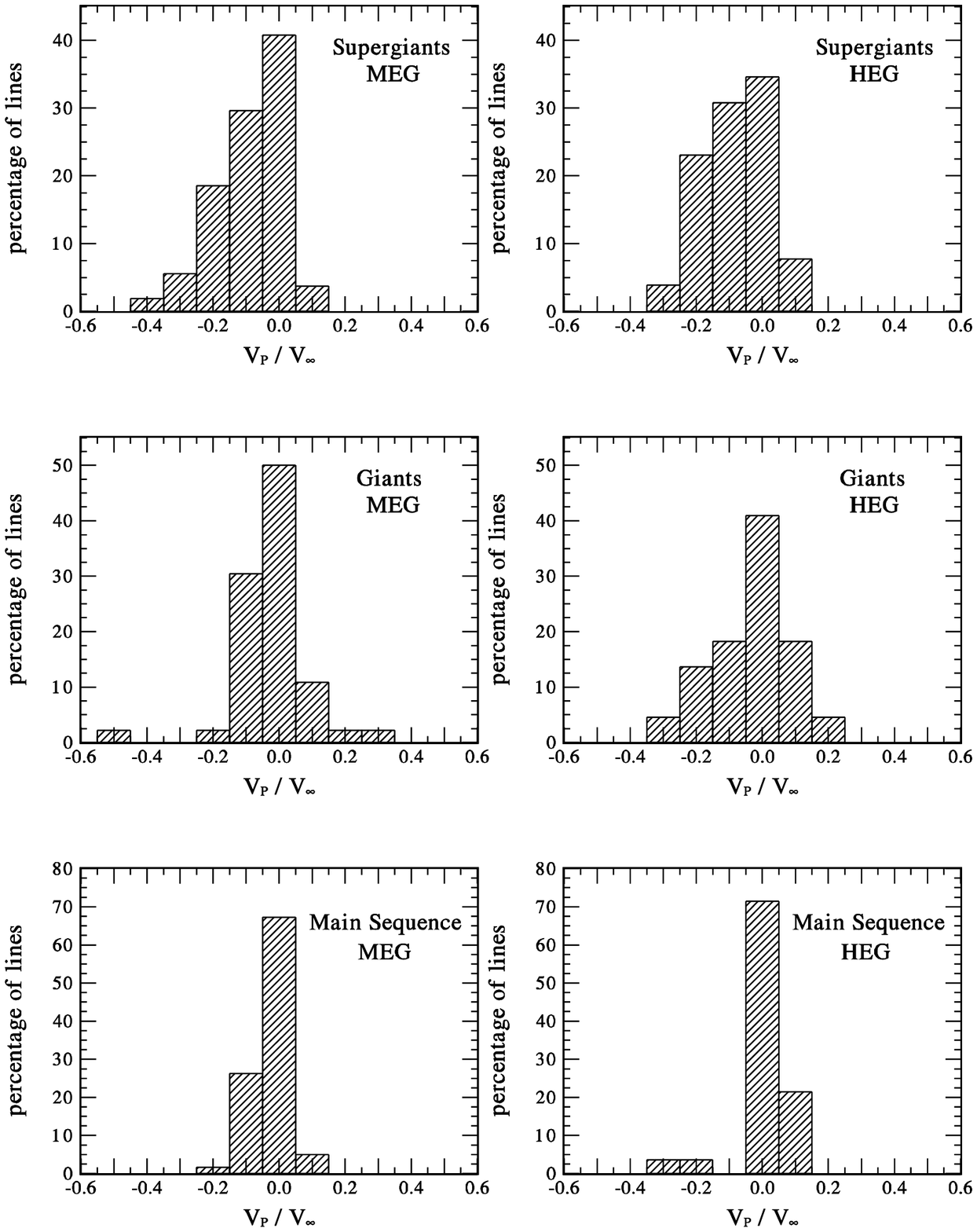}
\caption{The MEG and HEG $\VP / \vinf$ histograms illustrating the luminosity class
dependence of the percentage of lines within a given bin range. 
\label{fig:VPV8_LUM}}
\end{figure}

The MEG and HEG histograms showing the peak line-shift, \VP, dependence on the three
luminosity class groups are shown in Figure \ref{fig:VPV8_LUM} (normalized to \vinf).
Most notable, with respect to the maximum of the \VP\ distribution, is that each luminosity group
has a maximum that is nearly identical to the sample as a whole. That is, each class has a
maximum in its distribution occurring at very small velocities, which are less than the MEG
and HEG spectral resolutions. Next, in regards to the skewness or asymmetry of the distributions,
there is an increasing blue-ward asymmetry of \VP\ with luminosity group.  This indicates that a
small fraction of the lines do indeed exhibit finite blue-shifts. The blue-ward asymmetry extends
only to about -0.15 \vinf\ for MS stars and -0.20 \vinf\ for the giants. For the supergiants the \VP\
asymmetry distribution is well pronounced up to approximately -0.35 \vinf.  Nevertheless, this is
still far less than had been expected from early line calculation modeling by MacFarlane \etal\
(1991), and from the empirical shock model calculations of Owocki \& Cohen (2001).

The observed peak-shifts can be deduced from the uniform wind modeling expectations in several
ways:

\begin{itemize}

\item If the wind mass loss rate is less than the value inferred from radio observations (as
suggested by Fullerton et al. 2006), this reduction in wind density would allow more X-ray
radiation to emerge from the back side of the star (i.e., the red-shifted emission). This approach
was first demonstrated by Waldron \& Cassinelli (2001) in their analysis of \zori, and used by
Kramer et al. (2003) to model the X-ray line profiles and peak-shifts from \zpup. The Cohen et al.
(2006) re-analysis of \zori\ has also confirmed that this approach can provide a fit to these lines.
However, as mentioned before, there are some questions about the new \Mdot\ estimates
associated with the actual ionization fractions of phosphorus. At the present time we choose
to adopt the traditional mass loss rates until the major change in mass loss properties has been
verified and widely accepted.

\item If the X-rays can escape more easily from the sides of a shocked region, then the X-ray
plasma produced along a LOS perpendicular to the observer (i.e., from the wind on both sides of
the star) would dominate the emergent X-ray emission. This is the solution offered by Ignace \&
Gayley (2002), who used Sobolev escape probability theory to derive the line profile shapes.
However, it is questionable that the Sobolev theory they used is valid for non-monotonic cases
with discontinuities in the velocity, and jumps in the velocity gradient. Furthermore, the X-ray
formation region is almost surely not being accelerated at the same rate that the stellar UV
radiation is accelerating the wind that is colliding with the shocks. Perhaps the treatment could be
improved using the Rybicki \& Hummer (1978) approach for non-monotonic velocities.
Nevertheless, the results of Ignace \& Gayley do illustrate that by having the line radiation
escape out from the sides of a shock provides a plausible explanation of the symmetric line
problem. 

\item If the winds are clumpy, photon mean free paths can be increased and one can see more
easily to the back side of the star. This increased {\it porosity} effect has been investigated by
Feldmeier, Shlosman, \& Hamann (2002). It also addresses the reason why the radio fluxes could
be larger than inferred from a laminar wind case. Although, it appears that one should expect to
see more flat topped X-ray emission lines than are actually observed from this initial picture,
Oskinova, Feldmeier, \& Hamann (2006) have made improvements which appear to explain the
minimal blue-shifts, provided that these winds are highly clumped.

\item If the outflowing stellar winds are geometrically confined, then the observed line profiles
will be dependent on the observer's orientation relative to the geometric structure of the wind.
This approach was studied by Mullan \& Waldron (2006) by considering a two-component wind
structure (i.e., a polar and an equatorial wind), where the polar wind is slower and less dense than
the equatorial wind owing to the fact that the polar wind is hindered by surface magnetic
structures.  Their results indicated that the observed line-shifts will be dependent on the LOS
where a pole-on view would yield minimal blue-shifts. The main advantage of this model is that a
fairly large sector of the outflowing wind does not have to be clumped.

\end{itemize}

There are surely other more complicated scenarios that could explain the properties of the
observed line shifts, but these suggestions cover the recently studied ideas. Although several of
these can explain the observed lack of substantial blue-ward peak shifts, a consensus has not been
reached. Furthermore, this issue of un-shifted lines became even more problematic with the
HETGS X-ray emission line analysis for \cygate\ (Waldron et al. 2004). This star is believed to
have an \Mdot\ that is at least 5 times larger than any other previously studied OB star, which
implied that large blue-ward peak shifts should be seen. However, the observations show that the
X-ray lines from \cygate\ are similar to other OB stars in that only minimal peak shifts were
observed. A particularly perplexing question is why do substantial blue-ward peak shifts show up
only in \zpup?  This star has for many years been the prototypical early O-star in regards to its
wind properties, and observations of it played a major role in the development of line driven wind
theory.

\subsection{\EMX\ Dependence on Luminosity Class}

\begin{figure}
\includegraphics[height=20cm,angle=180]{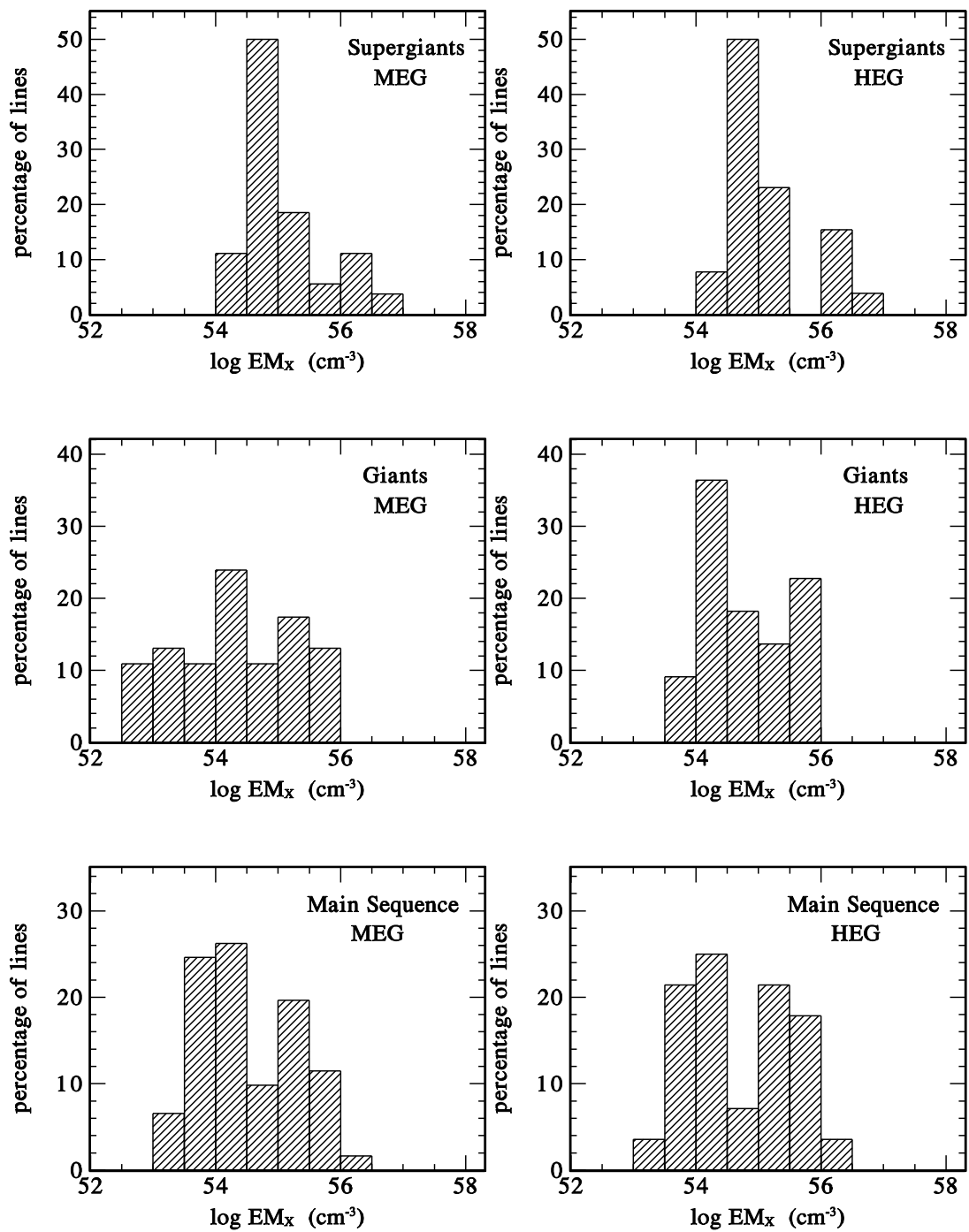}
\caption{The MEG and HEG log \EMX\ histograms illustrating the luminosity class dependence
of the percentage of lines within a given bin range. The \EMX\ have units of $cm^{-3}$. 
\label{fig:EM_LUM}}
\end{figure}

The MEG and HEG histograms showing the \EMX\ dependence on the three luminosity class
groups are shown in Figure \ref{fig:EM_LUM}. There are several interesting results shown in
these distributions.  
First we should recall that \EMX\ is a measure of the X-ray density squared times a volume
element, and \EMX\ scales as $( \Mdot\ / \vinf )^2 / \Rstar$. 
With this in mind, the results
shown in Figure \ref{fig:EM_LUM} suggest the following, using the data listed in Table
\ref{tab:PARM}. 1) Since the supergiants have very similar \Mdot, \vinf, and \Rstar\, then they
should have almost identical \EMX\ and indeed that is exactly what we see (the exceptions are
Cyg OB2 Nos. 8A \& 9). 2) The giants have the largest ranges in \Mdot\ and \vinf\, hence, what
we see in their \EMX\ distribution is clearly consistent with this spread in wind parameters. 3) The
MS stars show a double peaked distribution. The explanation is clear by inspection of their wind
parameters, which shows two extreme groupings of \Mdot\ and \vinf\ (i.e., early MS stars have
larger \Mdot\ as compared to late MS stars). In principle, these \EMX\ values which were
determined solely by fitting the observed line profiles, independent of any wind parameters, should
provide a means for establishing X-ray source densities if we know the radial location,
temperature, and geometric extent of the region (e.g., cooling length).  This is beyond the scope
of this paper but we plan to explore this possibility in a later paper.

\section{Line Ratio Diagnostics}

In astrophysical studies line emission ratios are commonly used as diagnostic tools.  In our study
of high spectral resolution X-ray astronomy, we focus on two line emission ratios that will be
used to establish X-ray source spatial locations and temperatures. 
Spatial information is obtained from the ratio of the He-like ion forbidden to intercombination
(\ftoi) emission lines. This ratio provides a diagnostic for estimating either the distance of the
dominant X-ray emission zone from the central EUV/UV radiation source (radiation dominated
case) or the electron density of the X-ray emission region (collisional dominated case). 
Although we commonly refer to the $i-$line as if it were a single line, it actually consists of two
lines that are unresolvable in both the MEG and HEG.  All theoretical \ftoi\ line ratio calculations
include the total emission from both lines. To determine X-ray temperatures, we use the
temperature sensitive line ratios from H-like ions to He-like ions (here abbreviated as the \HtoHe\
ratio). The \HtoHe\ line ratio increases dramatically with temperature as shown in Figure
\ref{fig:HHERATIO}. As discussed by Waldron et al. (2004), the temperatures derived from
\HtoHe\ ratios provide a measure of the ``average`` temperature of the He-like ions and the more
highly ionized H-like ions. 

Another temperature sensitive line ratio is the He-like G-ratio defined as $(i ~ + ~ f)/r$. This ratio
(introduced by Gabriel \& Jordan 1969) has been used extensively as a temperature diagnostic in
solar X-ray studies, and has been used in early studies of O-stars (e.g., Schulz et al. 2000;
Waldron \& Cassinelli 2001). The G-ratio dependence on temperature is opposite to that of the
\HtoHe\ ratio, the G-ratio decreases with increasing temperature and the slope versus temperature
is much weaker than is the case for the \HtoHe\ ratio. An advantage of the G-ratio is that it
provides a temperature of only one ion, the He-like ion, but, as shown by Waldron et al.
(2004), significant differences in the derived G-ratio temperatures as compared to the \HtoHe\
derived temperatures and their associated \TL\ were found. These discrepancies may be related to
``line-blending`` effects and/or resonance line scattering in the $r$-line as discussed by Porquet et
al. (2001). Complications in interpreting G-ratio derived temperatures have also been discussed in
studies of late-type stars (e.g., Ness et al. 2003).  Recently, Leutenegger et al. (2007) state that
they have found evidence for resonance line scattering among the low energy He-like ions in their
analysis of the \zpup\ $XMM-Newton$ RGS spectra.  Although we provide a tabulation of the
observed MEG and HEG G-ratios (see Table \ref{tab:GRATIO}), we believe that until we have a
clear understanding of the G-ratio idiosyncrasies, it is premature at this time to tabulate derived
G-ratio temperatures as it may lead to confusion and misinterpretations.

Line ratio diagnostics require good energy resolution of the observed lines.  Even with the high
energy resolution capabilities of the HETGS, we still must allow for possible contamination from
the blended dielectronic satellite lines and any other lines that are within the instrumental
resolution limits (these lines are easily identified in the APED data tables).  These line-blending
effects are most pronounced when dealing with ratios that use at least one or more of the He-like
\fir\ lines. For \HtoHe\ line ratios, the effects of line-blending are minimal, with the largest effect
occurring for neon. The line-blending effects on the \ftoi\ line ratio are also expected, but will not
be explored in this paper.  The main reason is that the \ftoi\ ratio is dependent on three parameters
(UV/EUV flux, spatial location, and X-ray temperature), whereas the \HtoHe\ ratios are
essentially only dependent on one parameter, the X-ray temperature.  This allows us to easily
tabulate the expected line ratios, including line-blending, that we use when comparing with
observed \HtoHe\ ratios. Although the \HtoHe\ ratios may also be somewhat dependent on wind
absorption and resonance line scattering effects (see Appendix), all \HtoHe\ ratio derived
temperatures (\THHe) presented in this paper only include line-blending effects. 

\subsection{Analysis of He-Like \ftoi\ Ratios}

The most widely used line ratio diagnostic to have emerged from \Chandra\ studies of OB stellar
X-ray emission is the He-like \ftoi\ line ratio. This ratio has proven to be a valuable observable
because, for the first time, we have a {\it direct means} of finding the predominant X-ray source
stellar wind locations. As we have seen, X-rays can arise from a wide range of wind radii.
However, as is the case in line formation in stellar atmospheres, there is a depth for which the
contribution function is maximal. Since the X-ray line emission varies as the square of the density,
deeper layers tend to contribute more strongly to the observable line strengths. It is information
about the predominant zone in radial distance that we can obtain information from using the \ftoi\
ratio. Ever since the beginning of OB stellar X-ray astronomy, determining the location of this
X-ray emitting plasma has been considered to be the most crucial quantity required for
understanding the X-ray emission process in OB stars. The \Einstein\ and \ROSAT\ satellites
provided only low resolution data, hence, the only information regarding the source location was
estimated from the change in X-ray attenuation across the major continuum jumps, such as the
Oxygen K-shell edge. Historically, this was sufficient to show that the X-rays from OB
supergiants could not all be coming from a thin coronal zone at the base of the winds (Cassinelli
et al. 1981).

The He-like \ftoi\ line ratios have been used in solar coronal studies as a diagnostic of the electron
{\it density} of the X-ray emitting medium as first demonstrated by Gabriel \& Jordan (1969). 
As the electron density is increases, a larger fraction of the ion is excited from the $2^3S$
level
(the ground state of the triplet sequence) to the $2^3P_J$ levels which results in a diminution of
the strength of the $2^3S \rightarrow 1^1S$ transition ($f-$line) and an increase in the strength of
the $2^3P_J \rightarrow 1^1S$ transition ($i-$line).
As discussed previously, the total $i-$line emission actually consists of two transition from the
$2^3P_J$ levels (J = 1 \& 2). Although the J = 0 level does not contribute to the $i-$line emission
since it is strictly forbidden, it is a contributing transition in the rate equations which establish the
relative populations of these levels (e.g., Gabriel \& Jordan 1969; Blumenthal \etal\ 1972).

In all O stars and early B stars the enhancement of the $i-$line relative to the $f-$line is caused by
photo-excitation from the $2^3S_1$ level by the strong UV/EUV photospheric emission. The
radiation required depends on the ion, and ranges from 1638 \AA\ (\OVII) to 674 \AA\ (\SXV)
(see Table \ref{tab:LINES}). The photospheric lines required to excite \OVII\, \NeIX\, and
\MgXI\ all have wavelengths long-ward of the Lyman limit (directly observable regions), whereas
the lines required to excite \SiXIII\ and \SXV\ have their wavelengths in the EUV, short-ward of
the Lyman limit (unobservable).  Hence, theoretical \ftoi\ ratios for \SiXIII\ and \SXV\ are
dependent on model atmosphere fluxes.  If one of these excitation line wavelengths happens to
coincide with a photospheric absorption line that for some reason is deeper than predicted by the
photospheric models, the distance inferred from the \ftoi\ ratio would be even closer. Conversely
if the overall EUV continuum is larger than expected from model atmospheres the formation
region is located farther out.  An example of these effects is shown by Leutenegger et al. (2006).

In the early \Chandra\ studies by Waldron \& Cassinelli (2001), Cassinelli et al. (2001), and Miller
et al. (2002), we felt assured that the we were not too far off in our radial distance estimate by the
fact that the radii derived from \ftoi\ (\Rfir) and from the X-ray continuum optical depth unity
radii (\Rtau) turned out to be nearly the same.  However, it is important to point out that this
correspondence is not exactly one-to-one, as discussed by Waldron et al.(2004). Furthermore, the
recent arguments by Fullerton et al. (2006), and others, that the mass loss rates of OB stars are
incorrect, are a source of concern and confusion as to why the \Rfir\ and \Rtau\ are in reasonably
good agreement. We will return to this later. 

Following the approach used by Blumenthal \etal\ (1972), the radial dependence of the \ftoi\ ratio
(commonly labeled as \RRR) can be determined by
\begin{equation}
\RRR (r) = ~ \frac{\RRR_O}{1~+~ n_e (r) / N_C~ +~ W(r)  \phi / \phi_C}
\label{eq:FIRatio}
\end{equation}
where $\phi$ is the photo-excitation rate from $2^3S_1 \rightarrow 2^3P_J$;
\begin{equation}
\phi~ =~ \frac{c^3}{8 \pi h \nu^3}~U_{\nu}
\label{eq:Phi}
\end{equation}
%
and $\RRR_O$ represents the low density and $\phi = 0$ limit, $n_e (r)$ is the X-ray electron
density of the X-ray emitting plasma, $N_C$ is the critical density, $\phi_C$ is the critical
photo-excitation rate, and
$U_{\nu}$ is defined as the {\it surface} photospheric radiation flux density which allows us to
factor out the radial dependent geometric dilution factor, $W(r) = 0.5(1 - (1-(R_*/r)^2)^{1/2})$. 
The values of $N_C$, $\phi_C$, and wavelengths of the $2^3S_1 \rightarrow 2^3S_J$ transitions
are given in Table \ref{tab:LINES}. Although $\phi_C$ is only dependent on atomic parameters,
$N_C$ is inversely proportional to the collision rate.  Hence, $N_C$ is also weakly dependent on
the temperature, and the values listed in Table \ref{tab:LINES} are evaluated at each ion's
expected maximum line emission X-ray temperature (\TL).

The format of eq. \ref{eq:FIRatio} allows us to see the distinction between collisional domination
and radiation
domination which is determined by the relative strengths of $\phi/\phi_C$ and $n_e (r) / N_C$. 
For OB stars, the radiation term is dominant throughout the wind except when one is interested in
conditions very close to the photosphere.  Assuming that the radiation term is dominant, we can
neglect the density term and obtain a relationship between radius, the stellar photospheric
radiation (i.e., the stellar effective temperature, \Teff), and the observed \ftoi\ ratio (\RRR), which
is given by
\begin{equation}
W(r) =  \frac{\phi_C}{\phi}(\frac{\RRR_O}{\RRR (r)} - 1)
\label{eq:W}
\end{equation}
The main advantage of this equation is that for any OB star with an observed \ftoi\ ratio we can
estimate a radial distance corresponding to the source of He-like ion emission, provided that we
know the corresponding radiative fluxes at the three $\lambda_{f-i}$ (see Table \ref{tab:LINES})
which can be determined using stellar atmospheric model spectra (e.g., Kurucz 1993; Hubeny \&
Lanz 1995). Leutenegger et al. (2006) found that there are model atmosphere dependent
differences in the \ftoi\ ratio. We find that these differences appear to be relatively minor and we
chose to use the Kurucz (1993) model atmospheres in our calculations.

The \ftoi\ ratio defined in eq. \ref{eq:FIRatio} is referred to here as the ``{\it localized}`` \ftoi\
ratio. 
To understand the difference between this {\it localized} \ftoi\ ratio approach and the more
complicated scenario, consider a {\it single} small isolated test packet of X-ray emitting
plasma, at some constant X-ray temperature (\TX), which can be placed at any wind radial
location. As this isolated test packet moves outward through the wind towards larger radii, the
\ftoi\ ratio increases because there is now less depletion of the $f-$line and less enhancement of
the $i-$line. This change is due to the radial decreasing strength of the radiation field (i.e., the
decreasing dilution factor). In general, the \ftoi\ ratio is also weakly dependent on \TX\ through
the critical density term since this term is inversely proportional to the collision rate, and
obviously, the strengths of the \fir\ lines are dependent on the fractional abundance of the He-like
ion as determined by the temperature of the X-ray emitting plasma.

Now consider a more general \ftoi\ case, the ``{\it distributed}`` \ftoi\, where there is a radial
distribution of X-ray sources throughout the wind.  Here, as compared to the
{\it localized} case, the total line fluxes of both the $f-$ and $i-$lines are determined by an
integration process (see Leutenegger et al. 2006), or equivalently, a summation of a very large (or
infinite) number of isolated test packets distributed throughout the wind. However, since the
strength of each line emissivity scales as the packet's density-squared, the contributions to $f-$line
deep in the wind are depleted by radiative excitation, and only the $i-$line accumulation benefits
from the density-squared effect since the $i-$line is much stronger than the $f-$line at low radial
distances.  On the other hand, the flux in the $f-$line only accumulates at larger radii where the
density is lower.  The resultant \ftoi\ ratio is then obtained from the ratio of the summed $i-$line
and $f-$line fluxes. The key parameter in this distributed \ftoi\ emission case is the 
radial location of the densest X-ray source that is capable of contributing to the 
observed $i-$line and $f-$line fluxes, i.e., the densest X-ray source that is not heavily
attenuated by the overlying stellar wind.

The main difference between these approaches is that the {\it distributed} \ftoi\ will increase faster
with radius than the {\it localized} \ftoi\ case as shown by Leutenegger et al. (2006). The
density-squared dependence of the line emissivity is the primary reason for this shift, although
there is also some change owing to the $r^2$ term in the volume integral. An even more
complex \ftoi\ scenario can be envisioned with the inclusion of a radial temperature distribution
and stellar wind absorption. This approach is beyond the scope of our current paper, and we plan
to address this issue in a subsequent paper.  

In this paper (as assumed in all our previous papers on this subject) all quoted values of the \ftoi\
ratio are based on the {\it localized} approach, assuming that the He-like ion \TX\ is always equal
to its respective \TL\ (i.e., the temperature where the given ion's X-ray line emissivity is at a
maximum which is known for every ion).  We know that a given set of He-like \fir\ lines must be
forming somewhere in the wind where $\TX = \TL$ and what the {\it localized} approach
provides is the most likely location where this occurs. 
Furthermore, detailed shock modeling such as that of Feldmeier (1995), indicate that the {\it
localized} approximation for treating \ftoi\ may indeed be appropriate since the dominant X-ray
regions are spatially well separated, and the largest contribution to the two lines will arise from
the highest density shocked plasmas that are capable of producing X-ray emission.
Regardless, the main advantage of this approach is that it provides results that are only dependent
of the adopted atomic physics and the UV/EUV model atmosphere.  This allows us to study our
collection of stars in a nearly identical way so as to reveal any significant differences associated
with basic stellar parameters. However, if the He-like emission is indeed distributed continuously
throughout the wind as discussed by Leutenegger et al. (2006), the resultant \Rfir\ for a given set
of He-like \fir\ lines represents the lower radial boundary of the distributed X-ray emission
integration.  Hence, for a wind distribution of X-ray sources there can be no emission below this
lower radial boundary. In general, these lower radial boundaries are expected to be lower than our
tabulated {\it localized} \Rfir\ values.

The observed \ftoi\ ratios and their associated derived \Rfir\ are given in Tables
\ref{tab:MGFI} (MEG results) and \ref{tab:HGFI} (HEG results).  The results for \OVII\ are not
given in the HEG table since this ion is outside the energy band of the HEG.  In addition,
although the He-like \ArXVII\ \fir\ lines were detected in \cygate\ (Waldron et al. 2004), these
results are not presented here since none of the other stars have sufficient \ArXVII\ $S/N$ to
carry
out a meaningful analysis. The most notable observation is that for the supergiants, where there is
a clear progression in the \Rfir\ being large for low energy ions and small for high energy ions.
This
behavior is what we have referred to as the $NSHIP$.  There is also another interesting
observation for the supergiants from examination of the \MgXI\ and \SiXIII\ \Rfir. We see that
essentially all of the \Rfir\ for \MgXI\ are located between 3 and 6 \Rstar, whereas, the \Rfir\ for
\SiXIII\ are located in a narrower range of 1.8 to 2.3 \Rstar. For both the giants and MS stars,
there is a similar dependence, but not as noticeable as for the supergiants.

An important observational feature that has emerged from analyses of He-like \fir\ lines is the
correlation between \Rfir\ and their associated X-ray continuum optical depth unity radii
(\Rtau) as first noticed by Waldron \& Cassinelli (2001).  A discussion on the calculation of \Rtau\
is given by Waldron et al. (2004). Figure \ref{fig:TAUFIR} shows a scatter plot of the
dependence of \Rfir\ on \Rtau\ for our luminosity groups (we refer to these plot types as scatter
plots because the data are shown with error bars in both the ordinate and abscissa directions). The
correspondence between the two radii is evident in the supergiants which was first shown by
Waldron \& Cassinelli (2002) using a small collection of OB stars.  
There is a key
distinction present in all luminosity groups in that there are very few data points that lie fully 
below the dashed line which represents the exact one-to-one correspondence, i.e., $\Rfir =
\Rtau$.
This is consistent with the idea that we do not see any observed line emission arising
from below the optical depth unity radii which is exactly what one would expect from basic
radiation transfer arguments, i.e., radiation can only escape from those radial locations where the
associated optical depths are $\leq 1$. 
For the supergiants, the most likely explanation of this correlation is that a very large number
of X-ray sources are distributed throughout these winds at essentially all radii, and the observed
emission line characteristics (line strength and wind locations, \Rfir) are primarily determined by
the dominant X-ray sources.  These dominant sources are those with the largest emission
measures (the density-squared dependence of the emissivity) which are no longer hindered by
significant wind absorption effects, and their maxima emissions arise from their associated X-ray
continuum optical depth unity radii (\Rtau). For the giants and MS stars, the X-ray emission is
seen to be occurring at wind radii that are larger than their associated \Rtau. Clearly, radiation can
always escape from any region that has a small optical depth, but the fact that there is evidence for
radiation emerging from radii $>$ \Rtau\ may be an indication that the number of wind distributed
X-ray sources in these stars are greatly reduced as compared to the supergiant winds.  For
example, consider a very simple case where there is ``one`` spherical expanding shock wave, then
the location of the X-ray emission will always be associated with the radial location of the shock
wave, independent of the location of \Rtau. By considering say, 1 to 5 expanding shock waves at
different X-ray temperatures, this may be able to explain the observed scatter shown in Figure
\ref{fig:TAUFIR} for the giants and MS stars. Although this is a highly unlikely scenario
primarily because such a structure would predict strong X-ray variability and we are not aware
that this is the case. A more likely explanation is that the \Mdot\ of the giants and MS stars may
actually be larger than is traditionally assumed, so that the \Rtau\ values may be larger. 

As evident in Figure \ref{fig:TAUFIR} there are four supergiant data points that do indicate
emission occurring below \Rtau.  One of these is the \doriA\ \NeIX\ emission, and the other three
are associated with the He-like ions of \cygate.  A possible explanation of this discrepancy for
\cygate\ may be related to the uncertainty in the mass loss rate as discussed by Waldron et al.
(2004).  In the original analysis of \doriA\ by Miller et al. (2002), the derived \NeIX\ \ftoi\
indicated a radial upper limit of $\sim$ 4 \Rstar, significantly larger than our upper limit shown in
Table \ref{tab:MGFI}. This discrepancy illustrates the importance of including line-blending
effects in the extraction of individual line fluxes. The \NeIX\ wavelength region is highly
contaminated by many lines. Although we believe that our quoted \NeIX\ results are correct, we
still do not have an answer as to why the \NeIX\ \Rfir\ is $\sim$ 1 \Rstar.  A possible answer may
be related to wind-wind interactions since \doriA\ is a well known binary system. It may be that
the majority of the observed \NeIX\ emission is actually occurring very close to the stellar surface
of \doriA\ binary companion, a B0.5 III star (Miller et al. 2002).

A controversy has recently arisen with regards to the \SiXIII\ \ftoi\ ratio for the late O
supergiant, \zori, as discussed by Leutenegger et al. (2006) and Cohen et al. (2006).  In the
original analysis of \zori, Waldron \& Cassinelli (2001) used the HEG \ftoi\ rather than the
MEG \ftoi\ in their analysis primarily due to possible problems in the interpretation of MEG \ftoi\
ratio. It is well known that the MEG ancillary response file (ARF) has a significant Si K-shell
edge (produced by the instrument silicon chip) within the MEG resolution limits of the \SiXIII\
$f-$line which is less prominent in the HEG ARF. Our tabulated \zori\ MEG \SiXIII\ \ftoi\ (see
Tables \ref{tab:MGFI} and \ref{tab:HGFI}) is consistent with the one derived by Leutenegger et
al., but both of these are larger than the value determined by Oskinova et al. (2006). However,
our HEG \SiXIII\ \ftoi\ is consistent with the MEG \SiXIII\ \ftoi\ given by Oskinova et al.
(2006).  Since most of the MEG and HEG determined \SiXIII\ \ftoi\ ratios are consistent in all the
other OB stars, it is unclear as to whether this \zori\ discrepancy in the \SiXIII\ \ftoi\ is related to
a specific problem with the \zori\ MEG ARF (different software versions), a problem associated
with the extracted \zori\ count spectrum, or maybe some other unknown problem. For example,
there are noticeable differences in the \SiXIII\ \fir\ lines when comparing the MEG+1 and MEG-1
dispersed spectra which could be evidence of contamination by an unknown X-ray source in either
the MEG+1 or MEG-1 dispersed spectrum in the energy vicinity of the \SiXIII\ \fir\ lines. Until
this disagreement between the MEG and HEG \SiXIII\ \ftoi\ for \zori\ is resolved, we choose to
adopt the HEG \ftoi\ for \zori\ as being the most appropriate value (as originally proposed by
Waldron \& Cassinelli 2001).

\subsection{X-ray Temperatures derived from the \HtoHe\ Line Ratio}

Deriving the temperature distribution versus radius of the X-rays sources in OB stars is a major
goal of this study. Establishing the X-ray temperatures (\TX) in OB stellar winds is a crucial
aspect in understanding the mechanisms responsible for the observed X-ray emission. Prior to the
availability of HETGS data, our knowledge of \TX\ was limited to one or two temperatures as
determined from fitting broad band X-ray spectral data. From HETGS data we now know that the
OB stellar X-ray emission is produced by a large range of \TX\ ($\sim$ 2 to 25 MK), and these 
temperatures can be used to probe the properties of the X-ray sources.  
If we also have detailed spatial information, we can then obtain a radial distribution of the
stellar wind temperature structure. In general, since each wind shock may also have a temperature
stratification which is determined by the extent of the post-shock cooling zone (e.g., Feldmeier et
al. 1997b), our derived radial distribution of \TX\ should be construed as representing the
dominant \TX\ at each associated radius. By knowing \TX\ we can also determine the pre-shock
velocity relative to the shock front (\UO), defined as $\UO = V - V_S$ where $V$ is the
pre-shock gas velocity, $V_S$ is the velocity of the shock front, and $V$ and $V_S$ are
measured in the rest frame of the star. From the Rankine-Hugoniot relation, the dependence
between \TX\ (or post-shock temperature) and \UO\ is given by 

\begin{equation}
\TX\ (MK) = 14 \left( \frac{\UO}{1000} \right)^2
\label{eq:TXofV} 
\end{equation}

The quantity \UO\ is the fundamental parameter required to determine the physical characteristics
of a shock.  In general, \UO\ can be used to determine the shock strength, post-shock
temperature, and relative post-shock speed ( $= 1/4 ~ \UO$). For a basic shock model
description, since the pre-shock gas velocity in the rest frame ($V$) is defined as equal to the
ambient wind velocity (hereafter denoted as $V_O$), \UO\ can also be used to determine the
speed of the shock front and a constraint on the shock front radial location provided we know the
radial dependence of \TX.  It is implied that all of these quantities ( \TX, \UO, $V$, $V_S$, and
$V_O$) are dependent on radius. 

The relevance of the \HtoHe\ emission line ratio as a diagnostic of the X-ray source temperatures
in OB stellar winds has been explored by Schulz et al. (2000), Miller et al. (2002), and Waldron et
al. (2004). The advantages of using \HtoHe\ ratios are: 1) the H-like and He-like lines are very
strong in HETGS data; 2) the line ratios display a strong dependence on the X-ray temperature as
shown in Figure \ref{fig:HHERATIO}, and; 3) line-blending effects on the \HtoHe\ ratio are
minimal, with the exception being the Neon \HtoHe\ ratio. We point out that our results shown in
Figure \ref{fig:HHERATIO} are slightly different from those shown by Miller et al. (2002) in that
their He-like line emission used in their \HtoHe\ ratio included all the \fir\ lines, while we choose
to only consider the He-like $r-$line to represent the He-like line emission (see discussion by
Waldron et al. 2004). 

The one possible disadvantage in the usage of the \HtoHe\ ratio with regard to studies of OB stars
is that the H-like and He-line ions could possibly form in different regions of the stellar wind
which implies that they may suffer from different amounts of wind absorption and/or resonance
line scattering.  
For example, Porquet et al. (2001) suggest that the observed G-ratio may be larger than
expected since the He-like $r-$line can be strongly affected by resonance line scattering escape
probability effects in contrast to the $i-$ and $f-$lines. 
Hence, the presence of resonance line scattering would lead to an underestimate of the actual
X-ray temperature. In the Appendix we explore the possible effects of X-ray continuum
absorption and resonance line scattering on the observed \HtoHe\ ratios. Our results suggest that
the temperatures derived from \HtoHe\ ratios are good indicators of the true X-ray temperatures.
Although the expected temperature dependent behavior of this ratio shown in Figure
\ref{fig:HHERATIO} is based on the MEG energy resolution limits to determine the range of
line-blending, we find that the HEG energy resolution produces only very minor differences.

We extract \HtoHe\ temperatures, \THHe, for all available OB H-He line pairs using the expected
temperature dependent \HtoHe\ ratios shown in Figure \ref{fig:HHERATIO} to determine the
range in temperature associated with the range in the observed ratios. Prior to the extraction of
\THHe, all lines in these observed ratios are corrected for ISM absorption, but no attempt is made
to estimate wind absorption and resonance line scattering. The observed \HtoHe\ ratios and their
inferred \THHe\ are listed in Tables \ref{tab:MGHHE}(MEG results) and \ref{tab:HGHHE}
(HEG results). For completeness, the observed MEG and HEG G-ratios are listed in Table
\ref{tab:GRATIO}. 

The \THHe\ data listed in Tables \ref{tab:MGHHE} and \ref{tab:HGHHE} illustrate several
important results: 1) these X-ray sources show a large range in X-ray temperature, from $\sim$ 2
to 23 MK; 2) all of the \THHe\ are found to be within the temperature range as specified by their
respective H and He \TL, i.e., the \TL\ range shown in Tables \ref{tab:MGHHE} and
\ref{tab:HGHHE} where \TL\ represents the temperature associated with the maximum line
emission of a particular ion (seeTable \ref{tab:LINES}); 
3) the consistency between \THHe\ and \TL\ for all ions can be interpreted as a verification
that the OB observed X-ray emission lines arise from a thermal plasma since \TL\ represents the
collisional ionization equilibrium temperature associated with a given line's maximum line
emission; 
4) for a given ion and a given luminosity class, the values of the derived \THHe\ are very similar,
and; 5) comparisons of the mean \THHe\ for each luminosity group show no significant
differences, except possibility the supergiant \MgXI\ mean \THHe\ which is somewhat larger than
the other luminosity groups, but this difference is traced to the higher \THHe\ of the Cyg OB2
stars.  Note that the mean \THHe\ for the MS stars does not include the results for \toriC, a well
known peculiar magnetic star. The most obvious difference between the \toriC\ \THHe\ for each
\HtoHe\ line pair is that they are larger than those of the other stars in our sample (see Tables
\ref{tab:MGHHE} \& \ref{tab:HGHHE}). In addition, the \toriC\ temperatures are only slightly
larger than those of \tsco\ which has also been confirmed to have magnetic structures (Donati et
al. 2006).

If the \HtoHe\ line pairs were sensitive to wind attenuation owing to continuum opacity, we
would have expected a wider dispersion in \THHe\ since our program stars contain a wide
diversity in stellar wind properties. However, there are a few stars in each luminosity group where
there is some noticeable difference among individual star \THHe\ with respect to the mean of the
group. This may signify certain effects such as, an overall higher temperature structure, wind
absorption, and resonance line scattering.  In general, since the \HtoHe\ temperature diagnostic
appears to predict temperatures that appear to be independent of wind structure (i.e., individual
stars), we suggest that this ratio should be considered to be the best available diagnostic for
establishing the X-ray source region temperatures. The overall consistency in these temperatures
is a very interesting result, in that, regardless of luminosity group, as well as individual stars, the
expected temperature for any given H-like to He-like line pair is always the same which implies a
global commonality among the X-ray temperature distributions for all OB stars.

\section{The Stellar Wind Distributions of \HWHM, \VP, \TX, and \UO}

In the previous sections, we have determined the line emission parameters (\HWHM, \VP, and
\EMX), spatial locations of the He-like ions (\Rfir), and X-ray temperatures (\THHe). 
We now combine this information to find the stellar wind spatial distribution of the line emission
parameters and temperatures. 
We will not explore the radial dependence of \EMX\ primarily because these are the {\it
observed} emission measures and not the {\it intrinsic} emission measures, plus there is some
uncertainty owing to the makeup of \EMX\, such as shock density and temperature, and
geometric factors, such as surface area and thickness. 
In our discussions of the radial dependence of \HWHM, \VP, and \UO\ it is beneficial to use
``velocity versus velocity`` type plots where the abscissa for any given radius, $R$, is defined as
$V_O (R) / \vinf$. The transformation between $R$ and $V_O$ is obtained through the standard
``$\beta$-law`` velocity, $V_O (R) / \vinf = (1 - R_O / R)^{\beta}$, assuming a $\beta = 0.8$. 

Figure \ref{fig:HWVP_FIR} shows scatter plots of the normalized \HWHM\ and \VP\ of the
He-like ions versus the normalized ambient wind velocities [$V_O(\Rfir)$], where all quantities
are evaluated at their \Rfir\ for all OB luminosity classes. From the basic shock model
description one would expect the \HWHM\ and \VP\ at any given radius (R) to be less than the
$V_O (R)$ associated with the line formation region since there is a significant decrease in the
flow velocity across a shock. For lines forming at large radii, the \HWHM\ are $<$ $V_O (\Rfir)$
as shown in the upper panel of Figure \ref{fig:HWVP_FIR}. However, for lines near the star,
although the \HWHM\ are $<$ \vinf\, these \HWHM\ are $> V_O$ at these low radial positions,
which presents a problem in that the lines formed near the star are too broad. This is also a
problem for the de-shadowing shock model (Feldmeier et al. 1997a) which predicts that the
post-shock velocities are comparable to the ambient wind velocities. Our results indicate that the
spread in \HWHM\ ranges from 0 to 0.6 \vinf\ at all radii. In the lower panel of Figure
\ref{fig:HWVP_FIR}, we see that the \VP\ values are barely shifted and the spread is only $\pm
0.2$ \vinf. Therefore, we find that observed distributions of both the \HWHM\ and \VP\ are
nearly independent of the wind location of the X-ray emitting plasma. This is an odd result that
has only become apparent from our multi-star study, i.e., there was no indication of such behavior
emerging from analyses of individual stars.


Figure \ref{fig:THHEFIR} shows a scatter plot of the dependence of \THHe\ on \Rfir\ for the
three luminosity groups. Our results confirm the initial results reported by Waldron (2005), the
\THHe\ of the supergiants show a well defined radial distribution that decreases outward from the
stellar surface as evident in both the MEG and HEG data. The supergiant data show a strong
correlation in which the highest temperatures (20 MK) only occur very close to the star, and the
lowest temperatures (2 MK) occur only in the outer wind regions. Although this behavior is not
as obvious in the giants and MS stars, it is clear that the highest temperatures are only located
near the surface, but low temperature can occur anywhere within the wind.
It is also interesting that the observed radial distribution of \THHe\ indicates that for all
luminosity classes there is a radial dependent maximum X-ray temperature which decreases with
radius. To our knowledge, this behavior has not been predicted by any shock model. The most
likely reason as to why the supergiants show such a tight correlation between temperature and
radius is that the supergiants have larger wind densities and column densities, hence, we cannot
see the low temperature regions that are present at small wind radii. This follows from the optical
depth unity argument.  For the giants and MS stars shown in Figure \ref{fig:THHEFIR}, one sees
a large range in temperature at small and intermediate radii with no cutoff as seen in the
supergiant case. This is probably also the case for the supergiants as well, but we cannot see the
lower temperatures at small and intermediate radii due to larger stellar wind absorption.
Furthermore, since all three luminosity groups show high energy ions existing very close to the
stellar surfaces, the $NSHIP$ is a common feature identified with all OB stars.

We now choose to examine the radial dependent distribution of the He-like ion values of \UO\
(the relative pre-shock velocity discussed in Sec. 5.2) with regards to the radial dependent
ambient wind velocity, $V_O$, where \UO\ and $V_O$ are evaluated at their \Rfir.
The values of \UO\ are obtained directly from eq. \ref{eq:TXofV} using the radial dependent
X-ray temperature structures [\THHe(\Rfir)] shown in Figure \ref{fig:THHEFIR}.  Our
discussion will incorporate a useful parameter, $\eta$, which in general should also be dependent
on radius, and is defined as

\begin{equation}
\eta (r) = \frac{\UO (r)}{V(r)} = 1 - \frac{V_S (r)}{V(r)}
\label{eq:eta} 
\end{equation}

This ratio measures the ratio of the pre-shock velocity relative to the shock front to the fixed
frame pre-shock velocity. For outflows ($V > 0$ and $V_S \geq 0$) $\eta$ has a maximum value
of 1 when $V_S = 0$ which implies that the shock front is stationary in the rest frame of the star,
and $\UO = V$, the maximum value of \UO.  For the case of in-falling gas clumps, i.e., $V_S <
0$, $\eta$ can be $> 1$, and the actual value of $\eta$ is determined by the magnitude of the 
clump in-fall velocity. 

In the following discussion we will address the results as relevant to the basic shock model
description for an outflowing gas, then, by definition, $V = V_O$ (the ambient wind velocity).
The scatter plot of \UO\ versus $V_O$ evaluated at the associated MEG determined \Rfir\ is
shown in Figure \ref{fig:SHOCKFIR} for all the OB luminosity classes.  The data display several 
interesting features: 1) there does not appear to be any obvious correlation between \UO\ and
$V_O$; 2) the majority of the data indicate a range in $\UO / \vinf$ from $\approx 0.2$ to
0.5; 3) the majority of the lines forming at large radii (i.e., with $V_O (\Rfir) / \vinf
> 0.6$) occur where $\eta < 0.5$, indicating weak relative pre-shock velocities which implies that
the shock front velocity ($V_S$) is becoming comparable to the pre-shock velocity ($V$), since
$V_S = V (1 - \eta)$, as $\eta \rightarrow 0$, then, $V_S \rightarrow V$; 4) at intermediate
ambient wind speeds ($\sim 0.5$ \vinf, a radius of $\sim 1.7$ \Rstar), $\eta \leq 1$ which means
that $V_S \rightarrow 0$, i.e., the shock front is stationary in the star's rest frame, and; 5) the
most interesting result is associated with those lines that form very near to the stellar surface
where $V_O < 0.2$ \vinf, or a radius $< 1.15$ \Rstar, with a large range in \UO\ from 0.2 to 0.8
\vinf, and these are the lines discussed earlier regarding the near-star-high-ion problem
($NSHIP$).

The data appear to support the idea that the majority of the He-like ion lines are consistent with a
basic shock model interpretation provided that these winds have shock front velocities that are
initially stationary at various points in the wind where $\eta = 1$ and then they accelerate
outwards where eventually the shock front velocities become comparable to the pre-shock
velocities and these weak shocks are no longer capable of producing detectable X-ray emission. 
This interpretation is consistent with our derived X-ray temperature structures shown in Figure
\ref{fig:THHEFIR}. It is also likely that similar conclusions may be attainable from other shock
models with appropriate parameter adjustments.

However, the He-like ion lines at low radial locations ($V_O(\Rfir) < 0.2$ \vinf) cannot be
explained by this basic shock model description. We explore two possible alternatives, and use the
observed data point with $\UO = 0.5$ and $V_O = 0.1$ as our sample for comparisons. For an
in-falling clump, we find an $\eta = 5$ and an in-fall velocity of $V_S = - 0.4$ \vinf.  It is unclear
as to whether such large in-fall velocities can occur at small radii. Now we examine whether the
rapid acceleration in the de-shadowing instability model is applicable.  As discussed earlier, for
outflowing gas, the maximum of $\eta$ is 1 which means that $V = \UO = 0.5 \vinf$, and this
acceleration must occur at a radius $\leq 1.1$ \Rstar. From inspection of several numerical
simulations (Owocki et al. 1988, MacFarlane \& Cassinelli, 1989, Feldmeier 1995, Feldmeier et
al. 1997a, Runacres \& Owocki 2002) we do not see any evidence for large pre-shock velocities
of $0.5 \vinf$ at such low stellar radii.


\section{Summary}

The main goal of this paper has been to present a large collection of observational results and
examine these in a uniform way so as to draw out general facts regarding OB stellar X-ray
emission characteristics. We have analyzed the \Chandra\ HETGS MEG and HEG data from 17
OB stars.  Although the HEG spectra are typically much weaker than their associated MEG
spectra, we have chosen to use both data sets to illustrate the rather good agreement between
their observed and derived attributes. As a slight departure from our original goal of analyzing
only ``normal OB stars`` we have also included \toriC\ in our sample, primarily to allow one to
compare the X-ray properties of a  know peculiar O-star with those of other OB stars. 

We use a well recognized Gaussian line-fitting method to extract the pertinent X-ray emission line
parameters which ensures that our results are easily reproducible. This model-independent line
fitting procedure has allowed us to determine the observed line emission parameters (\HWHM,
\VP, \EMX, and line flux) and line emission ratios (\ftoi\ and \HtoHe) in order to obtain
an easily verifiable description of these X-ray emission properties. We have primarily focused on
searching for luminosity class regularities in the data, and the radial distributions of several X-ray
derived parameters.

{\it $\HWHM / \vinf$ Histograms:} 
The $\HWHM / \vinf$ plots show peaks that are well below the wind terminal velocity for all
luminosity classes. We have argued that the differences from one luminosity class to another in
regards to the asymmetries in these distributions are due to the differences in wind column
densities. The fact that a large percentage of $\HWHM / \vinf < 0.5$ indicates that the majority of
the observed line emissions occur relatively deep in the wind where the wind flow is accelerating.  

{\it $\VP / \vinf$ Histograms:} The $\VP / \vinf$ plots show that the  majority of the lines are
symmetric, with very little line-shifts. There is some tendency for stars of all luminosity classes to
show a small but finite blue-ward asymmetry in their line shift distributions. Such an asymmetry
would occur if the X-rays are primarily arising from the near side of the star. 
Several ideas have been discussed, ranging from in-falling clumps, the orientation of elongated
clumps with respect to observer's LOS (line-of-sight), a two-component wind geometry
consisting of a polar wind and an equatorial wind which have different wind densities and velocity
structures, to a major decrease in mass loss rates as compared to traditional values. The actual
explanation is not yet clear.

{\it Emission Measure Histograms:} The supergiant \EMX\ values show a well defined peak at
log \EMX\ = 54.75 which is due to the fact that these stars all have similar \Mdot, \vinf, and
\Rstar. However, this is not the case for the giants and MS stars, and the resultant distributions
are more spread out.  In addition, the MS stars show a double peaked distribution in \EMX\
which is due to the fact that there is a broader range in wind properties along the MS spectral
class.

{\it X-ray Source Locations:} Using the He-like \ftoi\ line ratios we have derived the stellar
wind location, \Rfir, of the O, Ne, Mg, Si, and S He-like ion emission. The results clearly
emphasize the $NSHIP$ (near-star-high-ion problem) as evident from the \SXV\ derived \Rfir.
We stress that these \Rfir\ are based on the {\it localized} interpretation and a {\it distributed}
interpretation would result in a distribution that starts at an even smaller \Rfir.  Our result
confirms what has been shown for individual stars, all OB stars display the same basic
distributions with the He-like O and Ne ions located in the outer wind regions, the He-like Mg
and Si ions at intermediate locations, and the He-like S ions located near the stellar surface.  For
all supergiants, the MEG He-like Mg and Si \ftoi\ ratios predict a narrow radial region of 3.7 to
5.6 \Rstar, and 1.5 to 3.0 \Rstar\ respectively, whereas the He-like Ne and O \ftoi\ ratios predict
radial locations from 6.2 to 13.5 \Rstar.  The MEG He-like S \ftoi\ predict locations that are
essentially on the surface.  
In general, there is good agreement between the MEG and HEG derived \Rfir, with one
exception, the \SiXIII\ results for \zori\ (see discussion in Sec. 5.1). 


{\it X-ray Temperatures:} We think our most interesting results are in regards to the \TX\
radial distributions. We have derived X-ray temperatures (\THHe) using the \HtoHe\ line
ratio temperature diagnostic. First, we emphasize the \THHe\ should be considered as an average
temperature of the H-like and He-like ions (i.e., basically an average of the associated ion peak
\TL\ given in Table \ref{tab:LINES}). Overall our results are consistent with this interpretation,
e.g., by comparing the \TL\ range with the mean values listed in Tables \ref{tab:MGHHE} and
\ref{tab:HGHHE}. In general, for a given \HtoHe\ line pair, the resultant \THHe\ are found to be
essential the same regardless of the luminosity class. 
These results can provide valuable information regarding the shock formation processes that are
actually operating in OB stellar winds. However, there are some notable differences. The silicon
mean \THHe\ for all luminosity classes are at the lower end of the \TL\ range with a larger spread
in \THHe\ as compared to the other H-He line pairs. The spread in \THHe\ suggests that the
mechanism producing these higher temperatures may be dependent on the individual stellar
characteristics and probably related to resonance line scattering as discussed in the Appendix. The
three stars that have their silicon \THHe\ closer to the upper \TL\ limit are the two Cygnus OB2
stars
and the known peculiar star, \toriC.  As to whether these Cygnus OB2 stars are similar to \toriC\
remains to be determined. Ideally, one would like to have a more complete collection of sulfur
\THHe\ to explore the higher temperature behavior, but the data do not allow reasonable
signal-to-noise extractions. 

{\it The Correlation Between \Rfir\ and \Rtau:} Since the initial detailed studies of O supergiants
(e.g., Waldron \& Cassinelli 2001; Cassinelli et al. 2001), the basic idea that the \Rfir\ are
correlated with \Rtau\ has continually surfaced as a reasonable observational result.  To test this
correlation we have examined the luminosity class dependence.  
Our results show that this correlation is clearly evident in the supergiants. But, the data from
the giants and MS stars indicate that the X-ray emission arises primarily from radial locations
above their \Rtau\ surfaces. We suggest two possible explanations, either the giants and MS stars
have larger mass loss rates, or their number of distributed wind shocks are significantly reduced as
compared to those in supergiant winds (see discussion in Sec. 5.1). However, what is clearly
emerging from our results is that for all luminosity classes, {\bf \it there are very few observed
X-ray sources where \Rfir is less than \Rtau.} This supports our basic interpretation of this
correlation. One can only see those X-rays that are capable of escaping the wind, and since the
line emission is proportional to ${n_e}^2$, one is likely seeing the peak of the contribution
function as deep in the wind as possible. If the mass loss rates of these stars are indeed much
smaller than previously thought, one would have expected to see an uncorrelated scatter plot, i.e., 
we should also be seeing a scattering of X-ray emission from regions where $\Rfir < \Rtau$.
We do not see any plausible explanation as to why such a correlation should hold in a highly
porous or clumped wind since the X-ray continuum absorption is determined by the ``cool``
stellar wind opacity (as determined by bound-free transitions) which is linearly dependent on the
density in the X-ray energy range.  Hence, a reduced wind density or clumped wind structure
would predict \Rtau\ values much lower than those adopted in our study. In addition, this
correlation implies that we can now predict a particular He-like \ftoi\ ratio by simply calculating
the expected \Rtau. We therefore choose to present this as a challenge to the idea that the \Mdot\
values are so much lower, and perhaps also a problem for a clumped wind.

{\it Radial Distribution of \HWHM\ and \VP:} Since we know the radial locations of the
He-like \fir\ emitting ions we can explore the radial dependence of the He-like X-ray line
properties.  Almost all line \HWHM\ are found to be less than the radial ambient wind velocity as
expected from all shock model descriptions, but there is no apparent correlation between
\HWHM\ and the ambient wind velocity. However, several deeply embedded sources have
\HWHM\ that are significantly greater than the radial ambient wind velocity which implies a new
interesting problem, {\bf {\it the lines from deep in the flow are too broad!}} There is no obvious
radial dependence of \VP\, where the majority of observed $\VP / \vinf$ remain between
$\pm0.2$ regardless of wind location, a totally unexpected result. This radial dependence of \VP\
is an important problem that must be addressed before we can fully understand the source of the
X-ray emission from OB stars.

{\it Radial Distribution of X-ray Temperatures:} Although in principle we can only address the
radial locations of temperatures associated with the He-like \fir\ emission lines, we have argued
that the \HtoHe\ line ratios derived \THHe\ provide reasonable estimates of the average
temperature associated with the H-like and He-like ions. Waldron (2005) was the first to
show a correlation between the \THHe\ and \Rfir\ for OB supergiants.  We have now shown that
this correlation is consistent with both the MEG and HEG results. Our analysis has verified that
the temperature steadily decreases outward through the wind with the highest temperatures
occurring near the stellar surface. For the other luminosity classes, although the tight correlation
breaks down, we still find that the highest temperatures only occur near the stellar surface. 
Hence, there is no evidence of any high temperature in the outer wind regions which is perhaps
surprising since shock velocity jumps could in principle be at their largest at these large radial
locations. Our results support two interesting results. 1) There appears to be a well defined radial
dependent maximum X-ray temperature. 
This radial dependent maximum temperature can be extremely useful in determining basic
shock characteristics such as, the efficiency of the conversion of shock energy into X-ray
emission. 
2) The magnitude of the derived X-ray temperatures and associated radial distributions
are dependent on wind density.  In the dense winds of the supergiants we see little evidence of
low \TX\ at low radial locations, whereas, in the lower density winds of the giants and MS stars, a
large range in \TX\ exists at lower radii.  The implications are clear, X-ray emission is probably
existent throughout these winds at all temperatures provided they are below the radial dependent
maximum temperature. However, as the wind density gets larger, one cannot detect the low \TX\
at low radii due to wind absorption effects.  This is exactly what one would expect based on our
optical depth unity radii arguments.  
Furthermore, for all OB luminosity classes we find that the radial dependence of the pre-shock
velocities relative to the shock front (\UO) evaluated at \THHe, indicate that the majority of all
lines at large radial distances are consistent with the basic shock model interpretation.  These
results also support outward accelerating shock front velocities which are consistent with the
decreasing X-ray temperature structure.  However, a fundamental problem exists at low radial
locations ($ < 1.2$ \Rstar) where there are several lines, characterized by the high ion stages, 
with temperatures that appear to be too hot, and line profiles that have \HWHM\ that are too
broad. As to whether any wind shock mechanism can explain these too hot, and too broad X-ray
emission lines, needs to be investigated.

Some of the problems discussed here have recently been addressed in repeated analyses of the
X-ray emission from a few select O-stars such as \zpup\ and \zori\ using a semi-empirical shock
model description (e.g., Oskinova et al. 2006; Leutenegger et al. 2006; Cohen et al. 2006). Cohen
et al. (2006) argues that all emission lines from \zori\ can be explained by a shock model
description if the X-rays are distributed above 1.5 \Rstar, provided that the wind density is greatly
reduced or highly clumped.  Cohen et al. further argue that there is no problem with
understanding the X-ray emission from OB stars.  
However, we see two potential problems with these conclusions: 1) their modeling efforts assume
that everywhere $\TX = \TL$ regardless of the ion being considered, and as we have shown in
Section 5.2, this is not a good assumption, and more importantly; 2) the X-ray emission from
below 2 \Rstar\ is a notoriously difficult problem due to the line drag effect (Lucy 1984) or strong
source function gradients (Owocki \& Puls 1999) that affect the line-driven instability.  Hence, we
argue that until we have a better understanding of the shock properties below 2 \Rstar, such
conclusions are premature.

We have provided a summary of our results and
emphasized the critical problems associated with current interpretations. Our results have verified
that essentially all OB X-ray emission lines are un-shifted. We have introduce a new problem
associated with the X-ray temperatures and spatial locations which we have labeled as the {\it
``near-star high-ion problem``} ($NSHIP$) and we believe this problem is far more critical than
the un-shifted line problem. Resolution of this problem will lead to a better understanding of
the X-ray production mechanisms at work in OB stars. It seems that the wind shock model
may be responsible for the observed X-ray emission in the outer wind regions (although we still
must figure out why these lines are un-shifted), but we must consider alternatives for the
remaining X-ray emission. 
The un-shifted line problem has been addressed by either lowering the mass loss rates or
proposing a clumpy wind. Either approach produces the same desired effect, an overall reduction
in the stellar wind X-ray absorption.  This reduction in X-ray absorption allows for more X-rays
to emerge from the far side of the star (the red-ward emission) which makes it possible to explain
symmetrical line profiles. Although this provides an explanation for the observed line symmetry,
we are then left with a dilemma as to why such a strong correlation exist between \Rfir\ and 
\Rtau\ (as determined from the traditional mass loss rates) where the premise of this correlation is
based on fundamental radiation transfer arguments.


With regards to the observed X-ray emission not explainable by standard wind shocks, a
continuing possibility is that these deeply embedded X-ray sources are associated with magnetic
fields on or near the stellar surface. It is becoming increasing clear that these stars are theoretically
allowed to have surface magnetic structures (e.g., MacGregor \& Cassinelli 2003; Mullan \&
MacDonald 2005). In any case, if there is an alternative source of X-rays operating in these stars, 
it will be an exciting new feature associated with OB stellar astronomy.  

\acknowledgments

We would like to thank the anonymous referee for his detailed critique of our manuscript and
several informative suggestions. WLW acknowledges support by award GO2-3027A issued by
the \Chandra\ X-ray Observatory Center. JPC has been supported in part by award TM3-4001
issued by the \Chandra\ X-ray Observatory Center. \Chandra\ is operated by the Smithsonian
Astrophysical Observatory under NASA contract NAS8-03060. 

\appendix
\section{Wind Absorption and Resonance Line Scattering Effects}

\begin{figure}
\includegraphics[angle=-90,width=\linewidth]{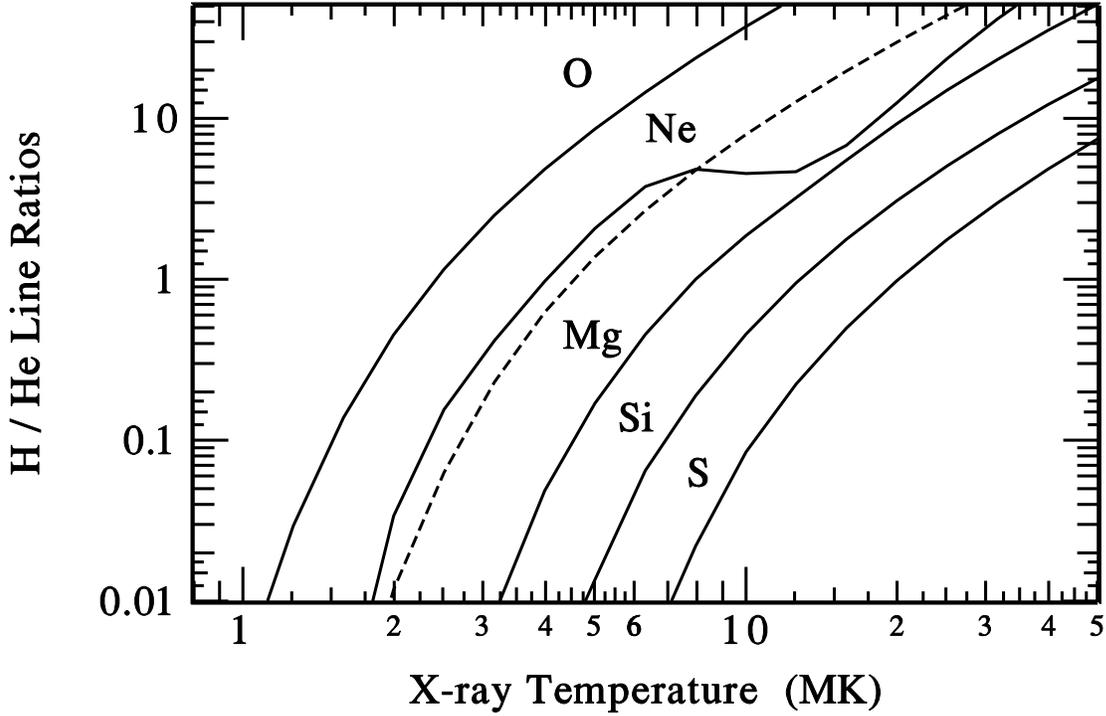}
\caption{The \HtoHe\ line ratio dependence on \TX\ for O, Ne, Mg, Si, and S as determined from
the APED data. For the He-like lines we only use $r$-line (see text). For each H-like and He-like
line region all lines within the instrumental wavelength resolution are included in determining the
corresponding total emission at a given temperature. Since the H-like \NeX\ line ($\sim$ 12.13
\AA) is a blend with a relatively strong \FeXVII\ ($\sim$ 12.12 \AA) and is unresolvable with the
HETGS, the dashed line shows the expected temperature dependence of the Ne H-He ratio if the
lines were resolved, whereas, the solid line represents the realistic observed ratio (i.e., the ratio of
the \FeXVII\ and \NeX\ line emission to the \NeIX\ $r$-line emission).  For \TX\ $>$ 8 MK, the
\NeIX\ $r$-line is enhanced by other Fe lines producing the drop in the \NeIX\ \HtoHe\ ratio.
\label{fig:HHERATIO}}
\end{figure}

To examine the possible effects of continuum absorption and escape probability effects, consider
the following simple illustration. If $H$ represents the observed \HtoHe\ ratio and $H_O$ 
represents the intrinsic temperature dependent \HtoHe\ ratio, then,
\begin{equation}
H~ =~ H_O~ \frac{1~ +~ ap_{He}}{1~ +~ ap_{H}}~exp~(~\tau_{He}~ -~ \tau_H~)
\label{eq:H-He}
\end{equation}
where $\tau_{He}$ and $\tau_H$ are respectively the X-ray continuum wind absorption optical
depths for the He-like and H-like lines, and $p_{He}$ and $p_H$ represent the associated
resonance line scattering $\tau$ using the approximate escape probability formalism developed by
Osterbrock (1974) (a is a constant equal to 0.58). First we examine the effects of continuum
absorption only (i.e., $p_{He}$ = $p_H$ = 0). Since the OB stellar wind opacities scale roughly
as $\lambda^3$, all H-He line pairs observable in HETGS data are expected to have
$\tau_{He}\geq \tau_H$ if both lines are formed at the same radial position.  Hence, for the case
of negligible resonance line scattering, $H$ would be expected to be larger than $H_O$ which
means that a temperature derived from $H$ will be greater than the actual value. There is one
exception to this scaling. For the oxygen H-He lines, the reverse is expected, i.e., $\tau_H \geq
\tau_{He}$ which would imply that the $H$ derived temperature is actually lower than the real
value (from Table \ref{tab:LINES} notice that the $\sigma_W$ of \OVIII\ is significantly larger
than $\sigma_W$ of \OVII). Now we consider the presence of finite resonance line scattering.  If
we again assume that both H-like and He-like lines are formed at the same density and
temperature, then $p_{He}$ = 2$p_H$ based on atomic physics considerations. Hence, an
additional increase in $H$ would also occur for the case of finite resonance line scattering.  For
example, consider an extreme case where $\tau_{He}$ - $\tau_H$ = 1 and $p_H$ = 1 ($p_{He}$
= 2), then $H$ = 3.7 $H_O$.  Now suppose we know that are observing a Si plasma at a
temperature of 10 MK ($H_O = 0.4$), then under these conditions the observed $H$ = 1.5
(Fig.\ref{fig:HHERATIO}).  The resultant observed H-He ratio would predict a temperature
$\sim$ 15 MK respectively.  However, a more likely scenario would be $\tau_{He}$ - $\tau_H$
= 0 since it seems that all observed line emissions seem to always arise from near their respective
X-ray continuum optical depth unity radii as shown in Figure \ref{fig:TAUFIR}.  For this case,
the $H$ would be equal to 1.4 $H_O$ and the predicted H-He temperature would be $\sim$ 11
MK, much closer to the actual temperature. Hence, we argue that these \HtoHe\ derived
temperatures provide very good estimates of the actual X-ray temperatures.

\onecolumn

\clearpage
\begin{deluxetable}{ccccccccc}
\tabletypesize{\scriptsize}
\tablecolumns{9}
\tablecaption{Adopted Stellar Parameters and HETGS Data Identification Numbers
\label{tab:PARM}}
\tablehead{
\colhead{Star}    & 
\colhead{Spectral}  & 
\colhead{HETGS}   & 
\colhead{d} &
\colhead{$\Teff$}     &
\colhead{$\Rstar$}  & 
\colhead{$\Mdot$} &
\colhead{$\vinf$}      & 
\colhead{$N_{WO}$}  \\
\colhead{ }                & 
\colhead{Type}      &
\colhead{Obs ID}  & 
\colhead{kpc} &
\colhead{K}              & 
\colhead{$\Rsun$} & 
\colhead{$10^{-6}\Msunyr$} &
\colhead{\kms}         & 
\colhead{$10^{22} cm^{-2}$}}
\startdata
\multicolumn{9}{l}{\textit{\underline{Supergiants (I, II)}}}  \\
\zpup    & O4 If          & 640     & 0.43 & 42400 & 16.5 & 2.40 & 2200 & 1.98 \\
\cygnin  & O5 f           & 2572   & 1.82 & 44700 & 34.0 & 12.70 & 2200 & 5.10 \\
\cygate  & O5.5 I(f)    & 2572   & 1.82 & 38500 & 27.9 & 13.50 & 2650 & 5.48 \\
\doriA   & O9.5 II      & 639     & 0.50  & 32900 & 17.0 & 1.07 & 2300 & 0.82 \\
\zoriA   & O9.7 Ib      & 610, 1524 & 0.50 & 30900 & 31.0 & 2.50 & 2100 & 1.15 \\
\eori    & B0 Ia           & 3753   & 0.46 & 28000 & 33.7 & 4.07 & 1500 & 2.42 \\
\multicolumn{9}{l}{\textit{\underline{Giants (IV, III)}}}  \\
HD150136 & O5 III(f)      & 2569 & 1.35 & 43000 & 16.0 & 3.98 & 3700 & 2.02 \\
\xper        & O7.5 III(n)(f) & 4512 & 0.40 & 36000 & 11.0 & 0.32 & 2600 & 0.33 \\
\iori          & O9 III         & 599, 2420 & 0.50 & 34000 & 17.8 & 1.10 & 2000 & 0.92 \\
\bcru        & B0.5 III       & 2575 & 0.15 & 27500 & 13.0 & 0.05 & 1600 & 0.07 \\
\multicolumn{9}{l}{\textit{\underline{Main Sequence}}}  \\
9 Sgr    & O4 V((f))      & 5398, 6285 & 1.60 & 43000 & 16.0 & 2.40 & 2950 & 1.53 \\
HD206267 & O6.5 V((f))    & 1888, 189 & 0.75 & 40500 & 12.7 & 0.63 & 3225 & 0.46 \\
15 Mon & O7 V((f))    & 5401, 6248, 6243 & 0.69 & 40100 & 9.7 & 0.50 & 2300 & 0.67 \\
\toriC   & O7 Vp          & 3, 4   & 0.55 & 38000 & 9.0 & 0.20 & 1650 & 0.40 \\
\zoph    & O9.5 Ve        & 2571, 4367 & 0.15 & 34000 & 8.0 & 0.13 & 1500 & 0.30 \\
\sori    & O9.5 V         & 3738 & 0.50 & 33000 & 9.0 & 0.08 & 1250 & 0.21 \\
\tsco    & B0 V           & 638, 2305 & 0.17 & 32000 & 6.2 & 0.03 & 2400 & 0.06 \\
\enddata

\tablecomments{The majority of stellar parameters are taken from Howarth \& Prinja (1989) and
Lamers \& Leitherer (1993), along with input from Koch \& Hrivnak (1981), Leitherer (1988),
Howarth et al. (1993), Cassinelli et al. (1994), and Waldron et al. (2004). Distances are taken
from Savage et al. (1977), Shull \& Van Steenberg (1985), and Bergh\"{o}fer et al. (1996).}
\tablecomments{$N_{WO}$ is the scale factor of the stellar wind column density defined as $N_{WO} =
\Mdot/4\pi \mu_{H} m_{H} \vinf \Rstar$ where $\mu_H$ and $m_H$ are respectively the mean
molecular weight and mass of H atom.}

\end{deluxetable}

\clearpage
\begin{deluxetable}{cccccccc}
\tablecolumns{8}
\tabletypesize{\footnotesize}
\tablecaption{Lines Used in Analysis, Wind and ISM Cross Sections, and Relevant $\ftoi$
Parameters \label{tab:LINES}}
\tablehead{
\colhead{Ion}               & 
\colhead{$\lamz$}                 & 
\colhead{$\TL$}   &   
\colhead{$\sigma_W$} & 
\colhead{$\sigma_{ISM}$}   & 
\colhead{$\lambda_{f-i}$} &   
\colhead{$N_C$}         & 
\colhead{$\phi_C$}               \\
\colhead{ } &   
\colhead{$\AA$}         & 
\colhead{MK}                       &   
\multicolumn{2}{c}{$10^{-22} cm^2$}  &   
\colhead{J = 0, 1, 2} &   
\colhead{$cm^{-3}$} &   
\colhead{photons $s^{-1}$}}
\startdata

\SXVI    & 4.727, 4.733    & 25.12 & 0.205 & 0.224 & \nodata & \nodata  & \nodata \\
\SXV(r)  & 5.039              & 15.85 & 0.240 & 0.247 & \nodata & \nodata  & \nodata \\
\SXV(i)  & 5.063, 5,066    & 12.59 & 0.242 & 0.249 & \nodata & \nodata  & \nodata  \\
\SXV(f)  & 5.102               & 15.85 & 0.247 & 0.255 & 673.9, 738.2, 756.0  & $1.9x10^{14}$ 
&   $9.2x10^5$  \\
\SiXIV    & 6.180, 6.186    & 15.85 & 0.401 & 0.433 & \nodata & \nodata & \nodata \\
\SiXIII(r) & 6.648              & 10.00 & 0.424 & 0.529 & \nodata & \nodata & \nodata \\
\SiXIII(i) & 6.685, 6.882    & 10.00 & 0.429 & 0.536 & \nodata & \nodata & \nodata \\
\SiXIII(f) & 6.740              & 10.00 & 0.438 & 0.496 & 815.2, 865.2, 878.4  &  $4.0x10^{13}$ 
&  $2.4x10^5$  \\
\MgXII & 8.419, 8.425       & 10.00 & 0.771 & 0.912 & \nodata & \nodata  & \nodata \\
\MgXI(r) & 9.169               & 6.31   & 0.897 & 1.148 & \nodata & \nodata & \nodata \\
\MgXI(i) & 9.228, 9.231    & 6.31    & 0.912 & 1.169 & \nodata & \nodata & \nodata \\
\MgXI(f) & 9.314               & 6.31   & 0.933 & 1.196 & 997.7, 1034.3, 1043.3 &
$6.2x10^{12}$  & $4.9x10^4$ \\
\NeX & 12.132, 12.138       & 6.31  & 1.825 & 2.290 & \nodata  & \nodata  & \nodata \\
\NeIX(r) & 13.447              & 3.98   & 2.138 & 3.021 & \nodata  & \nodata & \nodata \\
\NeIX(i) & 13.550, 13.553  & 3.98   & 2.125 & 3.082 & \nodata  & \nodata & \nodata \\
\NeIX(f) & 13.669               & 3.98  & 2.186 & 3.174 & 1247.8, 1273.2, 1277.7 & 
$6.4x10^{11}$  & $7.7x10^3$ \\
\FeXVII & 15.014               & 5.01  & 2.193 & 3.469 & \nodata & \nodata  & \nodata \\
\FeXVII & 16.780               & 5.01  & 2.917 & 4.672 & \nodata & \nodata  & \nodata \\
\OVIII & 18.967, 18.972     & 3.16  & 3.999 & 5.906 & \nodata & \nodata  & \nodata \\
\OVII(r) & 21.602               & 2.00  & 1.735 & 8.417 & \nodata & \nodata  & \nodata \\
\OVII(i) & 21.801, 21.804   & 2.00  & 1.736 & 8.631 & \nodata & \nodata  & \nodata \\
\OVII(f) & 22.098               & 2.00  & 1.796 & 8.959 & 1623.9, 1634.0, 1638.5 & 
$3.4x10^{10}$   & $7.3x10^2$ \\
\NVII & 24.779, 24.785      & 2.00  & 2.387 & 7.348 & \nodata & \nodata & \nodata \\

\enddata

\tablecomments{\lamz\ and \TL\ are taken from the APED line list. The $\lambda_{f-i}$ are the
EUV/UV wavelengths needed for the radiative excitation from the $2^3S_1\rightarrow$
$2^3P_J$ levels (J = 0, 1, 2). $N_C$ and $\phi_C$ are the critical density and critical
photo-excitation rate.  If $\phi$ = 0 then $N_C$ represents the density required to reduce the
$\ftoi$ ratio by one-half, whereas, if the density is $<< N_C$ then $\phi_C$ represents
the radiative excitation rate required to reduce the $\ftoi$ ratio by one-half.}

\end{deluxetable}

\clearpage
\begin{deluxetable}{ccccccccccc}
\tabletypesize{\tiny}
\rotate
\tablewidth{550pt}
\tablecolumns{11}
\tablecaption{Observed MEG \ftoi\ Line Ratios and $fir$-inferred Radii \label{tab:MGFI}}
\tablehead{\colhead{Star}
        &  \multicolumn{2}{c}{O VII}
        &  \multicolumn{2}{c}{Ne IX} 
        &  \multicolumn{2}{c}{Mg XI} 
        &  \multicolumn{2}{c}{Si XIII} 
        &  \multicolumn{2}{c}{S XV}
        \\  \colhead{ }
        &  \colhead{\ftoi}
        &  \colhead{\Rfir}
        &  \colhead{\ftoi}
        &  \colhead{\Rfir}
        &  \colhead{\ftoi}
        &  \colhead{\Rfir}
        &  \colhead{\ftoi}
        &  \colhead{\Rfir}
        &  \colhead{\ftoi}
        &  \colhead{\Rfir}}

\startdata
\multicolumn{4}{l}{\textit{\underline{Supergiants (I, II)}} } \\
 \zpup     &  $ 0.02\pm0.02$  & $ 7.67\pm3.29$  & $ 0.35\pm0.05$  & $10.56\pm0.81$ 
           & $ 0.33\pm0.03$  & $ 3.73\pm0.20$  & $ 1.16\pm0.11$  & $ 2.54\pm0.21$ 
           &  $ 0.72\pm0.24$ & $ \leq 1.22$   \\
 \cygnin   &    \nodata            &   \nodata           &   \nodata            & \nodata
           &   \nodata            &   \nodata           & $ 0.61\pm0.15$ & $ 1.78\pm0.27$ 
           &   \nodata            &   \nodata          \\
 \cygate   & \nodata              &  \nodata             & $ 0.51\pm0.24$ & $11.14\pm3.25$
           & $ 0.86\pm0.22$ & $ 5.61\pm0.99$ & $ 1.02\pm0.13$ & $ 1.91\pm0.18$ 
           & $ 0.75\pm0.23$ & $ \leq 1.18$         \\
 \doriA    &  $ 0.10\pm0.03$ & $13.32\pm2.36$ & $ \leq 0.001$       & $ \leq 1.01$
           &  $ 0.81\pm0.19$ & $ 4.43\pm0.69$  & $ 1.44\pm0.42$ & $ 1.53\pm0.43$ 
           &    \nodata            & \nodata               \\
 \zoriA    &  $ 0.12\pm0.02$ & $13.51\pm1.18$  & $ 0.27\pm0.05$ & $ 6.15\pm0.56$ 
           &  $ 1.10\pm0.14$ & $ 4.97\pm0.49$  & $ 2.62\pm0.56$ &  $ \geq 1.62$ 
           &  $ 0.31\pm0.26$  & $ \leq 1.02$        \\
 \eori     &  $ 0.10\pm0.03$  & $10.14\pm1.44$  & $ 0.51\pm0.08$ & $ 7.60\pm0.72$ 
           &  $ 1.21\pm0.24$  & $ 4.47\pm0.73$  & $ 1.97\pm0.52$  & $ 2.96\pm1.96$ 
           &   \nodata              &  \nodata                \\

\multicolumn{4}{l}{\textit{\underline{Giants (IV, III)}} } \\
 HD150136  & \nodata               & \nodata              & $ 0.12\pm0.04$  & $ 5.86\pm1.02$ 
           & $ 0.46\pm0.06$  & $ 4.51\pm0.37$ & $ 1.24\pm0.17$ & $ 2.81\pm0.37$ 
           & $ 2.50\pm1.13$  & $ \geq 1.76$       \\
 \xper     & $< 0.01$            & $< 4$                & $ 0.09\pm0.03$ & $ 4.10\pm0.74$ 
           &  $ 0.27\pm0.05$ & $ 2.58\pm0.28$ & $ 2.61\pm0.63$ & $ \geq 3.44$ 
           &   \nodata             & \nodata              \\
 \iori     &  $ 0.04\pm0.02$ & $ 8.63\pm2.14$ & $ 0.62\pm0.19$ & $11.14\pm2.13$ 
           &  $ 0.74\pm0.17$ & $ 4.33\pm0.67$ & $ 1.77\pm0.49$ & $ 2.80\pm1.32$ 
           &   \nodata             & \nodata               \\
 \bcru     &  $ \leq 0.03$        & $ \leq 5$             & \nodata                & \nodata
           & $ 0.28\pm0.17 $ & $ 1.70\pm0.55$  & $ 1.81\pm1.13$    & $ \geq 1.00$
           & \nodata               & \nodata                \\

\multicolumn{4}{l}{\textit{\underline{Main Sequence}} } \\
 9 Sgr     &  $ 0.30\pm0.12$ & $31.67\pm7.03$ & $ 0.02\pm0.02$ & $ 2.49\pm1.06$ 
           & $ 0.63\pm0.17$ & $ 5.46\pm0.92$  & $ 1.39\pm0.37$ & $ 3.31\pm0.97$ 
           & \nodata              &   \nodata              \\
 HD206267  & $ \leq 0.03$       & $ \leq 8.51$        & $ 0.48\pm0.19$ & $12.02\pm2.94$ 
           & $0.30\pm0.12$  & $ 3.32\pm0.76$  & $ 0.33\pm0.18$ & $ \leq 1.32$ 
           & \nodata              & \nodata                \\
15 Mon     & $ 0.07\pm0.05$  & $13.64\pm5.32$ & $ \leq 0.02$      & $ \leq 2.15$
           & $0.29\pm0.24$   & $2.96\pm1.59$   & $1.79\pm1.44$ & $ \geq 1.11$ 
           & \nodata               & \nodata               \\
 \toriC    & \nodata               & \nodata               & $ 0.23\pm0.08$ & $ 7.09\pm1.31$ 
           & $ 0.12\pm0.03$  & $ 1.80\pm0.21$  & $ 1.73\pm0.16$  & $ 3.26\pm0.49$ 
           &  $ 1.38\pm0.30$  & $ 1.53\pm0.43$   \\
 \zoph     & $ 0.00$               & $1.00$               & $ \leq 0.02$    & $ \leq 2.12$ 
           & $ 0.25\pm0.06$  & $ 2.31\pm0.28$  & $ 1.47\pm0.29$ & $ 1.76\pm0.38$ 
           & $ 0.64\pm0.41$  & $ \leq 1.02$        \\
 \sori     & $0.00$              & $1.00$              & $ 0.13\pm0.05$ & $ 4.50\pm0.87$ 
           & $ \leq 0.04$        & $ \leq 1.04$        & $ 0.34\pm0.24$ & $ \leq 1.02$ 
           & \nodata               & \nodata               \\
 \tsco     & $0.00$                & $1.00$            & $ \leq 0.01$ & $ \leq 1.32$ 
           & $ 0.36\pm0.05$  & $ 2.61\pm0.20$   &  $ 2.49\pm0.32$ & $ \geq 2.35$ 
           & $ 1.75\pm0.57$  & $ \geq 1.00$     \\

\enddata
\end{deluxetable}

\clearpage
\begin{deluxetable}{ccccccccc}
\tabletypesize{\scriptsize}
\rotate
\tablewidth{520pt}
\tablecolumns{9}
\tablecaption{Observed HEG \ftoi\ Line Ratios and $fir$-inferred Radii \label{tab:HGFI}}
\tablehead{\colhead{Star}
        &  \multicolumn{2}{c}{Ne IX} 
        &  \multicolumn{2}{c}{Mg XI} 
        &  \multicolumn{2}{c}{Si XIII} 
        &  \multicolumn{2}{c}{S XV}
        \\  \colhead{ }
        &  \colhead{\ftoi}
        &  \colhead{\Rfir}
        &  \colhead{\ftoi}
        &  \colhead{\Rfir}
        &  \colhead{\ftoi}
        &  \colhead{\Rfir}
        &  \colhead{\ftoi}
        &  \colhead{\Rfir}}

\startdata
\multicolumn{4}{l}{\textit{\underline{Supergiants (I, II)}} } \\
 \zpup    & $ 0.13\pm0.05$ & $ 6.15\pm1.16$ 
          & $ 0.35\pm0.06$  & $ 3.79\pm0.39$ & $ 0.98\pm0.16$ & $ 2.20\pm0.28$ 
          & $ 1.13\pm0.60$  & $ \leq 2.73$        \\
 \cygnin  & \nodata               & \nodata
          & $ \leq 1.30$        & $ \leq 8.81$       & $ \leq 0.99$       & $ \leq 2.06$
          & \nodata               & \nodata             \\ 
 \cygate  & $ \leq 0.09$        & $ \leq 4.33$
          &  $ 0.96\pm0.34$ & $ 6.12\pm1.58$  & $ 1.15\pm0.20$ & $ 2.11\pm0.33$ 
          &  $ 0.88\pm0.29$ & $ \leq 1.20$        \\
 \doriA   &   \nodata             & \nodata
          & $ 0.86\pm0.47$  & $ 4.61\pm1.77$  & $ 2.54\pm1.30$  & $ \geq 1.23$ 
          & \nodata               & \nodata               \\
 \zoriA   & $ 0.30\pm0.10$ & $ 6.44\pm1.20$ 
          & $ 0.62\pm0.18$ & $ 3.37\pm0.60$   & $ 1.51\pm0.52$ & $ \leq 1.56$ 
          & $ \leq 0.16$       & $ \leq 1.02$          \\
 \eori    & $ 0.30\pm0.12$ & $ 5.56\pm1.21$ 
          & $ 2.12\pm0.64$ & $11.27\pm5.94$  & $ 3.71\pm2.35$ & $ \geq 1.00$ 
          & \nodata                        & \nodata       \\

\multicolumn{4}{l}{\textit{\underline{Giants (IV, III)}} } \\
 HD150136 & $ 0.20\pm0.13$ & $ 7.47\pm2.82$ 
          & $ 0.35\pm0.08$ & $ 3.87\pm0.51$ & $ 1.16\pm0.27$ & $ 2.67\pm0.55$ 
          & \nodata              & \nodata               \\
 \xper    & \nodata              & \nodata
          & $ 0.33\pm0.10$ & $ 2.89\pm0.50$  & $ 0.31\pm0.11$ & $ \leq 1.02$ 
          & \nodata              & \nodata                \\
 \iori    & $ 0.55\pm0.43$ & $ 9.86\pm5.38$ 
          & $ 0.69\pm0.20$ & $4.14\pm0.78$    & $ \leq 0.02$      & $ \leq 1.02$ 
          & \nodata              & \nodata                 \\
 \bcru    & \nodata              & \nodata
          & \nodata              & \nodata              &  \nodata             & \nodata
          &  \nodata             &  \nodata              \\
\multicolumn{4}{l}{\textit{\underline{Main Sequence}} } \\
 9 Sgr    & $ \leq 0.11$        & $ \leq 5.85$ 
          & $ 0.54\pm0.22$  & $ 4.98\pm1.24$ &  \nodata             & \nodata 
          & \nodata               & \nodata               \\
 HD206267 & \nodata              & \nodata
          & \nodata              & \nodata              &  \nodata             & \nodata
          &  \nodata             &  \nodata              \\
15 Mon    & $ \leq 0.20$        & $ \leq 7.20$
          & $ 0.78\pm0.42$  & $ 5.84\pm2.20$  & \nodata             & \nodata
          & \nodata               & \nodata               \\
 \toriC   & $ \leq 0.01$         & $ \leq 1.20$
          & $ 0.41\pm0.10$   & $ 3.51\pm0.52$ & $ 1.10\pm0.14$ & $ 1.96\pm0.21$ 
          & $ 0.68\pm0.17$   & $ \leq 1.02$        \\
 \zoph    & $ 0.18\pm0.11$   & $ 5.38\pm1.87$ 
          & $ 0.27\pm0.10$  & $ 2.38\pm0.49$ & $ 1.37\pm0.49$ & $ 1.76\pm0.61$ 
          & \nodata               & \nodata               \\
 \sori    & $ 0.16\pm0.15$  & $ 4.38\pm2.87$ 
          & \nodata               & \nodata             & \nodata               & \nodata
          & \nodata               & \nodata              \\
 \tsco    & $ 0.05\pm0.03$  & $ 2.53\pm0.98$ 
          & $ 0.29\pm0.10$ & $ 2.31\pm0.41$  & $ 1.77\pm0.34$ & $ 1.69\pm0.49$ 
          & \nodata              & \nodata               \\

\enddata
\end{deluxetable}

\clearpage
\begin{deluxetable}{ccccccccccc}
\tabletypesize{\tiny}
\rotate
\tablewidth{550pt}
\tablecolumns{11}
\tablecaption{Observed MEG \HtoHe\ Line Ratios and Derived \THHe\ \label{tab:MGHHE}}
\tablehead{\colhead{Star}
        &  \multicolumn{2}{c}{Oxygen}
        &  \multicolumn{2}{c}{Neon} 
        &  \multicolumn{2}{c}{Magnesium} 
        &  \multicolumn{2}{c}{Silicon} 
        &  \multicolumn{2}{c}{Sulfur}
        \\  \colhead{ }
        &  \colhead{\HtoHe}
        &  \colhead{\THHe}
        &  \colhead{\HtoHe}
        &  \colhead{\THHe}
        &  \colhead{\HtoHe}
        &  \colhead{\THHe}
        &  \colhead{\HtoHe}
        &  \colhead{\THHe}
        &  \colhead{\HtoHe}
        &  \colhead{\THHe}}

\startdata
\TL\ range & \nodata  & 2.00 - 3.16   & \nodata & 3.98 - 6.31 
                  & \nodata & 6.31 - 10.00  & \nodata & 10.00 - 15.85 
                  & \nodata & 15.85 - 25.12 \\
\multicolumn{4}{l}{\textit{\underline{Supergiants (I, II)}} } \\
 \zpup     &  $ 1.57\pm0.20$ & $ 2.75\pm0.10$ & $ 0.84\pm0.05$ & $ 3.81\pm0.06$ 
           &  $ 0.40\pm0.03$ & $ 6.07\pm0.12$ & $ 0.16\pm0.02$ & $ 7.68\pm0.22$ 
           & \nodata               &  \nodata             \\
 \cygnin   &  \nodata              &  \nodata              & \nodata              & \nodata
           & $ 3.26\pm0.92$  & $12.48\pm1.53$ & $ 1.28\pm0.30$  & $13.93\pm1.25$ 
           & \nodata               & \nodata               \\
 \cygate   & \nodata               &  \nodata             &  \nodata              & \nodata
           & $ 1.23\pm0.15$  & $ 8.50\pm0.38$  & $ 1.08\pm0.09$  & $13.19\pm0.42$ 
           & $ 1.50\pm0.38$  & $23.05\pm2.42$   \\
 \doriA    & $ 1.34\pm0.13$  & $ 2.62\pm0.07$ & $ 0.80\pm0.09$  & $ 3.76\pm0.11$ 
           & $ 0.31\pm0.05$  & $ 5.73\pm0.24$ & $ 0.57\pm0.14$  & $10.64\pm0.80$ 
           &  \nodata              &  \nodata              \\
 \zoriA    & $ 1.14\pm0.07$  & $ 2.51\pm0.04$  & $ 0.48\pm0.04$  & $ 3.28\pm0.07$ 
           & $ 0.26\pm0.04$  & $ 5.50\pm0.18$  & $ 0.21\pm0.05$ & $ 8.08\pm0.53$ 
           &  \nodata              &  \nodata               \\
 \eori     & $ 1.18\pm0.10$  & $ 2.53\pm0.06$   & $ 0.57\pm0.05$ & $ 3.42\pm0.09$ 
           & $ 0.17\pm0.03$  & $ 5.01\pm0.19$   & $ 0.03\pm0.02$ & $ 5.07\pm0.97$ 
           & \nodata               &  \nodata               \\
mean \THHe & \nodata               & $2.60\pm0.07$     &  \nodata             & $3.57\pm0.08$
           & \nodata              & $7.22\pm0.66$     &   \nodata             & $9.76\pm0.78$
           & \nodata              & $23.05\pm2.42$       \\

\multicolumn{4}{l}{\textit{\underline{Giants (IV, III)}} } \\
 HD150136  &  \nodata              &  \nodata            & $ 1.68\pm0.18$ & $ 4.68\pm0.15$ 
           & $ 0.44\pm0.04$ & $ 6.20\pm0.14$ & $ 0.50\pm0.06$ & $10.27\pm0.36$ 
           &  $ 0.76\pm0.25$ & $17.71\pm2.10$   \\
 \xper     &  $ 2.74\pm0.43$ & $ 3.26\pm0.17$ & $ 0.92\pm0.09$ & $ 3.91\pm0.11$ 
           & $ 0.15\pm0.03$  & $ 4.87\pm0.17$ & $ 0.11\pm0.04$ & $ 6.99\pm0.54$ 
           &  \nodata              &  \nodata              \\
 \iori     & $ 1.35\pm0.13$ & $ 2.63\pm0.08$  & $ 0.18\pm0.03$ & $ 2.58\pm0.08$ 
           & $ 0.32\pm0.08$ & $ 5.73\pm0.35$ & $ 0.64\pm0.20$ & $11.00\pm1.09$ 
           &  \nodata             &  \nodata              \\
 \bcru     &  $ 1.19\pm0.18$ & $ 2.54\pm0.11$ & $ 0.44\pm0.09$ & $ 3.19\pm0.17$ 
           &  \nodata             &  \nodata              &  \nodata             & \nodata
           &  \nodata             &  \nodata               \\
mean \THHe &  \nodata             & $2.81\pm0.13$  & \nodata             & $3.59\pm0.13$
           &  \nodata             & $5.60\pm0.24$  &  \nodata            & $9.42\pm0.63$
           &  \nodata             & $17.71\pm2.10$      \\

\multicolumn{4}{l}{\textit{\underline{Main Sequence}} } \\
 9 Sgr     &  $ 2.63\pm0.55$ & $ 3.22\pm0.22$ & $ 0.45\pm0.05$ & $ 3.22\pm0.10$
           & $ 0.30\pm0.05$ & $ 5.68\pm0.24$ & $ 0.17\pm0.05$ & $ 7.72\pm0.50$
           & \nodata             &  \nodata              \\
 HD206267  &  \nodata             &  \nodata              & $ 1.36\pm0.27$ & $ 4.38\pm0.26$ 
           & $ 0.24\pm0.08$ & $ 5.35\pm0.43$  & $ 0.13\pm0.08$ & $ 7.06\pm1.05$ 
           & \nodata              &  \nodata              \\
15 Mon     & $ 0.74\pm0.11$ & $ 2.24\pm0.08$  & $ 0.12\pm0.04$ & $2.41\pm0.14$
           & $ 0.08\pm0.05$ & $ 4.25\pm0.54$  & \nodata              & \nodata
           & \nodata              & \nodata               \\
 \toriC    & \nodata              & \nodata              & $ 2.46\pm0.25$ & $ 5.32\pm0.21$ 
           & $ 2.36\pm0.18$ & $10.97\pm0.35$ & $ 1.76\pm0.09$ & $15.80\pm0.31$ 
           & $ 1.39\pm0.16$ & $22.49\pm1.06$  \\
 \zoph     &  $ 1.73\pm0.28$ & $ 2.83\pm0.14$ & $ 1.39\pm0.14$ & $ 4.42\pm0.13$
           & $ 0.51\pm0.08$ & $ 6.49\pm0.28$ & $ 0.38\pm0.08$ & $ 9.48\pm0.50$
           & \nodata             &  \nodata              \\
 \sori     &  $ 1.76\pm0.32$ & $ 2.84\pm0.16$ & $ 0.69\pm0.12$ & $ 3.60\pm0.17$ 
           & $ 0.04\pm0.03$  & $ 2.13\pm2.12$ &  \nodata             &  \nodata
           &  \nodata             &  \nodata              \\
 \tsco     &  $ 3.81\pm0.49$ & $ 3.66\pm0.17$ & $ 1.73\pm0.12$ & $ 4.72\pm0.10$ 
           & $ 0.84\pm0.07$  & $ 7.50\pm0.19$ & $ 0.61\pm0.06$ & $10.92\pm0.35$
           & $ 0.86\pm0.26$ & $18.57\pm2.07$        \\
mean \THHe &  \nodata             & $2.96\pm0.16$  &  \nodata             & $3.79\pm0.16$
           &  \nodata             & $5.23\pm0.86$  &  \nodata             & $8.80\pm0.65$
           &  \nodata             & $18.57\pm2.07$        \\
\enddata
\tablecomments{\HtoHe\ ratios are ISM corrected and \THHe\ are the derived X-ray temperatures in
MK. The results for \toriC\ are not included in the mean \THHe.}
\end{deluxetable}

\clearpage
\begin{deluxetable}{ccccccccc}
\tabletypesize{\scriptsize}
\rotate
\tablewidth{520pt}
\tablecolumns{9}
\tablecaption{Observed HEG \HtoHe\ Line Ratios and Derived \THHe\ \label{tab:HGHHE}}
\tablehead{\colhead{Star}
        &  \multicolumn{2}{c}{Neon} 
        &  \multicolumn{2}{c}{Magnesium} 
        &  \multicolumn{2}{c}{Silicon} 
        &  \multicolumn{2}{c}{Sulfur}
        \\  \colhead{ }
        &  \colhead{\HtoHe}
        &  \colhead{\THHe}
        &  \colhead{\HtoHe}
        &  \colhead{\THHe}
        &  \colhead{\HtoHe}
        &  \colhead{\THHe}
        &  \colhead{\HtoHe}
        &  \colhead{\THHe}}

\startdata
\TL\ range & \nodata & 3.98 - 6.31 
           & \nodata & 6.31 - 10.00  & \nodata & 10.00 - 15.85 
           & \nodata & 15.85 - 25.12 \\
\multicolumn{4}{l}{\textit{\underline{Supergiants (I, II)}} } \\
 \zpup     & $ 0.80\pm0.09$ & $ 3.76\pm0.12$
           & $ 0.28\pm0.04$ & $ 5.56\pm0.18$ & $ 0.10\pm0.03$ & $ 6.86\pm0.41$ 
           &  \nodata             & \nodata               \\
 \cygnin   & \nodata              & \nodata
           & \nodata              & \nodata              &  \nodata             & \nodata
           &  \nodata             &  \nodata              \\
 \cygate   &  \nodata             & \nodata
           & $ 1.30\pm0.24$ & $ 8.67\pm0.60$ & $ 1.08\pm0.16$ & $13.16\pm0.69$
           & $ 0.82\pm0.27$ & $18.25\pm2.21$  \\
 \doriA    & $ 0.18\pm0.06$ & $ 2.60\pm0.19$
           & $ 1.00\pm0.33$ & $ 7.90\pm0.87$ & $ 0.24\pm0.10$ & $ 8.34\pm0.92$
           &  \nodata             &  \nodata             \\
 \zoriA    & $ 0.39\pm0.06$ & $ 3.10\pm0.12$
           & $ 0.28\pm0.06$  & $ 5.55\pm0.29$ & $ 0.30\pm0.10$ & $ 8.81\pm0.81$ 
           &  \nodata              &  \nodata              \\
 \eori     & $ 0.48\pm0.09$ & $ 3.28\pm0.16$
           & $ 0.26\pm0.07$ & $ 5.47\pm0.37$ & $ 0.10\pm0.05$ & $ 6.81\pm0.80$ 
           &  \nodata             &  \nodata             \\
mean \THHe & \nodata              & $3.18\pm0.15$
           & \nodata              & $6.63\pm0.52$   & \nodata                & $8.80\pm0.75$
           & \nodata              & $18.25\pm2.21$     \\

\multicolumn{4}{l}{\textit{\underline{Giants (IV, III)}} } \\
 HD150136  & $ 0.54\pm0.16$ & $ 3.36\pm0.27$ 
           & $ 0.46\pm0.08$ & $ 6.29\pm0.28$ & $ 0.50\pm0.09$ & $10.28\pm0.59$ 
           &  \nodata             &  \nodata              \\
 \xper     &  $ 0.34\pm0.09$ & $ 3.00\pm0.19$
           & $ 0.22\pm0.07$ & $ 5.27\pm0.39$ & $ 0.23\pm0.10$ & $ 8.23\pm0.95$
           & \nodata              &  \nodata              \\
 \iori     & $ 0.86\pm0.23$ & $ 3.81\pm0.29$
           & \nodata              & \nodata              &  \nodata             & \nodata
           &  \nodata             &  \nodata              \\
 \bcru     & \nodata              & \nodata
           & \nodata              & \nodata              &  \nodata             & \nodata
           &  \nodata             &  \nodata              \\
mean \THHe & \nodata              & $3.39\pm0.25$
           & \nodata              & $6.00\pm0.34$   & \nodata                & $9.26\pm0.79$
           & \nodata              & \nodata               \\

\multicolumn{4}{l}{\textit{\underline{Main Sequence}} } \\
 9 Sgr     & $ 1.06\pm0.27$ & $ 4.06\pm0.31$
           & $ 0.22\pm0.08$ & $ 5.23\pm0.42$ & $ 0.37\pm0.13$ & $ 9.38\pm0.94$ 
           & \nodata              & \nodata               \\
 HD206267  & \nodata              & \nodata
           & \nodata              & \nodata              &  \nodata             & \nodata
           &  \nodata             &  \nodata              \\
15 Mon     & $ 0.27\pm0.11$ & $ 2.80\pm0.29$
           &  \nodata             & \nodata              &  \nodata             & \nodata
           &  \nodata             & \nodata    \\
 \toriC    &  \nodata             &  \nodata          
           & $ 3.07\pm0.38$ & $12.25\pm0.64$ & $ 1.68\pm0.13$ & $15.52\pm0.46$
           & $ 1.00\pm0.15$ & $19.74\pm1.09$      \\
 \zoph     & $ 1.14\pm0.22$ & $ 4.15\pm0.25$
           & $ 0.69\pm0.15$  & $ 7.04\pm0.47$  & $ 0.29\pm0.10$ & $ 8.74\pm0.81$
           &  \nodata              &  \nodata               \\
 \tsco     & $ 1.79\pm0.26$ & $ 4.76\pm0.21$
           & $ 0.84\pm0.12$  & $ 7.49\pm0.32$  & $ 0.68\pm0.10$ & $11.28\pm0.55$
           & $ 0.15\pm0.09$  & $10.89\pm1.71$     \\
mean \THHe & \nodata              & $3.94\pm0.27$
           & \nodata              & $6.59\pm0.41$   & \nodata                & $9.80\pm0.78$
           & \nodata              & $10.89\pm1.71$     \\
\enddata
\tablecomments{\HtoHe\ ratios are ISM corrected and \THHe\ are the derived X-ray temperatures in
MK. The results for \toriC\ are not included in the mean \THHe.}
\end{deluxetable}

\clearpage
\begin{deluxetable}{ccccccccccc}
\tabletypesize{\tiny}
\rotate
\tablewidth{520pt}
\tablecolumns{11}
\tablecaption{Observed MEG and HEG G-Ratios \label{tab:GRATIO}}
\tablehead{\colhead{Star}
        &  \multicolumn{2}{c}{Oxygen}
        &  \multicolumn{2}{c}{Neon} 
        &  \multicolumn{2}{c}{Magnesium} 
        &  \multicolumn{2}{c}{Silicon} 
        &  \multicolumn{2}{c}{Sulfur}
        \\  \colhead{ }
        &  \colhead{MEG}
        &  \colhead{HEG}
        &  \colhead{MEG}
        &  \colhead{HEG}
        &  \colhead{MEG}
        &  \colhead{HEG}
        &  \colhead{MEG}
        &  \colhead{HEG}
        &  \colhead{MEG}
        &  \colhead{HEG}}

\startdata
\multicolumn{4}{l}{\textit{\underline{Supergiants (I, II)}} } \\
 \zpup     &  $ 1.09\pm0.16$   & \nodata  & $ 0.48\pm0.04$  & $0.64\pm0.09$ 
           &  $ 0.82\pm0.05$  & $ 0.77\pm0.08$  & $ 0.96\pm0.06$  & $ 1.14\pm0.14$ 
           &  $ 2.45\pm0.67$  & $ 1.21\pm0.47$   \\
 \cygnin   & \nodata                & \nodata               &  \nodata               & \nodata
           & $ 0.50\pm0.21$   & \nodata               & $ 1.90\pm0.40$   & \nodata 
           &  \nodata               &  \nodata               \\
 \cygate   &  \nodata               &  \nodata              &  \nodata              & \nodata
           &  $ 0.40\pm0.06$  & $ 0.68\pm0.15$   & $ 0.90\pm0.08$ & $ 1.31\pm0.18$ 
           &  $ 1.87\pm0.46$  & $ 1.99\pm0.56$   \\
 \doriA    &  $ 0.79\pm0.09$  & \nodata                & $ 0.60\pm0.08$  & $ 0.44\pm0.16$ 
           &  $ 0.62\pm0.09$  & $0.84\pm0.30$    & $ 0.91\pm0.18$  & $ 0.67\pm0.19$ 
           & $ 2.11\pm1.51$   & $ 0.75\pm0.50$   \\
 \zoriA    &  $ 0.94\pm0.07$  & \nodata                & $ 0.54\pm0.05$ & $0.54\pm0.09$ 
           &  $ 1.03\pm0.09$  & $ 0.69\pm0.12$   & $ 1.21\pm0.17$  & $ 0.92\pm0.22$ 
           & $1.43\pm0.80$    & $\le 0.48$    \\
 \eori     &  $ 0.98\pm0.10$  & \nodata                & $ 0 .86\pm0.08$ & $ 0.62\pm0.12$ 
           &  $ 0.75\pm0.09$ & $ 0.97\pm0.19$    & $ 1.02\pm0.18$  & $ 0.43\pm0.13$ 
           &  \nodata              &  \nodata              \\

\multicolumn{4}{l}{\textit{\underline{Giants (IV, III)}} } \\
 HD150136  &   \nodata             &  \nodata             & $ 0.87\pm0.11$ & $ 1.01\pm0.32$ 
           & $ 0.64\pm0.05$ & $0.95\pm0.13$  & $ 0.76\pm0.07$ & $ 0.88\pm0.14$ 
           & $ 1.20\pm0.36$ & $ 0.70\pm0.32$    \\
 \xper     &  $ 0.53\pm0.13$ & \nodata               & $ 0.66\pm0.08$ & $ 0.77\pm0.21$ 
           &  $ 0.79\pm0.09$ & $ 1.19\pm0.23$  & $ 0.91\pm0.13$ & $ 1.33\pm0.34$ 
           & $ 1.02\pm0.66$  & \nodata       \\
 \iori     &  $ 0.91\pm0.11$ & \nodata                & $ 0.18\pm0.03$ & $0.31\pm0.13$ 
           & $ 0.90\pm0.14$ & $ 3.52\pm1.05$    & $ 1.85\pm0.42$ & $ 1.06\pm0.40$ 
           & \nodata              &  \nodata               \\
 \bcru     & $ 1.44\pm0.24$  & \nodata                 & $ 0.96\pm0.18$ & \nodata 
           & $ 1.03\pm0.37$ & \nodata                  & $ 1.34\pm0.61$ & \nodata 
           &  \nodata             & \nodata                 \\

\multicolumn{4}{l}{\textit{\underline{Main Sequence}} } \\
 9 Sgr     & $ 1.46\pm0.35$  & \nodata               & $ 0.45\pm0.06$ & $0.71\pm0.23$ 
           & $ 0.32\pm0.05$  & $0.77\pm0.19$    & $ 0.52\pm0.09$ & $0.67\pm0.19$ 
           & \nodata               &  \nodata                \\
 HD206267  & $0.18\pm0.14$   & \nodata                & $ 0.69\pm0.16$ & \nodata 
           & $ 0.70\pm0.15$  & \nodata                & $ 0.66\pm0.20$ & \nodata 
           &  \nodata              &  \nodata                \\
15 Mon     & $ 0.57\pm0.11$  & \nodata               & $0.94\pm0.18$  & $ 0.53\pm0.22$
           & $ 0.56\pm0.23$  & $ 2.95\pm1.59$  & $0.68\pm0.33$  & $ 0.62\pm0.40$
           &  \nodata              &  \nodata                \\
 \toriC    & $ 1.40\pm0.49$  & \nodata               & $ 0.51\pm0.08$ & $0.89\pm0.34$ 
           & $ 0.69\pm0.07$  & $ 0.95\pm0.15$  & $ 0.72\pm0.04$  & $0.92\pm0.08$ 
           & $ 0.73\pm0.10$  & $0.83\pm0.13$       \\
 \zoph     & $ 0.92\pm0.19$  & \nodata               & $ 0.67\pm0.08$  & $ 0.60\pm0.16$ 
           & $ 0.78\pm0.09$  & $ 1.00\pm0.22$   & $ 0.98\pm0.13$  & $ 0.84\pm0.20$ 
           & $ 1.99\pm1.06$  & \nodata    \\
 \sori     &  $ 1.09\pm0.24$ & \nodata                & $ 1.12\pm0.18$  & $ 0.58\pm0.22$ 
           &  $ 0.89\pm0.23$ & $ 1.06\pm0.53$  & $ 1.26\pm0.59$   & $ 0.74\pm0.41$
           &  \nodata              &  \nodata               \\
 \tsco     &  $ 1.59\pm0.25$  & \nodata              & $ 0.89\pm0.08$  & $ 0.78\pm0.15$ 
           &  $ 0.80\pm0.07$  & $0.58\pm0.10$  & $ 1.04\pm0.09$  & $ 1.14\pm0.15$ 
           &  $ 1.80\pm0.47$  & $ 0.53\pm0.22$         \\
\enddata
\tablecomments{G-Ratios are ISM corrected.}
\end{deluxetable}


\clearpage
\clearpage
\clearpage
\clearpage
\clearpage
\clearpage
\begin{figure}
\includegraphics[height=18cm,angle=180]{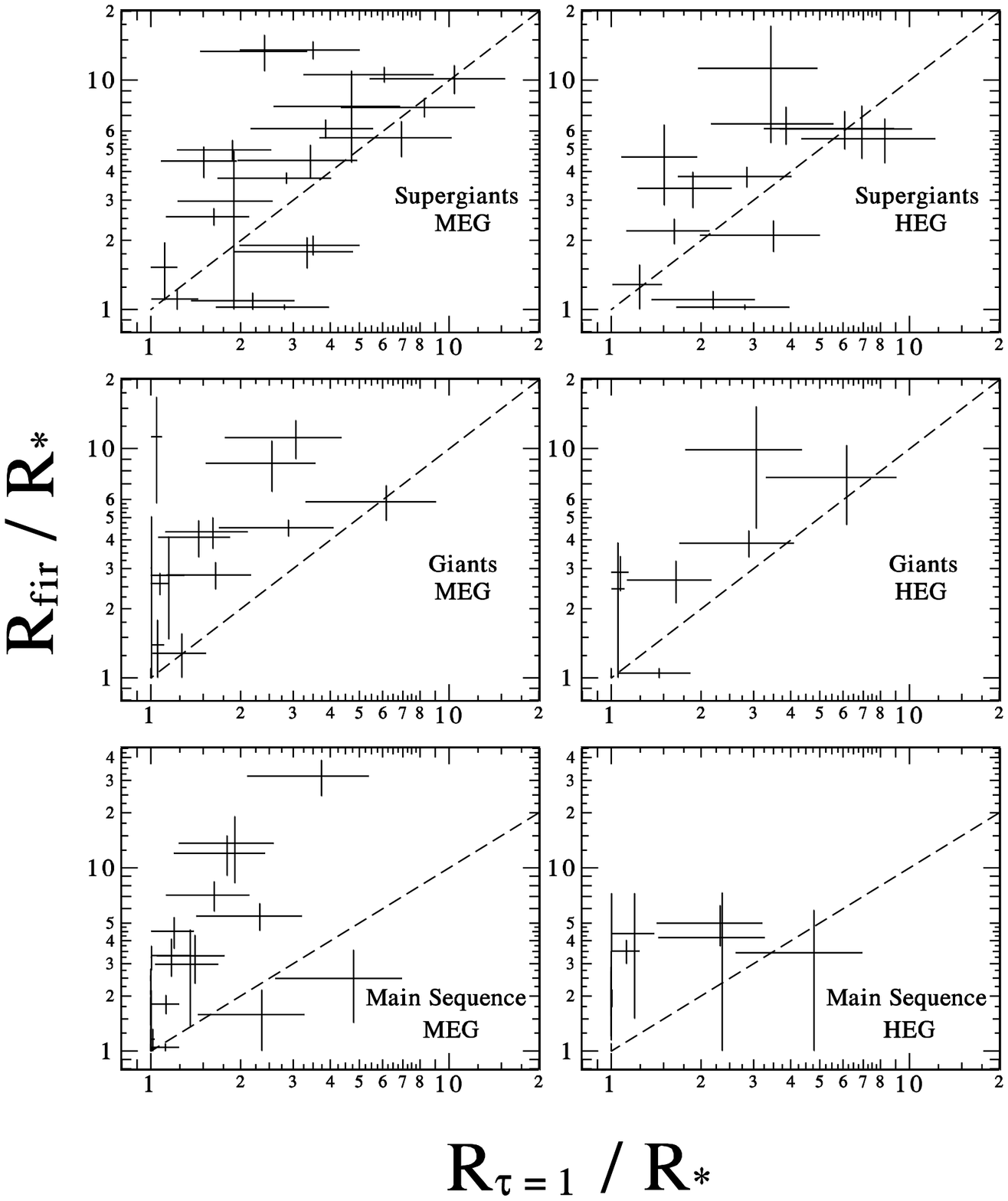}
\caption{MEG and HEG scatter plots showing the relation between \Rfir\ and \Rtau\ for all
luminosity classes. The error bars for \Rtau\ are determined for a X-ray continuum optical depth
of $1\pm0.5$ and the adopted value of \Rtau\ is the average of these limits.
\label{fig:TAUFIR}}
\end{figure}
\clearpage
\begin{figure}
\includegraphics[height=18cm,angle=180]{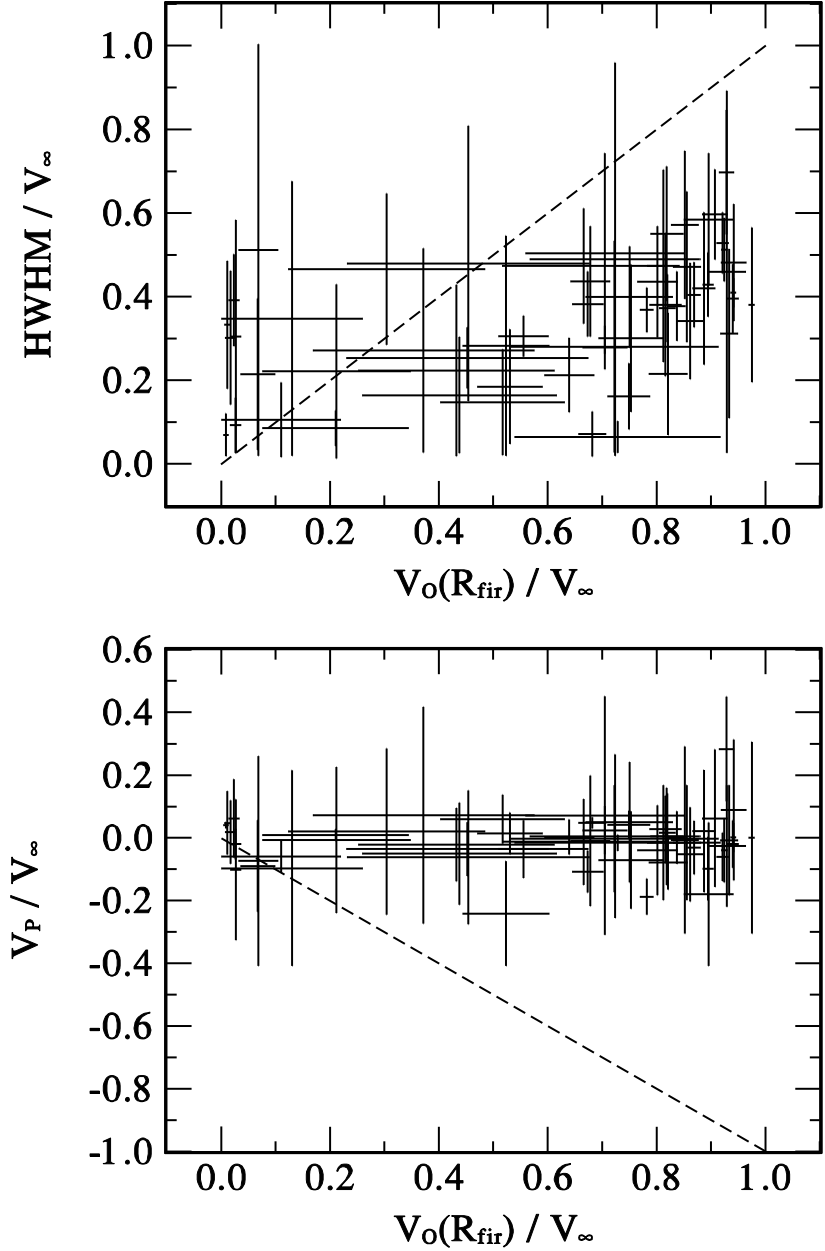}
\caption{
Scatter plots showing the dependence of the MEG derived He-like ion \HWHM\ (top panel)
and \VP\ (bottom panel) on the ambient wind velocity, $V_O$, for all OB luminosity classes.  All
quantities are determined for each at their \Rfir\ value and normalized to their star's \vinf.  The
dashed-lines represent special cases when $\HWHM = V_O$ and $\VP = - V_O$
(negative means the expected maximum blue-shift velocity). The \HWHM\ range from 0 to 0.6
\vinf\ with no clear indication of any dependence on $V_O$.  The \HWHM\ $<$ 0.2 \vinf\ pose a
problem since they are $>$ the local ambient velocities at these low radial positions. All \VP\ are
concentrated about zero velocity with no indication of any large blue-shifted lines.
\label{fig:HWVP_FIR}}
\end{figure}
\clearpage
\begin{figure}
\includegraphics[height=18cm,angle=180]{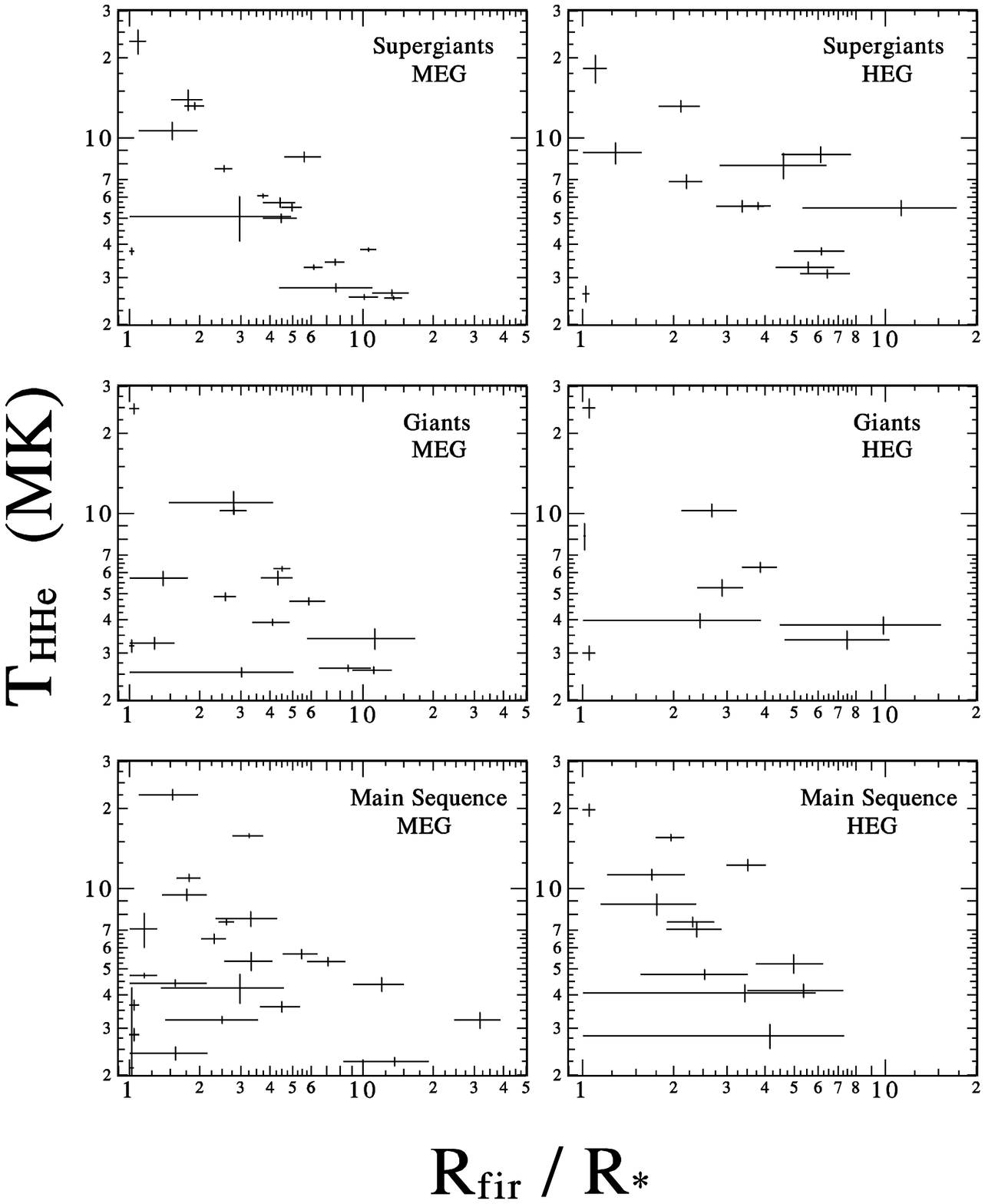}
\caption{MEG and HEG scatter plots showing the dependence of \TX\ (MK) (determined from
\HtoHe\ line ratios) on the He-like \fir-inferred radii, \Rfir\ (determined from He-like \ftoi\ line
ratios), for all OB luminosity classes.
\label{fig:THHEFIR}}
\end{figure}
\clearpage
\begin{figure}
\includegraphics[angle=-90,width=\linewidth]{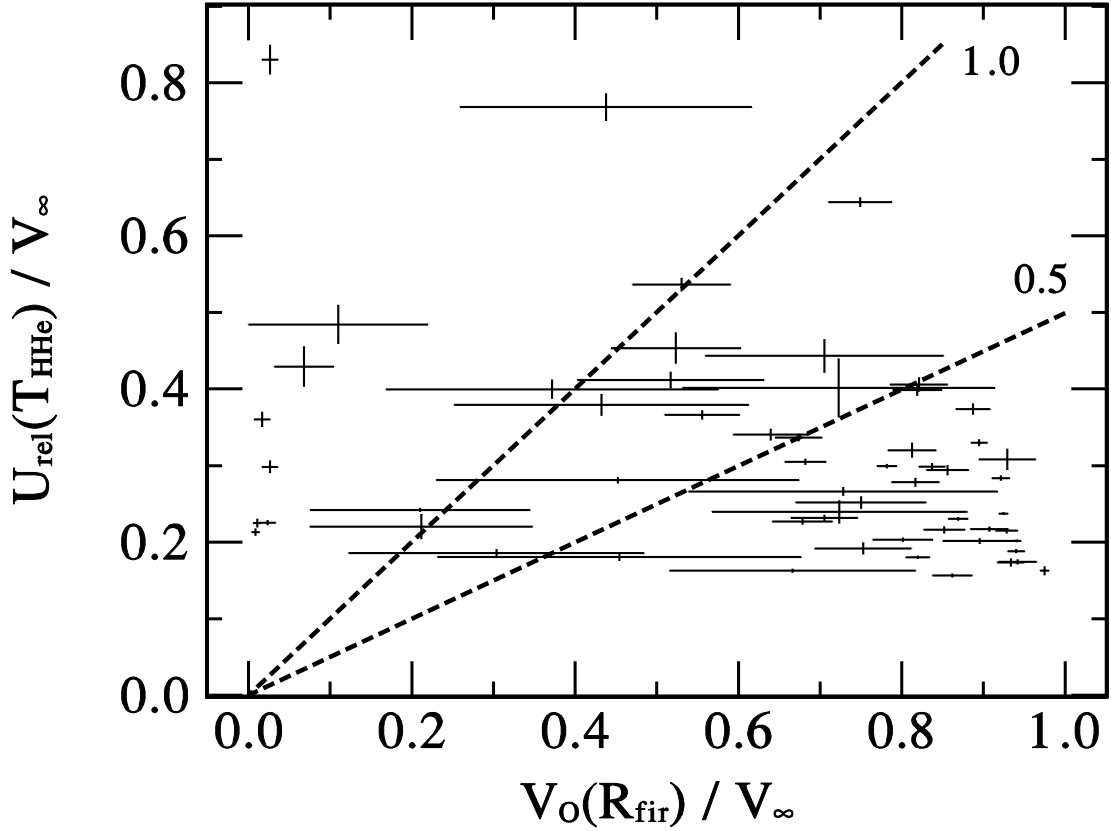}
\caption{Scatter plot of \UO\ (pre-shock velocity relative to the shock front) versus $V_O$
(the ambient wind velocity) for all OB luminosity classes.  The values of \UO\ are determined
from the MEG derived values of \THHe\ using eq. \ref{eq:TXofV}. Both \UO\ and $V_O$ for
each star are evaluated at their \Rfir\ value and normalized to their star's \vinf.  The
dashed-lines represent two special cases where $\UO = V_O$ and $\UO = 0.5 V_O$.  A detailed
discussion is given in  Section 6.
\label{fig:SHOCKFIR}}
\end{figure}

\clearpage

\appendix
\section{ERRATUM: ``An Extensive Collection of Stellar Wind X-ray Source Region Emission
Line Parameters, Temperatures, Velocities, and Their Radial Distributions as Obtained from
Chandra Observations of 17 OB Stars'' (ApJ, 608, 456 [2007] )}

The major objective of the paper was to provide a detailed tabulation of the observed HETGS
X-ray emission line flux ratios. We presented the MEG and HEG He-like $f/i$ line ratios, the
H-like to He-like (\HtoHe) line ratios, and the He-like G-ratios. The stellar wind spatial locations
of the X-ray sources were derived from the $f/i$ ratios and their associated X-ray temperatures
were obtained from the \HtoHe\ ratios (\THHe). This information was used to verify the
correlations between \Rfir\ and \Rtau\ (Figure 6) and \THHe\ and \Rfir\ (Figure 8). However, we
have realized that some of our tabulated uncertainties for these line ratios were underestimated,
primarily for those lines with low $S/N$ data. Hence, the primary purpose of this erratum is to
provide a tabulation of the corrected line ratios and their uncertainties.  

The details of our line fitting procedure are discussed in Section 3.3.  All uncertainties were
determined using standard $\chi^2$ statistics (e.g., Bevington 1969).  First, we would like to
clarify a statement in Section 3.3 (second paragraph), which states that  all parameter
uncertainties were determined from 90\% confidence regions, but in actuality all uncertainties
were established using 68\% confidence regions. With regards to the main point of this erratum,
we found that our algorithm for determining the $\chi^2$ covariance matrix which is used to
determine the uncertainties of the fitting parameters had an indexing error in the coding logic
which produced errors in some of the off-diagonal terms. From our detailed examination of the
code, we found that certain cases were especially vulnerable to this coding error, in particular,
those cases where the $\chi^2$ normalization ranges were large (i.e., low $S/N$ data). This
code correction has also produced changes in some of the line ratios and their derived quantities
(e.g., \Rfir\ and \THHe).     

The algorithm has been corrected and the affected Tables (Tables 3, 4, 5, 6, and 7) have been
updated and re-produced in this erratum. We also corrected a few entries that were originally
tabulated incorrectly, and some data were removed as they did not satisfy our $S/N$ criterion,
i.e., the HEG \SXV\ $f/i$ data for \zori, the MEG \SXV\ $f/i$ data for \zoph, and the HEG
\MgXI\ and \SiXIII\ $f/i$ data for Cyg OB2 No. 9.  As discussed in Section 3.1, we stated that  if
a reasonable flux had been established, these results would be used only for estimating line ratios
that provide interesting limits. However, the meaning of a ``reasonable`` flux limit was unclear;
the  criterion used is that the observed total net counts from all three He-like $fir$ lines must have
a $S/N \geq 3$. We also need to clarify the significance of blank entries in our Tables. The blanks
just indicate that the given line ratio has either an unphysical result that produces an anomalously
large uncertainty, or did not satisfy our He-like $S/N \geq 3$ criterion. An unphysical result
example is when the fitting procedure predicts a He-like $i$-line flux that is too small.  This
occurs primarily in low $S/N$ high energy He-like $fir$ lines where the effects of line overlap can
lead to a poor determination of the $i$-line. 

In addition, we would like to clarify why the relative uncertainties in \Rfir\ are typically smaller
than the corresponding $f/i$ relative uncertainties.  As shown in eq. (2) of the paper, for $\phi /
\phi_C > 1$ and $n_e << N_C$ (valid throughout the wind except when extremely close to the
star), the $f/i$ ratio is inversely proportional to the dilution factor, W(r), which changes rapidly
for small changes in radius.  Hence, dramatic changes in the $f/i$ ratio can occur for only small
changes in radius, which explains the differences seen in the relative uncertainties. Plus, another
key point that was not mentioned in the original paper is that there is a lower limit on the $f/i$
ratio for the case where $n_e << N_C$ determined by setting W = 0.5. This implies that any
observed $f/i$ ratio below this limit indicates that density effects may be important. We are
currently investigating this possibility.

The format of the data presented in these new tables have been changed slightly: 1) all best-fit
line ratios ($f/i$ and \HtoHe) and their uncertainties are tabulated, regardless of the size of the
uncertainty; 2) as before, all derived \Rfir\ and \THHe\ represent an average of their respective
ranges predicted by the $f/i$ and \HtoHe\ line ratios, and their associated uncertainties are equal
to half the difference in these ranges; 3) for those cases where the $f/i$ uncertainty is $>$ the
best-fit $f /i$ ratio, the derived \Rfir\ average is determined using a lower limit of $\Rfir = \Rstar$,
and an upper limit on \Rfir\ determined by the $f /i$ + uncertainty; 5) for those cases where the
minimum $f/i$ ($f/i$ - uncertainty) ratio predicts a finite \Rfir, but the upper limit in the $f/i$
range is at or greater than its asymptotic value (i.e., the low-density and zero UV flux limit), the
upper limit on \Rfir\ is undetermined (i.e., $\Rfir \rightarrow \infty$), and these \Rfir\ values are
tabulated as lower limits, and; 5) for those cases where the \HtoHe\ uncertainty is $ > $ the
\HtoHe\ ratio, upper limits on \THHe\ are presented.

The two key figures of the paper (Figures 6 and 8) have also been re-produced.  In these figures
only data with finite limits on \Rfir\ and \THHe\ are plotted, i.e., data with just lower limits on
\Rfir\ and upper limits on \THHe\ are not shown. For clarity, we also chose not to display any
\Rfir\ data where the uncertainty is $ > 10 \Rstar$. The impact of these changes in the other
figures that depend on the new derived \Rfir\ and uncertainties (Figure 7 and 9) are found to show
minimal differences from the original results. However, we did find an erroneous high temperature
data point in Figure 9 at $V_O (\Rfir)/ \vinf \approx 0$ and $U_{rel}(\THHe) / \vinf \approx
0.82$ and it should be ignored.  This same data point at low \Rfir\ was also in the original Figure
8 for the giants in both the MEG and HEG plots, and it has been removed from the re-produced
Figure 8.  The source of this data point was traced to the star $\gamma$ Vel, originally
considered in our analysis, but was dropped from our study due to its highly unusual X-ray
spectra which were deemed inappropriate for this study of ``normal'' OB stars. We have
confirmed that no other data points from $\gamma$ Vel were present in any of the original plots.

We would also like to add a comment concerning the importance of obtaining high $S/N$ HEG
data as illustrated by comparing the \SXV\ MEG and HEG determined $f/i$ ratios for \toriC. 
This is a clear example of how line overlap, either caused by the physical line width or the energy
resolution capabilities of the instrument, can lead to larger uncertainties.  Although for this case
both the MEG and HEG do have comparable $S/N$ data, the MEG data had significant line
overlap produced by the best-fit line width and lower energy resolution of the MEG, whereas, in
the HEG data the \SXV\ $fir$ were resolved leading to a significant reduction in the uncertainty.

In general, comparisons of these new derived quantities with the original tabulated data show that
the largest differences are seen in the uncertainties, primarily the uncertainty results for \OVII\ and
\SXV\, and any other low $S/N$ lines. In addition, there are minor changes evident in some of the
$f/i$ ratios, the \HtoHe\ ratios, and the G-ratios. In addition, the newly tabulated data and the
reproduced Figures 6 and 8 indicate that the fundamental results discussed in the paper have not
changed, and these corrections have not altered any of the conclusions discussed in the paper.

We wish to thank Maurice Leutenegger for his communication concerning the specific details of
our line fitting approach.  This inquiry motivated us to re-examine all aspects of our line fitting
algorithms whereupon we found the above mentioned coding mistake in the line flux
uncertainties.

Bevington, P. R. 1969, ``Data Reduction and Error Analysis for the Physical Sciences'',
McGraw-Hill, Inc.
\\


\setcounter{table}{2}

\clearpage
\begin{deluxetable}{ccccccccccc}
\tabletypesize{\tiny}
\rotate
\tablewidth{550pt}
\tablecolumns{11}
\tablecaption{Observed MEG \ftoi\ Line Ratios and $fir$-inferred Radii \label{tab:MGFI}}
\tablehead{\colhead{Star}
        &  \multicolumn{2}{c}{O VII}
        &  \multicolumn{2}{c}{Ne IX} 
        &  \multicolumn{2}{c}{Mg XI} 
        &  \multicolumn{2}{c}{Si XIII} 
        &  \multicolumn{2}{c}{S XV}
        \\  \colhead{ }
        &  \colhead{\ftoi}
        &  \colhead{\Rfir}
        &  \colhead{\ftoi}
        &  \colhead{\Rfir}
        &  \colhead{\ftoi}
        &  \colhead{\Rfir}
        &  \colhead{\ftoi}
        &  \colhead{\Rfir}
        &  \colhead{\ftoi}
        &  \colhead{\Rfir}}

\startdata
\multicolumn{4}{l}{\textit{\underline{Supergiants (I, II)}} } \\
 \zpup      & $ 0.03\pm0.11$       & $ 11.21\pm10.21$  
               & $ 0.38\pm0.13$         & $10.85\pm2.15$ 
               & $ 0.34\pm0.05$         & $ 3.76\pm0.30$  
               & $ 1.17\pm0.17$         & $ 2.56\pm0.33$ 
               & $ 0.60\pm0.62$         & $ 1.24\pm0.24$      \\
 \cygnin   &  \nodata                     &  \nodata       
                & \nodata                     &  \nodata
                & \nodata                     &  \nodata           
                & $ 0.60\pm0.20$        &  $ 1.49\pm0.28$ 
                &   \nodata                   &   \nodata                      \\
 \cygate    & \nodata                     &  \nodata        
                & $ 0.51\pm1.29$        & $15.32\pm14.32$
                & $ 0.79\pm0.65$        & $ 5.22\pm3.20$ 
                & $ 1.02\pm0.22$        & $ 1.91\pm0.31$ 
                & $ 0.74\pm0.34$        & $ 1.07\pm0.07$             \\

 \doriA     & $ 0.11\pm0.08$        & $12.98\pm5.45$ 
                & $ 0.00\pm0.02$        & $ 1.43\pm0.43$
                & $ 0.84\pm0.27$        & $ 4.49\pm0.98$  
                & $ 1.45\pm0.54$        & $ 1.63\pm0.59$ 
                &    \nodata                  & \nodata                       \\
 \zoriA     & $ 0.13\pm0.03$        & $13.85\pm1.60$  
                & $ 0.32\pm0.10$        & $ 6.76\pm1.16$ 
                & $ 1.10\pm0.18$        & $ 4.97\pm0.64$  
                & $ 3.48\pm1.34$        & $ \geq 1.80$ 
                & $ 0.77\pm1.21$        & $ 1.33\pm0.33$                       \\
 \eori        & $ 0.10\pm0.05$        & $10.23\pm2.56$  
                & $ 0.51\pm0.11$        & $ 7.54\pm0.99$ 
                & $ 1.24\pm0.35$        & $ 4.61\pm1.09$  
                & $ 1.98\pm0.68$        & $ \geq 1.0$ 
                &   \nodata                   &  \nodata                       \\

\multicolumn{4}{l}{\textit{\underline{Giants (IV, III)}} } \\
 HD150136  & $ 0.00\pm0.10$        & $ 9.80\pm8.80$              
                    & $ 0.12\pm0.10$        & $ 5.30\pm2.96$ 
                    & $ 0.45\pm0.16$        & $ 4.41\pm0.90$ 
                    & $ 1.26\pm0.40$        & $ 2.97\pm0.91$ 
                    & $ 1.46\pm2.25$        & $ \geq 1.0$                  \\
 \xper           & $ 0.00\pm0.09$        & $ 7.62\pm6.62$            
                    & $ 0.09\pm0.08$        & $ 3.59\pm2.19$ 
                    & $ 0.26\pm0.10$        & $ 2.52\pm0.52$ 
                    & $ 2.35\pm1.06$        & $ \geq 1.90$ 
                    &   \nodata                   & \nodata                        \\
 \iori             & $ 0.03\pm0.08$        & $ 7.64\pm6.64$ 
                    & $ 0.58\pm1.06$        & $12.29\pm11.29$ 
                    & $ 0.72\pm0.23$        & $ 4.24\pm0.90$    
                    & $ 1.79\pm0.90$        & $ \geq 1.15$ 
                    &   \nodata                   & \nodata                        \\
 \bcru           & $ 0.01\pm0.04$        & $ 4.03\pm3.03$   
                    & $ 0.00\pm0.06$        & $ 1.67\pm0.67$
                    & $ 0.32\pm0.26$         & $ 1.79\pm0.79$  
                    & $ 1.43\pm1.27$         & $ \geq 1.0 $
                    & \nodata                      & \nodata                        \\

\multicolumn{4}{l}{\textit{\underline{Main Sequence}} } \\
 9 Sgr            & $ 0.22\pm0.44$      & $25.75\pm24.75$   
                     & $ 0.05\pm0.11$       & $ 4.06\pm3.06$ 
                     & $ 0.84\pm1.96$       & $19.14\pm18.14$
                     & $ 1.43\pm1.31$       & $ \ge 1.0$ 
                     & \nodata                    &   \nodata                  \\
 HD206267   & $ 0.00\pm0.56$       & $22.12\pm21.12$  
                     & $ 0.56\pm0.45$       & $12.61\pm7.08$ 
                     & $ 0.30\pm0.21$       & $ 3.17\pm1.38$  
                     & $ 0.34\pm0.32$       & $ 1.26\pm0.26$ 
                     & \nodata                    & \nodata                       \\
15 Mon        & $ 0.05\pm0.26$        & $16.00\pm15.00$ 
                    & $ 0.01\pm0.11$        & $ 3.35\pm2.35$
                    & $ 0.22\pm0.62$        & $ 3.56\pm2.56$   
                    & $ 1.16\pm1.85$        & $ \geq 1.0$ 
                    & \nodata                     & \nodata                       \\
 \toriC          & $ 0.00\pm0.12$        & $ 8.96\pm7.96$
                    & $0.22\pm0.21$          & $ 5.92\pm4.22$ 
                    & $ 0.12\pm0.09$         & $ 1.72\pm0.72$   
                    & $ 1.73\pm0.27$         & $ 3.40\pm0.84$ 
                    & $ 1.32\pm0.66$         & $ 2.94\pm1.94$                  \\
 \zoph          & $ 0.00\pm0.16$         & $ 9.25\pm8.25$
                   & $ 0.01\pm0.05$         & $ 2.12\pm1.12$ 
                   & $ 0.26\pm0.08$          & $ 2.36\pm0.38$  
                   & $ 1.45\pm0.34$          & $ 1.76\pm0.44$ 
                   &  \nodata                      &  \nodata            \\
 \sori           & $ 0.00\pm0.03$           & $ 3.92\pm2.92$
                   & $ 0.13\pm0.06$           & $ 4.37\pm1.19$ 
                   & $ 0.02\pm0.10$           & $ 1.29\pm0.29$
                   & $ 0.21\pm0.37$           & $ < 1.02$ 
                   & \nodata                        & \nodata                         \\
 \tsco          & $ 0.00\pm0.01$            & $ 2.37\pm1.37$
                  & $ 0.00\pm0.03$            & $ 1.61\pm0.61$ 
                  & $ 0.36\pm0.07$            & $ 2.59\pm0.27$ 
                  & $ 2.33\pm0.40$            & $ \geq 1.76$ 
                  & $ 1.74\pm0.84$            & $ \geq 1.0$     \\

\enddata
\end{deluxetable}

\clearpage
\begin{deluxetable}{ccccccccc}
\tabletypesize{\scriptsize}
\rotate
\tablewidth{520pt}
\tablecolumns{9}
\tablecaption{Observed HEG \ftoi\ Line Ratios and $fir$-inferred Radii \label{tab:HGFI}}
\tablehead{\colhead{Star}
        &  \multicolumn{2}{c}{Ne IX} 
        &  \multicolumn{2}{c}{Mg XI} 
        &  \multicolumn{2}{c}{Si XIII} 
        &  \multicolumn{2}{c}{S XV}
        \\  \colhead{ }
        &  \colhead{\ftoi}
        &  \colhead{\Rfir}
        &  \colhead{\ftoi}
        &  \colhead{\Rfir}
        &  \colhead{\ftoi}
        &  \colhead{\Rfir}
        &  \colhead{\ftoi}
        &  \colhead{\Rfir}}

\startdata
\multicolumn{4}{l}{\textit{\underline{Supergiants (I, II)}} } \\
 \zpup       & $ 0.13\pm0.10$       & $ 5.74\pm2.55$ 
                & $ 0.36\pm0.09$       & $ 3.84\pm0.53$ 
                & $ 0.97\pm0.20$       & $ 2.19\pm0.35$ 
                & $ 0.79\pm0.65$       & $ 1.42\pm0.42$             \\
 \cygnin    & \nodata                    & \nodata
                &  \nodata                   & \nodata          
                & \nodata                    & \nodata
                & \nodata                    & \nodata                    \\ 
 \cygate    & \nodata                    & \nodata
                & $ 0.91\pm0.61$       & $ 5.95\pm2.97$  
                & $ 1.15\pm0.33$       & $ 2.15\pm0.53$ 
                & $ 0.87\pm0.37$       & $ 1.13\pm0.13$        \\
 \doriA     & $ 0.00\pm0.22$       &  $ 3.52\pm2.52$
                & $ 0.79\pm0.93$       & $ 4.64\pm3.64$
                & $ 2.52\pm1.91$       & $ \geq 1.0$ 
                & \nodata                    & \nodata                      \\

 \zoriA      & $ 0.31\pm0.17$       & $ 6.40\pm2.12$ 
                 & $ 0.62\pm0.24$       & $ 3.32\pm0.82$   
                 & $ 1.82\pm1.11$       & $ \ge 1.0$ 
                 & \nodata                    & \nodata                       \\
 \eori         & $ 0.30\pm0.25$       & $ 5.09\pm2.80$ 
                 & $ 2.08\pm0.84$       & $15.73\pm11.24$  
                 & \nodata                    & \nodata 
                 & \nodata                    & \nodata                        \\

\multicolumn{4}{l}{\textit{\underline{Giants (IV, III)}} } \\
 HD150136 & $ 0.27\pm0.60$     & $ 9.77\pm8.77$ 
                   & $ 0.34\pm0.12$     & $ 3.78\pm0.77$ 
                   & $ 1.15\pm0.58$     & $ 2.85\pm1.27$ 
                   &  \nodata                 & \nodata                         \\
 \xper          & $ 0.00\pm0.09$     & $ 2.58\pm1.58$
                   & $ 0.35\pm0.16$     & $ 2.92\pm0.78$  
                   & $ 0.30\pm0.26$     & $ 1.06\pm0.06$ 
                   & \nodata                  & \nodata                         \\
 \iori            & $ 0.57\pm1.94$     & $22.84\pm21.84$ 
                   & $ 0.68\pm0.33$      & $ 4.10\pm1.30$
                   & $ 2.50\pm2.73$      & $ \geq 1.0$ 
                   & \nodata                   & \nodata                          \\
 \bcru          & $ 0.10\pm0.53$       & $ 4.70\pm3.70$
                   & $ 0.55\pm0.54$      &  $ 2.47\pm1.47$
                   &  \nodata                  & \nodata
                   &  \nodata                  &  \nodata                          \\
\multicolumn{4}{l}{\textit{\underline{Main Sequence}} } \\
 9 Sgr          & $ 0.05\pm0.50$       & $ 7.40\pm6.40$ 
                   & $ 0.54\pm0.41$       & $ 4.76\pm2.46$ 
                   &  \nodata                   & \nodata 
                   & \nodata                    & \nodata                        \\
 HD206267 & \nodata                    & \nodata
                   & $ 1.02\pm1.00$       & $ 7.56\pm6.52$
                   & $ 1.97\pm1.71$       & $ \geq 1.02$
                   &  \nodata                   &  \nodata                        \\
15 Mon      & $ 0.00\pm0.39$       & $ 5.83\pm4.83$
                   & $ 0.76\pm0.94$      & $6.12\pm5.12$
                   & \nodata                    & \nodata
                   & \nodata                    & \nodata                         \\
 \toriC         & $ 0.00\pm0.14$       & $ 3.31\pm2.31$
                   & $ 0.38\pm0.20$       & $ 3.24\pm1.01$ 
                   & $ 1.11\pm0.19$       & $ 1.97\pm0.28$ 
                   & $ 0.70\pm0.25$       & $ 1.02\pm0.02$              \\

 \zoph         & $ 0.22\pm0.33$        & $ 5.75\pm4.75$ 
                  & $ 0.27\pm0.14$        & $ 2.37\pm0.68$ 
                  & $ 1.37\pm0.57$        & $ 1.83\pm0.73$ 
                  & \nodata                     & \nodata                      \\
 \sori          & $ 0.17\pm0.43$        & $ 5.80\pm4.80$ 
                  & $ 0.00\pm0.12$        & $ 5.80\pm4.80$  
                  & \nodata                     & \nodata
                  & \nodata                     & \nodata                      \\
 \tsco         & $ 0.03\pm0.09$        & $ 2.59\pm1.59$ 
                  & $ 0.29\pm0.13$        & $ 2.26\pm0.56$  
                  & $ 1.77\pm0.37$        & $ 1.72\pm0.54$ 
                  & \nodata                     & \nodata                        \\

\enddata
\end{deluxetable}

\clearpage
\begin{deluxetable}{ccccccccccc}
\tabletypesize{\tiny}
\rotate
\tablewidth{550pt}
\tablecolumns{11}
\tablecaption{Observed MEG \HtoHe\ Line Ratios and Derived \THHe\ \label{tab:MGHHE}}
\tablehead{\colhead{Star}
        &  \multicolumn{2}{c}{Oxygen}
        &  \multicolumn{2}{c}{Neon} 
        &  \multicolumn{2}{c}{Magnesium} 
        &  \multicolumn{2}{c}{Silicon} 
        &  \multicolumn{2}{c}{Sulfur}
        \\  \colhead{ }
        &  \colhead{\HtoHe}
        &  \colhead{\THHe}
        &  \colhead{\HtoHe}
        &  \colhead{\THHe}
        &  \colhead{\HtoHe}
        &  \colhead{\THHe}
        &  \colhead{\HtoHe}
        &  \colhead{\THHe}
        &  \colhead{\HtoHe}
        &  \colhead{\THHe}}

\startdata
\TL\ range & \nodata  & 2.00 - 3.16   & \nodata & 3.98 - 6.31 
                  & \nodata & 6.31 - 10.00  & \nodata & 10.00 - 15.85 
                  & \nodata & 15.85 - 25.12 \\
\multicolumn{4}{l}{\textit{\underline{Supergiants (I, II)}} } \\
 \zpup     & $ 1.50\pm0.27$        & $ 2.68\pm0.13$ 
               & $ 0.96\pm0.10$        & $ 3.92\pm0.12$ 
               & $ 0.37\pm0.05$        & $ 5.86\pm0.20$ 
               & $ 0.16\pm0.03$        & $ 7.55\pm0.42$ 
               & $ 0.09\pm0.15$        & $ < 13 $                \\
 \cygnin   &  \nodata                    &  \nodata   
               &  \nodata                     & \nodata 
               & $ 0.82\pm0.53$        & $  7.12\pm1.61$
               & $ 1.01\pm0.33$        & $12.63\pm1.49$
               & \nodata                     & \nodata                               \\

 \cygate   & \nodata                        &  \nodata             
               &  \nodata                       & \nodata
               & $ 0.91\pm0.22$           & $  7.59\pm0.63$  
               & $ 0.91\pm0.12$           & $12.34\pm0.57$ 
               & $ 2.25\pm1.44$           & $26.28\pm8.12$        \\
 \doriA    & $ 1.21\pm0.13$           & $ 2.53\pm0.07$ 
               & $ 0.90\pm0.13$           & $ 3.84\pm0.17$ 
               & $ 0.29\pm0.07$           & $ 5.53\pm0.31$ 
               & $ 0.58\pm0.18$           & $10.57\pm1.01$ 
               &  \nodata                       &  \nodata                     \\
 \zoriA   & $ 0.99\pm0.06$            & $ 2.38\pm0.05$  
              & $ 0.66\pm0.06$            & $ 3.52\pm0.09$ 
              & $ 0.25\pm0.05$            & $ 5.32\pm0.23$  
              & $ 0.19\pm0.07$            & $ 7.79\pm0.70$ 
              & $ 1.30\pm4.13$            & $ < 41 $                     \\
 \eori      & $ 1.02\pm0.09$             & $ 2.41\pm0.07$   
              & $ 0.72\pm0.07$             & $ 3.60\pm0.10$ 
              & $ 0.18\pm0.04$             & $ 4.96\pm0.25$
              & $ 0.03\pm0.10$             & $ < 7.2$ 
              & $ 0.13\pm1.07$             & $ < 22 $                     \\
mean \THHe    & \nodata                & $  2.50\pm0.11$     
                        & \nodata                & $  3.72\pm0.15$
                        & \nodata                & $  6.06\pm0.83$ 
                        & \nodata                & $10.18\pm1.71$
                        & \nodata                & $26.28\pm8.12$       \\

\multicolumn{4}{l}{\textit{\underline{Giants (IV, III)}} } \\
 HD150136  & $ 1.94\pm1.06$       & $2.83\pm0.51$  
                    & $ 1.27\pm0.24$       & $ 4.25\pm0.22$ 
                    & $ 0.40\pm0.08$       & $ 6.01\pm0.34$ 
                    & $ 0.49\pm0.09$       & $10.11\pm0.52$ 
                    & $ 1.18\pm1.41$       & $ < 30$         \\
 \xper           &  $ 1.49\pm0.36$      & $ 2.67\pm0.18$ 
                    & $ 0.74\pm0.11$       & $ 3.63\pm0.16$ 
                    & $ 0.15\pm0.05$       & $ 4.77\pm0.34$ 
                    & $ 0.14\pm0.09$       & $ 7.08\pm1.11$ 
                    &  \nodata                   &  \nodata                       \\
 \iori             & $ 1.27\pm0.16$       & $ 2.56\pm0.08$  
                    & $ 0.36\pm0.13$       & $ 2.97\pm0.29$ 
                    & $ 0.28\pm0.10$       & $ 5.44\pm0.48$ 
                    & $ 0.67\pm0.32$       & $10.92\pm1.82$ 
                    &  \nodata                   &  \nodata                       \\

 \bcru           &  $ 1.12\pm0.18$       & $ 2.47\pm0.11$ 
                    &  $ 0.41\pm0.10$       & $ 3.09\pm0.20$ 
                    &  $ 0.21\pm0.26$       & $ < 6.4$
                    &  \nodata                   & \nodata
                    &  \nodata                   &  \nodata                  \\
mean \THHe &  \nodata                  & $ 2.63\pm0.28$  
                     & \nodata                   & $ 3.48\pm0.42$
                     &  \nodata                  & $ 5.41\pm0.41$  
                     &  \nodata                  & $ 9.37\pm1.29$
                     &  \nodata                  &  \nodata      \\

\multicolumn{4}{l}{\textit{\underline{Main Sequence}} } \\
 9 Sgr           & $ 0.84\pm0.33$      & $ 2.28\pm0.24$ 
                    & $ 0.43\pm0.09$      & $ 3.13\pm0.18$
                    & $ 0.27\pm0.13$      & $ 5.38\pm0.65$ 
                    & $ 0.14\pm0.07$      & $ 7.21\pm0.83$
                    & \nodata                   &  \nodata                       \\
 HD206267  & $ 0.47\pm0.23$      & $ 1.94\pm0.23$  
                    & $ 0.32\pm0.09$      & $ 2.91\pm0.22$ 
                    & $ 0.14\pm0.08$      & $ 4.62\pm0.57$  
                    & $ 0.15\pm0.20$      & $ < 9.2 $ 
                    & \nodata                   &  \nodata                        \\
15 Mon       & $ 0.71\pm0.15$       & $ 2.18\pm0.11$  
                   & $ 0.18\pm0.07$       & $ 2.52\pm0.21$
                   & $ 0.12\pm0.13$       & $ < 5.4$  
                   & $ 0.14\pm2.37$       & $ < 18$
                   & \nodata                    & \nodata                          \\
 \toriC         & $ 3.09\pm1.84$       & $  3.28\pm0.73$  
                   & $ 2.53\pm0.36$       & $  5.33\pm0.26$ 
                   & $ 2.06\pm0.22$       & $10.31\pm0.43$ 
                   & $ 1.68\pm0.11$       & $15.44\pm0.43$ 
                   & $ 1.40\pm0.26$       & $22.36\pm1.65$             \\
 \zoph         & $ 1.28\pm0.25$       & $  2.56\pm0.14$ 
                  & $ 1.25\pm0.14$        & $ 4.23\pm0.13$
                  & $ 0.48\pm0.09$        & $ 6.28\pm0.32$ 
                  & $ 0.36\pm0.09$        & $ 9.28\pm0.71$
                  & $ 0.17\pm0.50$        & $ < 18$                            \\
 \sori          & $ 1.56\pm0.32$        & $ 2.71\pm0.16$ 
                  & $ 0.65\pm0.12$        & $ 3.50\pm0.17$ 
                  & $ 0.23\pm0.34$        & $ < 6.7$ 
                  & $ 0.17\pm0.73$        & $ < 13 $
                  &  \nodata                    &  \nodata              \\

 \tsco             &  $ 3.28\pm0.44$             & $ 3.44\pm0.15$ 
                      & $ 1.79\pm0.14$             & $ 4.73\pm0.13$ 
                      & $ 0.84\pm0.09$             & $ 7.41\pm0.26$ 
                      & $ 0.58\pm0.07$             & $10.62\pm0.37$
                      & $ 0.89\pm0.38$             & $18.53\pm2.98$         \\
mean \THHe  &  \nodata                         & $ 2.52\pm0.44$  
                      &  \nodata                         & $ 3.50\pm0.68$
                      &  \nodata                         & $ 5.92\pm0.94$  
                      &  \nodata                         & $ 9.03\pm1.19$
                      &  \nodata                         & $18.53\pm2.98$         \\
\enddata
\tablecomments{\HtoHe\ ratios are ISM corrected and \THHe\ are the derived X-ray
temperatures in MK. The results for \toriC\ are not included in the mean \THHe.}
\end{deluxetable}

\clearpage
\begin{deluxetable}{ccccccccc}
\tabletypesize{\scriptsize}
\rotate
\tablewidth{520pt}
\tablecolumns{9}
\tablecaption{Observed HEG \HtoHe\ Line Ratios and Derived \THHe\ \label{tab:HGHHE}}
\tablehead{\colhead{Star}
        &  \multicolumn{2}{c}{Neon} 
        &  \multicolumn{2}{c}{Magnesium} 
        &  \multicolumn{2}{c}{Silicon} 
        &  \multicolumn{2}{c}{Sulfur}
        \\  \colhead{ }
        &  \colhead{\HtoHe}
        &  \colhead{\THHe}
        &  \colhead{\HtoHe}
        &  \colhead{\THHe}
        &  \colhead{\HtoHe}
        &  \colhead{\THHe}
        &  \colhead{\HtoHe}
        &  \colhead{\THHe}}

\startdata
\TL\ range & \nodata & 3.98 - 6.31 
           & \nodata & 6.31 - 10.00  & \nodata & 10.00 - 15.85 
           & \nodata & 15.85 - 25.12 \\
\multicolumn{4}{l}{\textit{\underline{Supergiants (I, II)}} } \\
 \zpup       & $ 1.21\pm0.16$          & $  4.19\pm0.15$
                 & $ 0.26\pm0.05$          & $ 5.38\pm0.22$ 
                 & $ 0.08\pm0.04$         & $  6.44\pm0.59$ 
                 & $ 0.08\pm0.29$         & $ < 15$                      \\
 \cygnin     & \nodata                      & \nodata
                 & \nodata                      & \nodata   
                 &  \nodata                     & \nodata
                 &  \nodata                     &  \nodata                       \\
 \cygate     &  \nodata                     & \nodata
                 & $ 0.84\pm0.25$         & $  7.39\pm0.73$ 
                 & $ 0.94\pm0.19$         & $12.39\pm0.90$
                 & $ 0.80\pm0.41$         & $17.66\pm3.45$           \\
 \doriA      & $ 0.68\pm0.32$         & $ 3.50\pm0.50$
                 & $ 0.95\pm0.45$         & $ 7.61\pm1.22$ 
                 & $ 0.25\pm0.17$         & $ 8.13\pm1.62$
                 & $ 0.22\pm0.79$         & $ < 20 $             \\
 \zoriA     & $ 0.63\pm0.10$           & $ 3.46\pm0.14$
                & $ 0.28\pm0.08$           & $ 5.49\pm0.35$ 
                & $ 0.25\pm0.16$           & $ 8.10\pm1.49$ 
                & \nodata                        & \nodata                \\
 \eori        & $ 0.53\pm0.11$           & $ 3.32\pm0.16$
                & $ 0.26\pm0.11$           & $ 5.32\pm0.56$ 
                & $ 0.10\pm0.14$           & $ < 8.4$ 
                &  \nodata                      &  \nodata                  \\
mean \THHe & \nodata                  & $  3.62\pm0.27$
                     & \nodata                  & $  6.24\pm0.84$   
                     & \nodata                  & $  8.77\pm1.61$
                     & \nodata                  & $17.66\pm3.45$     \\

\multicolumn{4}{l}{\textit{\underline{Giants (IV, III)}} } \\
 HD150136   & $ 0.67\pm0.36$     & $ 3.46\pm0.56$ 
                     & $ 0.42\pm0.10$     & $ 6.04\pm0.41$ 
                     & $ 0.53\pm0.17$     & $10.27\pm0.99$ 
                     & $ 0.16\pm0.30$     & $ < 16$             \\
 \xper            &  $ 0.39\pm0.12$    & $ 3.04\pm0.26$
                     & $ 0.17\pm0.09$     & $ 4.81\pm0.61$ 
                     & $ 0.26\pm0.27$     & $ < 11$
                     & \nodata                  &  \nodata              \\
 \iori              & $ 0.80\pm0.30$     & $ 3.68\pm0.40$
                     & $ 0.64\pm0.53$     & $ 6.39\pm1.91$
                     & $0.39\pm0.44$      & $ < 12$
                     &  \nodata                 &  \nodata              \\
 \bcru            & $ 0.20\pm0.12$     & $ 2.53\pm0.38$
                     & $ 0.15\pm0.74$     & $ < 7.6$  
                     &  \nodata                 & \nodata
                     &  \nodata                 &  \nodata              \\
mean \THHe & \nodata                  & $ 3.18\pm0.46$
                     & \nodata                  & $ 5.74\pm0.57$   
                     & \nodata                  & $10.27\pm0.99$
                     & \nodata                  & \nodata               \\

\multicolumn{4}{l}{\textit{\underline{Main Sequence}} } \\
 9 Sgr           & $ 0.80\pm0.34$      & $ 3.67\pm0.45$
                    & $ 0.34\pm0.27$      & $ 5.40\pm1.33$ 
                    & $ 0.53\pm0.68$      & $ < 14$ 
                    & \nodata                   & \nodata               \\
 HD206267  & $ 0.93\pm0.95$      & $ < 5 $
                    & $ 0.36\pm0.47$      & $ < 7.5 $
                    & $ 0.13\pm0.87$      & $ < 13 $
                    &  \nodata                  &  \nodata              \\

15 Mon        & $ 0.40\pm0.42$      & $ < 3.8$
                    & $ 0.89\pm2.23$      & $ < 13$ 
                    &  \nodata                  & \nodata
                    &  \nodata                  & \nodata    \\
 \toriC          & $ 2.92\pm1.25$      & $ 5.68\pm1.06$          
                    & $ 2.78\pm0.49$      & $11.68\pm0.93$ 
                    & $ 1.68\pm0.15$      & $15.43\pm0.58$
                    & $ 1.00\pm0.19$      & $19.53\pm1.44$      \\
 \zoph          & $ 1.01\pm0.22$      & $ 3.95\pm0.26$
                   & $ 0.67\pm0.17$       & $ 6.92\pm0.51$  
                   & $ 0.32\pm0.16$       & $ 8.82\pm1.28$
                   &  \nodata                   &  \nodata               \\
\sori            & $ 0.37\pm0.13$       & $ 2.99\pm0.28$
                   & $ 0.26\pm0.39$       & $ < 7 $  
                   &  \nodata                   & \nodata
                   &  \nodata                   & \nodata    \\
 \tsco          & $ 1.88\pm0.31$         & $  4.80\pm0.28$
                  & $ 0.85\pm0.14$         & $  7.44\pm0.40$  
                  & $ 0.67\pm0.11$         & $11.10\pm0.58$
                  & $ 0.15\pm0.24$         & $ < 15$    \\
mean \THHe    & \nodata                & $  3.86\pm0.60$
                        & \nodata                & $  6.59\pm1.01$   
                        & \nodata                & $  9.96\pm1.21$
                        & \nodata                &  \nodata     \\
\enddata
\tablecomments{\HtoHe\ ratios are ISM corrected and \THHe\ are the derived X-ray
temperatures in MK. The results for \toriC\ are not included in the mean \THHe.}
\end{deluxetable}

\clearpage
\begin{deluxetable}{ccccccccccc}
\tabletypesize{\tiny}
\rotate
\tablewidth{520pt}
\tablecolumns{11}
\tablecaption{Observed MEG and HEG G-Ratios \label{tab:GRATIO}}
\tablehead{\colhead{Star}
        &  \multicolumn{2}{c}{Oxygen}
        &  \multicolumn{2}{c}{Neon} 
        &  \multicolumn{2}{c}{Magnesium} 
        &  \multicolumn{2}{c}{Silicon} 
        &  \multicolumn{2}{c}{Sulfur}
        \\  \colhead{ }
        &  \colhead{MEG}
        &  \colhead{HEG}
        &  \colhead{MEG}
        &  \colhead{HEG}
        &  \colhead{MEG}
        &  \colhead{HEG}
        &  \colhead{MEG}
        &  \colhead{HEG}
        &  \colhead{MEG}
        &  \colhead{HEG}}

\startdata
\multicolumn{4}{l}{\textit{\underline{Supergiants (I, II)}} } \\
 \zpup     & $ 1.10\pm0.18$    & \nodata              
               & $ 0.43\pm0.04$    & $ 0.65\pm0.07$ 
               & $ 0.82\pm0.04$    & $ 0.78\pm0.06$  
               & $ 0.96\pm0.06$    & $ 1.14\pm0.12$ 
               & $ 0.83\pm0.49$    & $ 1.14\pm0.42$   \\
 \cygnin   & \nodata                 & \nodata              
               &  \nodata                & \nodata
               & $ 0.49\pm0.18$    & \nodata    
               & $ 1.95\pm0.46$    & \nodata 
               &  \nodata                &  \nodata               \\
 \cygate   &  \nodata                &  \nodata    
               &  \nodata                & \nodata
               &  $ 0.40\pm0.07$   & $ 0.68\pm0.15$   
               &  $ 0.89\pm0.08$   & $ 1.32\pm0.21$ 
               &  $ 3.16\pm1.70$   & $ 1.98\pm0.57$   \\

 \doriA    &  $ 0.79\pm0.07$     & \nodata     
               &  $ 0.60\pm0.07$     & $ 0.44\pm0.19$ 
               &  $ 0.62\pm0.06$     & $ 0.81\pm0.31$    
               &  $ 0.90\pm0.14$     & $ 0.67\pm0.14$ 
               &  $ 2.02\pm2.19$     & $ 0.76\pm0.66$   \\
 \zoriA    &  $ 0.96\pm0.05$     & \nodata    
               &  $ 0.49\pm0.04$     & $ 0.57\pm0.06$ 
               &  $ 0.97\pm0.08$     & $ 0.78\pm0.10$  
               &  $ 1.22\pm0.16$     & $ 1.08\pm0.26$ 
               &  $ 2.50\pm5.05$     & \nodata    \\
 \eori       &  $ 1.00\pm0.08$     & \nodata  
               & $ 0 .86\pm0.06$     & $ 0.63\pm0.10$ 
               &  $ 0.75\pm0.07$     & $ 0.93\pm0.14$   
               &  $ 1.03\pm0.16$     & $ 0.43\pm0.09$ 
               &  $ 1.50\pm1.09$     &  \nodata              \\

\multicolumn{4}{l}{\textit{\underline{Giants (IV, III)}} } \\
 HD150136  & $ 1.30\pm0.68$    &  \nodata             
                    & $ 0.87\pm0.15$    & $ 0.98\pm0.48$ 
                    & $ 0.65\pm0.10$    & $ 0.96\pm0.14$  
                    & $ 0.76\pm0.08$    & $ 0.88\pm0.18$ 
                    & $ 1.10\pm0.72$    & $ 0.68\pm0.49$    \\
 \xper           & $ 0.52\pm0.12$    & \nodata 
                    & $ 0.67\pm0.08$    & $ 0.73\pm0.18$ 
                    & $ 0.80\pm0.12$    & $ 1.18\pm0.21$  
                    & $ 0.92\pm0.14$    & $ 1.52\pm0.99$ 
                    & $ 0.97\pm0.77$    & \nodata       \\
 \iori             & $ 0.91\pm0.10$    & \nodata        
                    & $ 0.18\pm0.06$    & $ 0.32\pm0.09$ 
                    & $ 0.91\pm0.13$    & $ 3.58\pm2.20$  
                    & $ 1.86\pm0.61$    & $ 1.05\pm0.37$ 
                    & \nodata                 &  \nodata               \\
 \bcru           & $ 1.42\pm0.18$    & \nodata   
                    & $ 0.92\pm0.12$    & $ 0.49\pm0.16$ 
                    & $ 1.09\pm0.34$    & $ 3.03\pm2.56$  
                    & $ 1.31\pm0.66$    & \nodata 
                    &  \nodata                & \nodata                 \\

\multicolumn{4}{l}{\textit{\underline{Main Sequence}} } \\
 9 Sgr          & $ 1.10\pm0.39$     & \nodata  
                   & $ 0.44\pm0.08$     & $0.73\pm0.27$ 
                   & $ 0.25\pm0.09$     & $0.75\pm0.16$ 
                   & $ 0.54\pm0.12$     & $0.73\pm0.31$ 
                   & \nodata                  &  \nodata                \\

 HD206267  & $ 0.20\pm0.09$    & \nodata   
                    & $ 0.63\pm0.13$    & \nodata 
                    & $ 0.66\pm0.12$    & $ 2.57\pm2.14$   
                    & $ 0.67\pm0.19$    & $ 3.66\pm3.11$ 
                    &  \nodata                &  \nodata                \\
15 Mon       & $ 0.60\pm0.10$     & \nodata    
                   & $ 0.88\pm0.17$     & $ 0.41\pm0.36$
                   & $ 0.53\pm0.16$     & $ 3.33\pm4.54$   
                   & $ 0.60\pm0.26$     & $ 0.79\pm0.61$
                   &  \nodata                 &  \nodata                \\
 \toriC         & $ 1.34\pm0.78$     & \nodata    
                   & $ 0.52\pm0.07$     & $ 0.89\pm0.37$ 
                   & $ 0.68\pm0.06$     & $ 0.92\pm0.15$ 
                   & $ 0.72\pm0.04$     & $ 0.92\pm0.07$ 
                   & $ 0.70\pm0.10$     & $ 0.84\pm0.11$       \\
 \zoph          & $ 0.93\pm0.16$     & \nodata   
                   & $ 0.67\pm0.06$      & $ 0.51\pm0.09$ 
                   & $ 0.78\pm0.07$      & $ 0.98\pm0.16$  
                   & $ 0.99\pm0.11$      & $ 0.83\pm0.14$ 
                   & $ 0.35\pm0.21$      & \nodata    \\
 \sori           & $ 1.16\pm0.21$      & \nodata        
                   & $ 1.12\pm0.14$      & $ 0.56\pm0.14$ 
                   & $ 0.91\pm0.18$      & $ 1.03\pm0.44$ 
                   & $ 1.18\pm0.56$      & $ 0.74\pm0.36$
                   &  \nodata                  &  \nodata               \\
 \tsco          & $ 1.57\pm0.20$       & \nodata   
                  & $ 0.90\pm0.06$       & $ 0.90\pm0.13$ 
                  & $ 0.79\pm0.05$       & $ 0.57\pm0.06$ 
                  & $ 0.93\pm0.06$       & $ 1.14\pm0.12$ 
                  & $ 1.83\pm0.56$       & $ 0.54\pm0.16$         \\
\enddata
\tablecomments{G-Ratios are ISM corrected.}
\end{deluxetable}

\setcounter{figure}{5}

\clearpage
\begin{figure}
\includegraphics[height=18cm,angle=180]{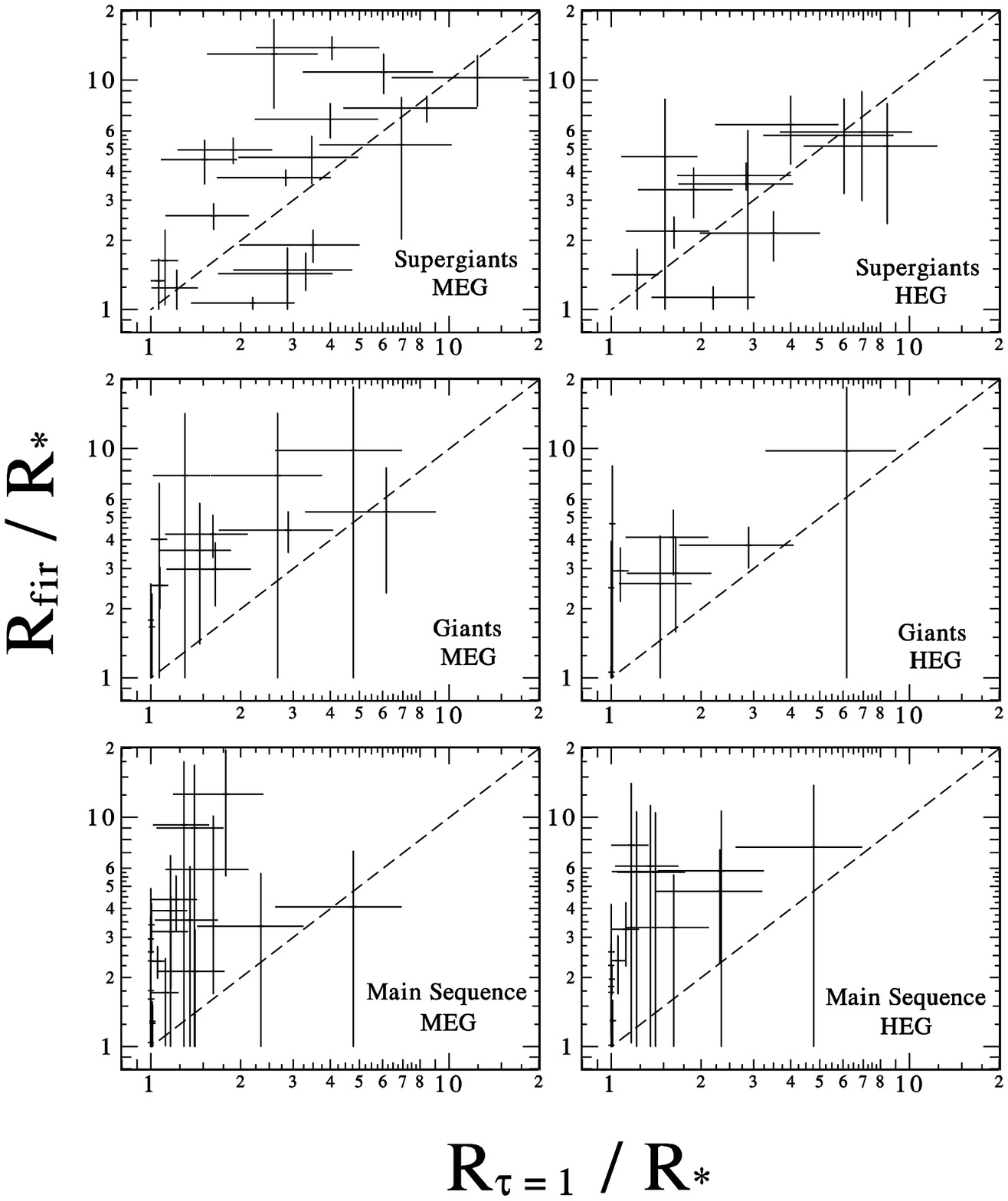}
\caption{MEG and HEG scatter plots showing the relation between \Rfir\ and \Rtau\ for all
luminosity classes. The error bars for \Rtau\ are determined for a X-ray continuum optical depth
of $1\pm0.5$ and the adopted value of \Rtau\ is the average of these limits. \Rfir\ data that only
have lower limits or an uncertainty $ > 10 \Rstar$ are not shown (see Tables 3 and 4). 
\label{fig:TAUFIR}}
\end{figure}

\setcounter{figure}{7}

\clearpage
\begin{figure}
\includegraphics[height=18cm,angle=180]{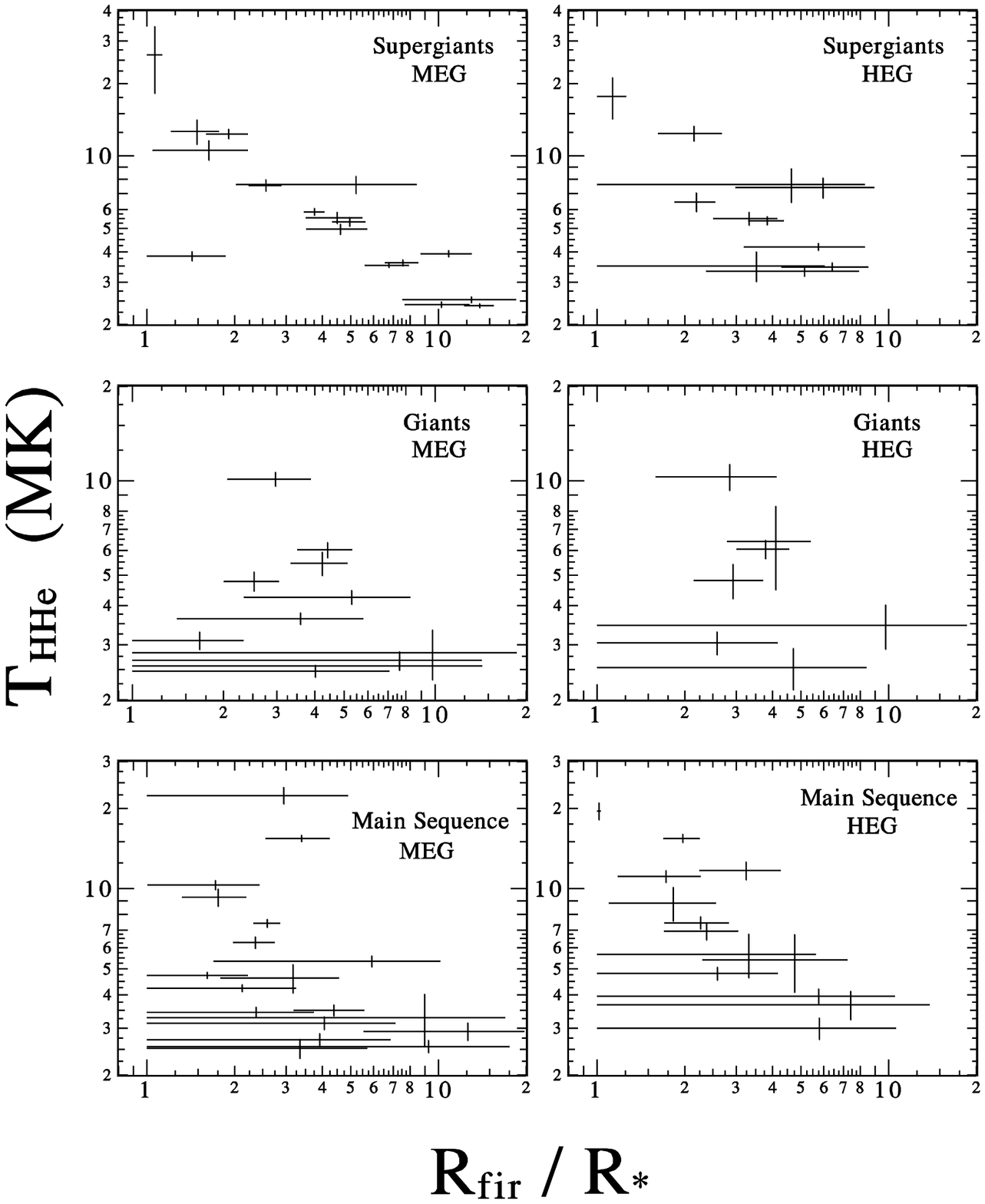}
\caption{MEG and HEG scatter plots showing the dependence of \TX\ (MK) (determined from
\HtoHe\ line ratios) on the He-like \fir-inferred radii, \Rfir\ (determined from He-like \ftoi\ line
ratios), for all OB luminosity classes. \Rfir\ data that only have lower limits or an uncertainty $ >
10 \Rstar$ and upper limit \TX\ data are not shown (see Tables 3, 4, 5, and 6).
\label{fig:THHEFIR}}
\end{figure}

\end{document}